\documentclass[aps,floatfix,superscriptaddress,nofootinbib,,tightenlines,longbibliography,preprintnumbers,11pt]{revtex4-1}
\usepackage{babel}
\usepackage{natbib}
\usepackage{amssymb,amsthm,amsfonts,braket,amsmath,graphicx,color,dsfont, mathrsfs}
\usepackage{array}
\newcolumntype{C}[1]{>{\centering\let\newline\\\arraybackslash\hspace{2pt}}m{#1}}
\usepackage{bm}
\usepackage[export]{adjustbox}
\usepackage{enumerate}
\usepackage{adjustbox}
\usepackage{xfrac}
\usepackage{tikz}
\usepackage{quantikz}
\usepackage{url}
\usepackage{xcolor}
\usepackage{multirow}
\usepackage{rotating,booktabs}
\usepackage[verbose]{placeins}
\usepackage{ragged2e}
\usepackage[titletoc]{appendix}
\usepackage{slashed}
\usepackage{comment}
\usepackage{outlines}
\usepackage[colorlinks=true,citecolor=blue,linkcolor=blue
]{hyperref}
\usepackage{soul}
\usepackage{enumitem}
\usepackage{float}
\setlist[enumerate]{label*=\arabic*.}
\usepackage{amsmath}
\usepackage{nameref}

\usepackage[capitalise,nameinlink]{cleveref}
\Crefname{lemma}{Lemma}{Lemmas}
\Crefname{proposition}{Proposition}{Propositions}
\Crefname{definition}{Definition}{Definitions}
\Crefname{theorem}{Theorem}{Theorems}
\Crefname{conjecture}{Conjecture}{Conjectures}
\Crefname{corollary}{Corollary}{Corollaries}
\Crefname{example}{Example}{Examples}
\Crefname{item}{Property}{Properties}
\Crefname{remark}{Remark}{Remarks}
\Crefname{exercise}{}{Exercise}

\newtheorem{theorem}{Theorem}

\newtheorem{exercise}[theorem]{Exercise}

\usepackage{tcolorbox}
\tcbuselibrary{breakable}
\tcbuselibrary{skins}
\tcbset{skin=enhanced,lines before break=4}
\usepackage{xcolor}
\definecolor{mygray}{gray}{0.935}

\begin{document}

\dimen\footins=5\baselineskip\relax

\preprint{UMD-PP-025-06}

\title{
{\small{TASI/CERN/KITP Lecture Notes on:}}
\\
Toward Quantum Computing Gauge Theories of Nature
}

\author{Zohreh Davoudi}
\email{davoudi@umd.edu}
\affiliation{Department of Physics and Maryland Center for Fundamental Physics (MCFP), University of Maryland, University of Maryland, College Park, MD 20742 USA}
\affiliation{Joint Center for Quantum Information and Computer Science (QuICS), National Institute of Standards and Technology (NIST) and University of Maryland, College Park, MD 20742 USA}
\affiliation{The NSF Institute for Robust Quantum Simulation, University of Maryland, College Park, Maryland 20742, USA}

\begin{abstract}
A hallmark of the computational campaign in nuclear and particle physics is the lattice-gauge-theory program. It continues to enable theoretical predictions for a range of phenomena in nature from the underlying Standard Model. The emergence of a new computational paradigm based on quantum computing, therefore, can introduce further advances in this program. In particular, it is believed that quantum computing will make possible first-principles studies of matter at extreme densities, and in and out of equilibrium, hence improving our theoretical description of early universe, astrophysical environments, and high-energy particle collisions. Developing and advancing a quantum-computing based lattice-gauge-theory program, therefore, is a vibrant and fast-moving area of research in theoretical nuclear and particle physics.

These lecture notes introduce the topic of quantum computing lattice gauge theories in a pedagogical manner, with an emphasis on theoretical and algorithmic aspects of the program, and on the most common approaches and practices, to keep the presentation focused and useful. Hamiltonian formulation of lattice gauge theories is introduced within the Kogut-Susskind framework, the notion of Hilbert space and physical states is discussed, and some elementary numerical methods for performing Hamiltonian simulations are discussed. Quantum-simulation preliminaries and digital quantum-computing basics are presented, which set the stage for concrete examples of gauge-theory quantum-circuit design and resource analysis. A step-by-step analysis is provided for a simpler Abelian gauge theory, and an overview of our current understanding of the quantum-computing cost of quantum chromodynamics is presented in the end. Examples and exercises augment the material, and reinforce the concepts and methods introduced throughout.

These lecture notes were initially presented at the 2024 Theoretical Advanced Study Institute in Particle Theory (TASI) school on ``The Frontiers of Particle Theory'' in Boulder, Colorado in June 2024. They were later expanded for lectures presented at the CERN school on ``Continuum Foundations of Lattice Gauge Theories'' in Geneva, Switzerland in July 2024, and at the Kavli Institute for Theoretical Physics (KITP) program on ``What is Particle Theory?'' in Santa Barbara, California in January 2025. Subsequent improvements were enabled by lectures presented at the Department of Energy's LGT4HEP Traineeship program in November 2024, the Origins Cluster at the Technical University of Munich in Garching, Germany in April 2025, and the joint Doctoral Training Program (DTP) and Training in Low-energy Nuclear Physics (TALENT) course at the European Center for Theoretical Studies in Nuclear Physics and Related Areas (ECT*) in Trento, Italy in June 2025.

\end{abstract}

\maketitle


\tableofcontents

\section{Introduction
\label{sec:introduction}
}
\noindent
Computation takes a central part in theoretical nuclear and particle physics, enabling studies of systems from the tiniest scales (subatomic world) to astronomical scales (neutron stars and cosmos). Gauge theories constitute the backbone of particle physics, the Standard Model~\cite{weinberg2004making,oerter2006theory,aitchison2012gauge}, and potentially models beyond the Standard Model~\cite{brivio2019standard,langacker2017standard,svetitsky2018looking,usqcd2019lattice}. First-principles studies of nature rooted in the Standard Model are of paramount importance, and involve some of the largest and most complex computations in any scientific discipline. Such computations, enabled by lattice-gauge-theory methods~\cite{wilson1974confinement,creutz1983monte,montvay1994quantum,rothe2012lattice,gattringer2009quantum}, have long taken advantage of the newest and most advanced computing technology in the market, from hardware architectures to software and algorithms. They have enabled a plethora of confirmations, predictions, and explorations to date, covering phenomena in single-hadron physics, multi-hadron physics, properties of matter in thermal equilibrium, and beyond-the-Standard-Model studies~\cite{aoki2024flag,usqcd2019hot,davoudi2021nuclear,davoudi2022report,kronfeld2022lattice}. Such methods, nonetheless, rely on Monte-Carlo sampling techniques in imaginary time. They are hindered by statistical noise overshadowing the signal in several scenarios~\cite{troyer2005computational,goy2017sign,nagata2022finite,gattringer2016approaches,cohen2015taming}: when a finite baryon density is introduced (in studies of matter's phase diagram under the strong interactions), or when real-time dynamical phenomena and observables are studied (in the universe evolution or in the aftermath of high-energy particle colliders). It is, therefore, conceivable that one may have to adopt radically different computing strategies to make progress in these settings, as even the Moore's law increase in supercomputing power will not be able to fight off the resource requirements of these problems, which are exponential in system size, and hence intractable.

Quantum-information-processing advances have brought excitement to scientific disciplines, and promise tackling previously intractable computational problems. Lattice gauge theorists, therefore, have naturally taken this possibility seriously, and are developing knowledge, tools, and algorithms to use quantum-simulation and quantum-computing technologies to advance the frontier of this program~\cite{banuls2020simulating,klco2022standard,bauer2023quantumnature,bauer2023quantum,di2024quantum,catterall2022report,beck2023quantum}. Quantum computers are believed to be suitable for simulating quantum systems: such computers use quantum degrees of freedom, and leverage quantum operations rooted in superposition and entanglement, to store information and process them. In other words, having access to the full Hilbert space of quantum-mechanical bits, or qubits, provides exponentially larger storage capacity compared to classical bits. Moreover, the ability to transform one state in this Hilbert space to another using quantum operations, or gates, in an efficient and scalable manner (i.e., polynomially costly in system size), makes highly complex operations tractable.

Hamiltonian simulation is the most natural framework for simulating physical models, including gauge theories of nature. The reason is that quantum simulators/computers can encode states, and can implement unitary operations, including the real-time evolution operator of quantum mechanics~\cite{lloyd1996universal}. There are two non-trivial steps toward quantum simulating  gauge theories. First is developing a Hamiltonian framework for gauge theories, which can be complicated, since such a formulation needs to be finite-dimensional, be efficiently mapped to a finite set of quantum degrees of freedom, and recover the continuum infinite-dimensional physics correctly. In fact, one faces many such Hamiltonian-formulation choices, and the optimal choice needs to be decided based on various theoretical and computational factors. Second is turning those Hamiltonian formulations into the quantum-computer language, translate the unitary operations of interest to a set of finer implementable quantum operations on the device, and to figure out what quantities can be accessed in these computations and how. These quantum-circuit constructions can vary depending on the hardware simulation mode, available elementary operations and their computational cost, and the complexity of the Hamiltonians and degrees of freedom.

These lecture notes are an attempt to introduce all the above considerations, through concrete examples and exercises, while providing a broader view of the field. They are, by no means, complete but are hopefully sufficiently focused and pedagogical. Studying these notes should equip the reader with at least one set of Hamiltonian gauge theories and their basic properties, and with some standard algorithms and methods to simulate them using quantum computers. The focus will be on digital quantum computations, rather than analog quantum simulation, but the distinction between these simulation modes will be sufficiently explained. Hamiltonian formulation of gauge theories will be introduced in Sec.~\ref{sec:Hamiltonian} and quantum computation of the developed Hamiltonians will be discussed in Sec.~\ref{sec:simulation}. We conclude in Sec.~\ref{sec:summary} with a summary and outlook, including enumerating the topics that did not make it to these notes, plus additional references for interested reader.

\section{Hamiltonian formulation of gauge theories
\label{sec:Hamiltonian}
}
\begin{figure}[t!]
    \centering
    \includegraphics[scale=0.75]{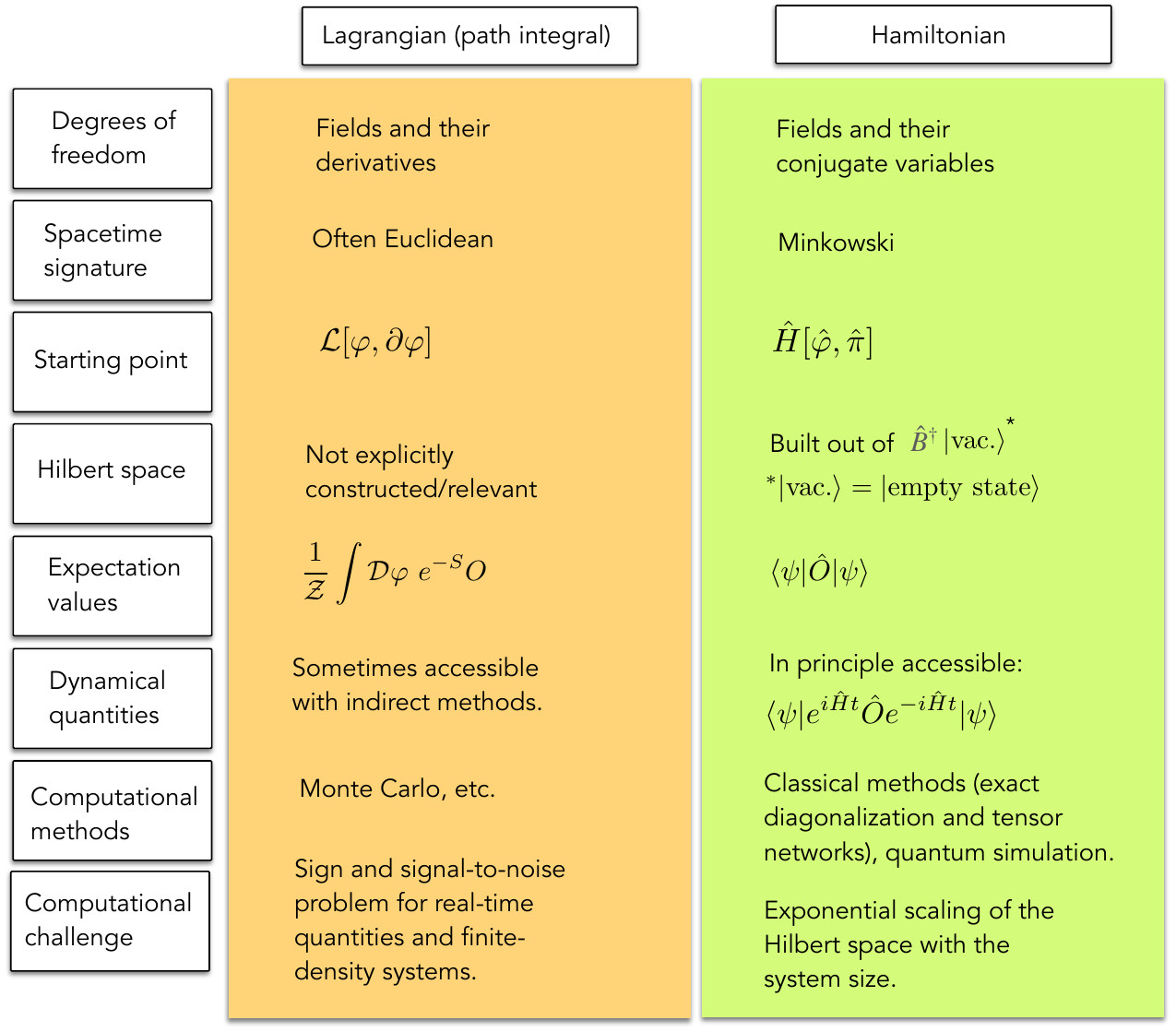}
    \caption{Key differences between the Lagrangian (path-integral) and Hamiltonian-based simulations of gauge theories. $\mathcal{L}$ denotes the Lagrangian density, $S$ the (Euclidean) action, $\hat{H}$ the Hamiltonian operator, $\varphi$ ($\hat{\varphi}$) the field variable (field operator), $\pi$ ($\hat{\pi}$) the conjugate-field variable (conjugate-field operator), $O$ ($\hat{O}$) a generic variable (operator), $\mathcal{Z}$ the partition function, i.e., $\mathcal{Z}=\int \mathcal{D} \varphi e^{-S}$, $\hat{B}^\dagger$ a generic creation operator for states in the Hilbert space, and $\ket{\psi}$ a generic state vector.}
    \label{fig:L-vs-H}
\end{figure}

\noindent
While the conventional lattice-gauge-theory program is built upon the Lagrangian/path-integral formulation of gauge theories, the quantum-simulation program is most naturally implemented via the Hamiltonian formulation. Features of each formulation are summarized in Fig.~\ref{fig:L-vs-H}.

The starting point of any quantum simulation, therefore, is to construct a Hamiltonian operator, then build the Hilbert space on which the Hamiltonian and other operators act on. This Hilbert space needs to be mapped to a quantum computer, and the action of operators needs to be encoded as quantum-gate operations. Since there are many basis choices to represent the Hamiltonian, and that all these obtain the same physical observables, we should decide what Hamiltonian formulation and what basis states to work with before proceeding to develop quantum algorithms for them. For gauge theories, in particular, the landscape of options is diverse, and those options amount to different computational complexities for Hamiltonian simulation. Let us start discussing at least one Hamiltonian framework, which is also one of the most commonly used to date, and enumerate a few basis choices for gauge theories of relevance to nature. Other formulations will not be covered in these lectures, but if you are curious, you can check out brief descriptions of each formulation in Ref.~\cite{bauer2023quantum}.
%

\subsection{Kogut-Susskind Hamiltonian of lattice gauge theories
\label{sec:KS}
}
The goal is to simulate a gauge theory in continuous spacetime. However, simulation resources are finite in practice, and we need finite countable degrees of freedom. A first step toward a finite countable Hilbert space is to define the quantum fields on a discrete spatial finite volume. The time is merely a parameter in the Hamiltonian formulation and is kept continuous. As with the Lagrangian formulation of lattice gauge theories, there are more than one way to define a discretized quantum field theory, all of which should yield the same continuum theory. Nonetheless, each discretization scheme may lead to different results at any finite value of lattice spacing. Primary examples are a variety of lattice fermions~\cite{kaplan2009chiral}, and compact versus non-compact formulations of lattice gauge theories.

A quite popular choice when it comes to Hamiltonian simulation is the Kogut-Susskind ``staggered'' formulation~\cite{kogut1975hamiltonian} with a compact formulation of gauge degrees of freedom. Let us start with a simple Abelian case and derive the staggered-lattice formulation, then introduce non-Abelian theories.

\subsubsection{U(1) lattice gauge theory
\label{sec:U(1)}
}

Consider a theory of free Dirac fermions in the continuum, before introducing any gauge fields. The Hamiltonian density can be derived from the Lagrangian density of Dirac fermions using a standard Legendre transform. It has the form:
\begin{align}
\mathcal{H}_f(x)=-i\overline{\psi}(x)\gamma^i \partial_i \psi(x) + m \overline{\psi}(x) \psi(x).
\label{eq:H-free-Dirac}
\end{align}
Here, $\psi$ is the multi-component Dirac field, $\gamma^i$ are spatial Gamma matrices (in Minkowski space), and $m$ is the fermion mass. To reduce clutter, here and in the rest of these notes, we drop the operator-hat notation unless when it helps with clarity.

To be concrete, let us consider the simpler case of fermions in $1+1$ dimensions (D), in which case we assign:
\begin{align}
\gamma^0=\begin{pmatrix}
1 & 0 \\
0 & -1
\end{pmatrix}=\sigma^{\mathbf{z}},~
\gamma^1=\begin{pmatrix}
0 & 1 \\
-1 & 0
\end{pmatrix}=i\sigma^{\mathbf{y}}.
\label{eq:xx}
\end{align}
With the two-component Dirac field defined by $\psi=\begin{pmatrix}
\psi_e \\
\psi_o
\end{pmatrix}$, the Hamiltonian density becomes:
\begin{align}
\mathcal{H}_f^{1+1}=-i\psi^\dagger\gamma^0\gamma^1 \partial_1 \psi + m \psi^\dagger \gamma^0 \psi=-i(\psi_e^\dagger \partial_1 \psi_o+\psi_o^\dagger \partial_1 \psi_e)+m(\psi_e^\dagger \psi_e-\psi_o^\dagger \psi_o),
\label{eq:xx}
\end{align}
where for brevity, the spacetime dependence of fields is dropped.

Let us now define the theory on a spacial lattice with the lattice spacing $a$, and employ the finite-difference form $\partial_1\psi_{e/o} \to \frac{\psi_{e/o}(x)-\psi_{e/o}(x-a)}{a}$. We further perform a field rescaling $\psi_{e/o} \to \sqrt{a} \, \psi_{e/o}$ to render the fields dimensionless. The Hamiltonian, thus, reads:
\begin{align}
H_f^{1+1}=\frac{i}{2a}\sum_x\left[\psi_e^\dagger(x-a) \psi_o(x)+\psi_o^\dagger(x-a) \psi_e(x) - \text{H.c.}\right]+m\sum_x\left[\psi_e^\dagger(x) \psi_e(x)-\psi_o^\dagger(x) \psi_o(x)\right].
\label{eq:xx}
\end{align}
The equations of motion for $\psi_e$ and $\psi_o$ are, respectively,\footnote{The $+/-$ notation in the first term in the left-hand side correspond to the $0/1$ eigenvalues of $\psi_{e/o}(x)^\dagger \psi_{e/o}(x)$.}
\begin{subequations}
\begin{align}
&\frac{d}{dt}\psi_e(x)=i[H_f^{1+1},\psi_e(x)]=\mp \frac{\psi_o(x+a)-\psi_o(x-a)}{2a}\mp m \psi_e(x),\\
&\frac{d}{dt}\psi_o(x)=i[H_f^{1+1},\psi_o(x)]=\mp \frac{\psi_e(x+a)-\psi_e(x-a)}{2a}\pm m \psi_o(x).
\end{align}
\label{eq:xx}
\end{subequations}
The idea of the Kogut-Susskind formulation is that instead of using $H_f^{1+1}$ in terms of $\psi=\begin{pmatrix}
\psi_e \\
\psi_o
\end{pmatrix}$, one can introduce a Hamiltonian in terms of a single-component fermion field $\varphi$. Explicitly, $\varphi$ represents either $\psi_o$ or $\psi_e$ according to the following map:
\begin{align}
\begin{cases}
\varphi(x)=\psi_e(x)~\text{if}~x~\text{even}, \\
\varphi(x)=\psi_o(x)~\text{if}~x~\text{odd}.
\end{cases}
\label{eq:phi-to-psi-mappin}
\end{align}
The new Hamiltonian in terms of the $\varphi$ field can be written as:
\begin{align}
H_\text{KS}^{1+1}=\frac{i}{2a}\sum_x\left[\varphi^\dagger(x) \varphi(x+a)-\varphi^\dagger(x+a) \varphi(x)\right]+m\sum_x(-1)^{x/a}\varphi^\dagger(x) \varphi(x).
\label{eq:xx}
\end{align}
This Hamiltonian yields the exact same equations of motion for field $\varphi$:
\begin{align}
&\frac{d}{dt}\varphi(x)=i[H_\text{KS}^{1+1},\varphi(x)]=\mp \frac{\varphi(x+a)-\varphi(x-a)}{2a}\mp m (-1)^{x/a} \varphi(x),
\label{eq:xx}
\end{align}
as the discretized $H_f^{1+1}$ yields for fields $\psi_{e/o}$, considering the mapping established in Eq.~\eqref{eq:phi-to-psi-mappin}. So while it looks like we left out some degrees of freedom (i.e., $\psi_o$ at even sites and $\psi_e$ at odd sites), the two theories in the continuum limit have the same spectrum. This is the reason as to why the staggering procedure removes the fermion doubling in a (1+1)D Hamiltonian formulation~\cite{kaplan2009chiral}. In higher dimensions, staggering does not fully remove the doublers, leaving us still with a number of the so-called ``tastes''.\footnote{Fermion doubling refers to the fact that naively discretizing Dirac fermions on a D-dimensional spacetime lattice, generates $2^{D}-1$ extra fermion species as one takes the continuum limit. Discretization on only a spatial lattice, as is the case in the Hamiltonian formulation, results in $2^{d}-1$ extra species, where $d=D-1$ is the space dimension. Solutions to the doubling problem exist~\cite{kaplan2009chiral}, and can be ported to the Hamiltonian formulation. Kogut-Susskind solution fixes this issue on a 1D spatial lattice but only partially fixes it in higher dimensions.} In the Kogut-Susskind formulation, the lattice spacing $a$ is effectively replaced by $2a$, since the theory is translationally invariant by translation of $2a$ units.

Let us now introduce gauge-invariant interactions for this staggered formulation. In the standard way, the non-local fermion bilinear needs to be parallel transported by a gauge-link operator:
\begin{align}
U_\mu(x)=e^{iagA_\mu(x)},~0 \leq agA_\mu(x) < 2\pi.
\label{eq:xx}
\end{align}
This operator originates from point $x$ along the link that joins points $x$ and $x+a\hat{\mu}$ along the direction $\hat{\mu}$. Due to gauge redundancy, we are free to fix the gauge fully or partially. A suitable choice is to partially fix the gauge such that $A_0=0$ at all sites (hence $U_0=1$) at all links. This is called a temporal gauge and was adopted by Kogut and Susskind. It is a natural and convenient choice: $A_0$ is a static quantity in Yang-Mills theories and can be set to a constant. Hence, we do not need to deal with the temporal component of the gauge field anymore. This choice in (1+1)D leaves us with only one gauge-(field) link component, ($A_1$) $U_1$, which we denote as ($A$) $U$ for simplicity. The (1+1)D Hamiltonian in presence of gauge-matter interactions reads:
\begin{align}
H_\text{KS}^{1+1}&=\frac{i}{2a}\sum_x\big[\varphi^\dagger(x) U(x) \varphi(x+a)-\varphi^\dagger(x+a) U^\dagger(x)\varphi(x)\big]+m\sum_x(-1)^{x/a}\varphi^\dagger(x) \varphi(x)\\
&\equiv H_\text{KS,h}^{1+1}+H_\text{KS,m}^{1+1}.
\label{eq:xx}
\end{align}

In the canonical quantization of gauge theories, the conjugate variable to the operator $agA$ is the electric field, $E(x)$:
\begin{align}
[agA(x),E(x')]=i\delta_{x,x'},
\label{eq:U(1)-algebra}
\end{align}
with dimensionless $A$ and $E$ fields. Since $agA$ is an angular variable, $E$ is the generator of cyclic translations in variable $agA$, hinting at a quantum-rotor representation of this theory. We come back to this point in Sec.~\ref{sec:basis} where we discuss the Hilbert space of the Kogut-Susskind theory.

\begin{figure}[t!]
    \centering
    \includegraphics[scale=0.625]{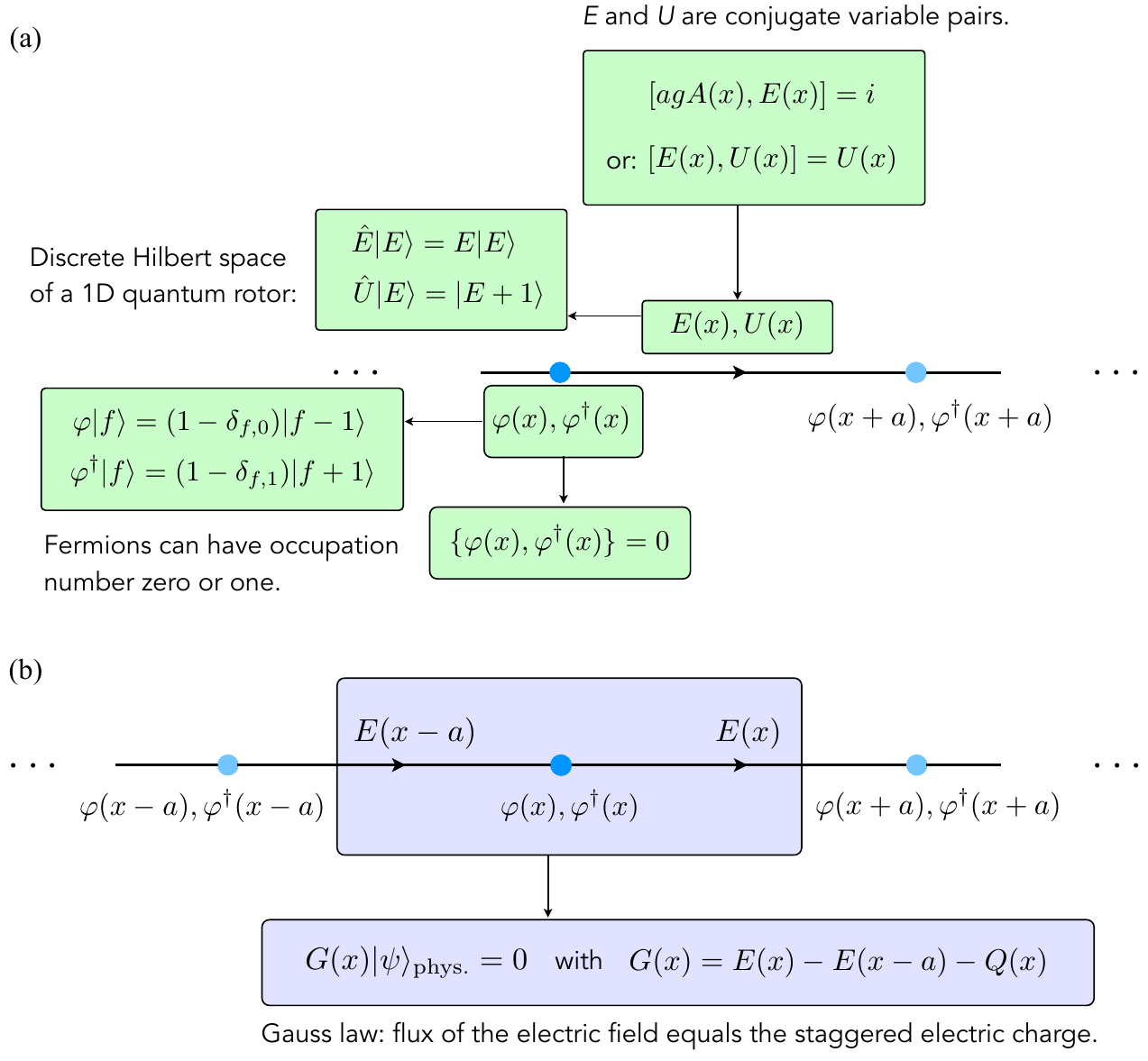}
    \caption{The building blocks of a (1+1)D U(1) lattice gauge theory in the Kogut-Susskind formulation. (a) Degrees of freedom and the associated operators, operator algebra, and operators action on the fermionic and electric-field basis states, on two representative sites of the staggered lattice. (b) The Abelian Gauss laws that needs to hold at each lattice sites. Staggered-charge operator $Q(x)$ is defined in Eq.~\eqref{eq:electric-charge}.}
    \label{fig:u1-lattice}
\end{figure}

The continuum quantum-electrodynamic (QED) Hamiltonian involves a term associated with the energy stored in the electric field. The corresponding lattice Hamiltonian is:
\begin{align}
H_\text{KS,E}^{1+1}=\frac{g^2a}{2}\sum_x E(x)^2.
\label{eq:xx}
\end{align}
In summary, the full Kogut-Susskind Hamiltonian in (1+1)D is:
\begin{align}
H_\text{KS}^{1+1}=H_\text{KS,h}^{1+1}+H_\text{KS,m}^{1+1}+H_\text{KS,E}^{1+1}.
\label{eq:KS-1p1}
\end{align}
The building blocks and key relation in the (1+1)D Kogut-Susskind U(1) lattice gauge theory are depicted in Fig.~\ref{fig:u1-lattice}. 

The Kogut-Susskind Hamiltonian can be generalized to higher dimensions. First, $H_\text{KS,h}$ remains almost the same as in (1+1)D, except there will be site-dependent phases $s_j(\bm{x})$ associated with fermion hopping along the $j$-th Cartesian directions (which depends on the particular choice of Gamma matrices):
\begin{align}
H_\text{KS,h}^{d+1}=\frac{i}{2a}\sum_{\bm{x},j}s_j(\bm{x})\left[\varphi^\dagger(\bm{x}) U_j(\bm{x}) \varphi(\bm{x}+a\hat{\bm{x}}_j)-\varphi^\dagger(\bm{x}+a\hat{\bm{x}}_j) U_j^\dagger(\bm{x})\varphi(\bm{x})\right].
\label{eq:xx}
\end{align}
Here, $j \in \{1,\cdots,d\}$, $\bm{x}=(x_1,\cdots,x_d)$, and $\hat{\bm{x}}_j=x_j/|\bm{x}|$. The mass Hamiltonian receives slight modification in its staggering factor:
\begin{align}
H_\text{KS,h}^{d+1}=m\sum_{\bm{x}} (-1)^{\sum_jx_j/a} \varphi^\dagger(\bm{x}) \varphi(\bm{x}).
\label{eq:xx}
\end{align}
The electric-field Hamiltonian follows the same form, except one needs to sum over electric-field contributions along all links:
\begin{align}
H_\text{KS,E}^{d+1}=\frac{g^2}{2a^{d-2}}\sum_{\bm{x},j} E_j(\bm{x})^2.
\label{eq:H-KS-E}
\end{align}
We have kept all the fields to be dimensionless, hence compensating by appropriate powers of lattice spacing to retain the mass dimensionality of the Hamiltonian.\footnote{The coupling $g$ has mass dimensionality $1$, $\frac{1}{2}$, and $0$ in (1+1)D, (2+1)D, and (3+1)D, respectively.}
Finally, and importantly, in ($d$+1)D with $d>1$, there is an additional contribution associated with the energy stored in the magnetic field. One form that recovers the continuum magnetic energy $\propto B^2$ is a ``plaquette'' Hamiltonian:
\begin{align}
H_\text{KS,B}^{d+1}=\frac{a^d}{2a^4g^2}\sum_{\bm{x},i,j} \text{Tr}[2-\mathcal{P}_{i,j}(\bm{x})-\mathcal{P}_{i,j}^\dagger(\bm{x})].
\label{eq:plaquette}
\end{align}
The plaquettes are the elementary squares on a square lattice, and the plaquette operator $\mathcal{P}_{i,j}(\bm{x})$ initiated at point $\bm{x}$ in the $(i,j)$-plane is defined as: $\mathcal{P}_{i,j}(\bm{x})=U_i(\bm{x})U_j(\bm{x}+a\hat{\bm{x}}_i)U_i^\dagger(\bm{x}+a\hat{\bm{x}}_j)U_j^\dagger(\bm{x})$. The trace is taken over the gauge-color space and so is trivial for the case of a U(1) lattice gauge theory. To connect this lattice Hamiltonian to the continuum one, one realizes the relation $\mathcal{P}_{i,j}(\bm{x})=e^{ia^{(3-d)/2}gB_{i,j}(\bm{x})}$ where $B_{i,j}(\bm{x})$ is the curl of the vector gauge field $A_i$ around the plaquette starting and ending at $\bm{x}$ in the $(i,j)$-plane. Expanding Eq.~\eqref{eq:plaquette} in small $a$ and taking the limit $a \to 0$ recovers the continuum magnetic Hamiltonian.

In summary, the ($d$+1)D Kogut-Susskind Hamiltonian for the U(1) lattice gauge theory consists of the four contributions above:
\begin{align}
H_\text{KS}^{d+1}=H_\text{KS,h}^{d+1}+H_\text{KS,m}^{d+1}+H_\text{KS,E}^{d+1}+H_\text{KS,B}^{d+1}.
\label{eq:KS-dp1}
\end{align}
Before trying to make sense of states and symmetries in this Abelian theory, let us introduce the Kogut-Susskind Hamiltonian for some relevant non-Abelian theories as well.

\subsubsection{SU(2) lattice gauge theory
\label{sec:SU(2)}
}
The non-Abelian case, luckily, is not radically different from what we have already constructed for the U(1) case. Consider the SU(2) lattice gauge theory coupled to one flavor of quarks. The fields $\varphi$, $\varphi^\dagger$, $U$ and $E$, notationally, are the same as in the Abelian case, nonetheless, they should be understood as following. At each site, the fermion field consists of two SU(2) color components:
\begin{align}
\varphi \to \begin{pmatrix}
\varphi_1 \\
\varphi_2
\end{pmatrix},~\varphi^\dagger \to (\varphi_1^\dagger~\varphi_2^\dagger),
\label{eq:xx}
\end{align}
assuming a fundamental representation. On each link, the gauge link is described by a $2 \times 2$ matrix:
\begin{align}
U \to \begin{pmatrix}
U_{1,1} & U_{1,2} \\
U_{2,1} & U_{2,2}
\end{pmatrix},
\label{eq:xx}
\end{align}
and the electric field is realized as three left and three right electric fields associated with each of the three ``gluons'' in the SU(2) theory:
\begin{align}
E \to E_L^a,E_R^a,~a \in \{1,2,3\}.
\label{eq:xx}
\end{align}
\begin{figure}[t!]
    \centering
    \includegraphics[scale=0.625]{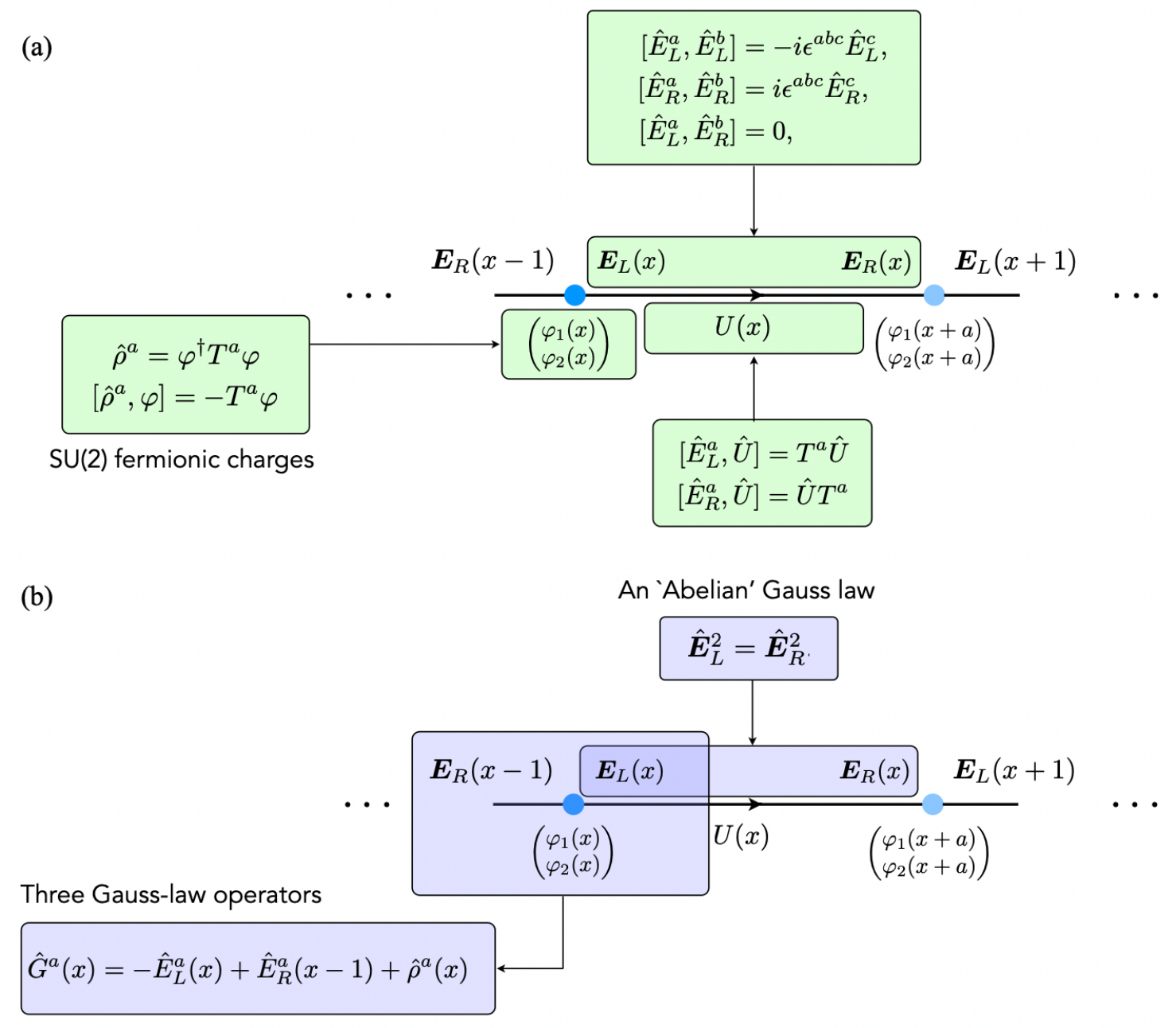}
    \caption{The building blocks of a (1+1)D SU(2) lattice gauge theory in the Kogut-Susskind formulation. (a) Degrees of freedom and the associated operators and operator algebra on two representative sites of the staggered lattice. (b) The Abelian and non-Abelian Gauss laws that need to hold on lattice links and lattice sites, respectively.}
    \label{fig:su2-lattice}
\end{figure}

The (anti)commutation algebra for these fields goes as follows. For fermions:
\begin{align}
&\{\varphi_\alpha(\bm{x}),\varphi_\beta^\dagger(\bm{x}')\}=\delta_{\bm{x},\bm{x}'}\delta_{\alpha,\beta},~\{\varphi_\alpha(\bm{x}),\varphi_\beta(\bm{x}')\}=0,~\{\varphi_\alpha^\dagger(\bm{x}),\varphi_\beta^\dagger(\bm{x}')\}=0,
\label{eq:xx}
\end{align}
with $\alpha,\beta \in \{1,2\}$, while for the gauge bosons:
\begin{align}
[E^a_L(\bm{x}),U(\bm{x}')]=T^aU(\bm{x})\delta_{\bm{x},\bm{x}'},~[E^a_R(\bm{x}),U(\bm{x}')]=U(\bm{x})T^a\delta_{\bm{x},\bm{x}'},
\label{eq:xx}
\end{align}
and
\begin{align}
&[E^a_L(\bm{x}),E_L^b(\bm{x}')]=-i\epsilon^{a,b,c}E_L^c(\bm{x})\delta_{\bm{x},\bm{x}'},~[E^a_R(\bm{x}),E_R^b(\bm{x}')]=i\epsilon^{a,b,c}E_R^c(\bm{x})\delta_{\bm{x},\bm{x}'},~[E^a_L(\bm{x}),E_R^b(\bm{x}')]=0.
\label{eq:xx}
\end{align}
The latter are the commutation algebra of the body- and space-frame angular momenta of a rigid rotor upon the identification $E_L \equiv -J_b$ and $E_R \equiv J_s$. We come back to this point in Sec.~\ref{sec:basis} where we attempt to build the Hilbert space of this theory.

The Kogut-Susskind Hamiltonian will have symbolically the same form as the U(1) case. Nonetheless, it will contain more terms compared to the U(1) Hamiltonian. For example, the mass term consists of contributions from the two components of the fermion field, since $\varphi^\dagger \varphi = \varphi_1^\dagger \varphi_1+\varphi_2^\dagger \varphi_2$. The interaction term contains terms of the form $\varphi_\alpha^\dagger U_{\alpha,\beta} \varphi_\beta$ and its Hermitian conjugate. The electric-field term is proportional to $(E_L)^2=(E^1_L)^2+(E^2_L)^2+(E^3_L)^2=(E_R)^2$. The magnetic Hamiltonian involves a trace over a $2 \times 2$ SU(2) matrix, which thus involves several terms.

\subsubsection{SU(3) lattice gauge theory
\label{sec:SU(3)}
}
Having worked out the case of a SU(2) lattice gauge theory, it is straightforward to recognize what needs to be modified to construct a SU(3) theory. At each site, the fermion field consists of three SU(3) color components:
\begin{align}
\varphi \to \begin{pmatrix}
\varphi_1 \\
\varphi_2 \\
\varphi_3
\end{pmatrix},~\varphi^\dagger \to (\varphi_1^\dagger~\varphi_2^\dagger~\varphi_3^\dagger),
\label{eq:xx}
\end{align}
assuming a fundamental representation. On each link, the gauge link is described by a $3 \times 3$ matrix:
\begin{align}
U \to \begin{pmatrix}
U_{1,1} & U_{1,2} & U_{1,3} \\
U_{2,1} & U_{2,2} & U_{2,3} \\
U_{3,1} & U_{3,2} & U_{3,3}
\end{pmatrix},
\label{eq:xx}
\end{align}
and eight left and eight right electric fields associated with each of the eight ``gluons'' in the SU(2) theory:
\begin{align}
E \to E_L^a,E_R^a,~a \in \{1,2,\cdots,8\}.
\label{eq:xx}
\end{align}
The (anti)commutation algebra is the same as in the SU(2) case upon the replacements $\tau^a \to \lambda^a$ [SU(3) Gell-Mann matrices] and $\epsilon^{a,b,c} \to f^{a,b,c}$ [SU(3) structure constants]. The Hamiltonian retains its form symbolically, but now clearly contains more terms when expanded out in terms of the SU(3) components of the fields.

\subsection{Basis choices, Hilbert space, local constraints
\label{sec:basis}
}
To compute observables, we need to identify states and the action of Hamiltonian on states. It is useful to start from a set of basis states and identify the action of Hamiltonian on such states first. Obviously, there are many choices. Here, we present a few common options for each of the gauge theories we enumerated so far. This step will set us up to proceed with learning how to compute observables in these theories using Hamiltonian-simulation strategies. 

\subsubsection{U(1) lattice gauge theory
\label{sec:U(1)-basis}
}
The staggered fermions are single-component fermions at each site. One can pick the on-site fermion-occupation-number eigenstates as the basis states: $\hat{N}_f({\bm{x}})\ket{f}_{\bm{x}} \equiv \hat{\psi}^\dagger ({\bm{x}})\hat{\psi}({\bm{x}})\ket{f}_{\bm{x}}=f_{\bm{x}}\ket{f}_{\bm{x}}$ with $f \in \{0,1\}$. Then:
\begin{align}
\begin{cases}
\psi(\bm{x})\ket{f}_{\bm{x}}=(1-\delta_{f_{\bm{x}},0})\ket{f-1}_{\bm{x}},\\
\psi^\dagger(\bm{x})\ket{f}_{\bm{x}}=(1-\delta_{f_{\bm{x}},1})\ket{f+1}_{\bm{x}}.
\end{cases}
\label{eq:xx}
\end{align}
Let us now recall that he Hilbert space of a U(1) gauge link is that of a quantum rotor, given the commutation algebra in Eq.~\eqref{eq:U(1)-algebra}. A common basis choice is the irreducible representation (irrep) or electric-field basis, which is unbounded but discrete. In this basis:
\begin{align}
\begin{cases}
\hat E(\bm{x})\ket{E}_{\bm{x}}=E_{\bm{x}}\ket{E}_{\bm{x}},\\
\hat U(\bm{x})\ket{E}_{\bm{x}}=\ket{E+1}_{\bm{x}}, \\
\hat U^\dagger(\bm{x})\ket{E}_{\bm{x}}=\ket{E-1}_{\bm{x}}.
\end{cases}
\label{eq:xx}
\end{align}
The relation for $U$ and $U^\dagger$ are obtained straightforwardly from the relation $[E(\bm{x}),U(\bm{x})]=U(\bm{x})$. Here, $E_{\bm{x}} \in \mathbb{Z}$, nonetheless in practice, the electric eigenbasis needs to be truncated to host up to a large but finite eigenvalue: $|E_{\bm{x}}| \leq \Lambda$. The truncated on-link Hilbert space has, therefore, $2\Lambda+1$ elements. While the commutation relation between $E$ and $U$ remains intact with this truncation, the Abelian relation $[U(\bm{x}),U^\dagger(\bm{x})]=0$ breaks down when acting on $\ket{\pm \Lambda}$. The reason is that the raising/lowering operators acting on the truncated states are modified as:
\begin{align}
\begin{cases}
\hat U(\bm{x})\ket{E}_{\bm{x}}=(1-\delta_{E_{\bm{x}},\Lambda})\ket{E+1}_{\bm{x}}, \\
\hat U^\dagger(\bm{x})\ket{E}_{\bm{x}}=(1-\delta_{E_{\bm{x}},-\Lambda})\ket{E-1}_{\bm{x}}.
\end{cases}
\label{eq:xx}
\end{align}
Such a modification, therefore, introduces an error that needs to be systematically controlled and removed at the end of the computation. In fact, analytical relations have been developed to study the error made due to this truncation, see e.g., Refs.~\cite{jordan2012quantum,tong2022provably}. 

The above relations define the action of the Hamiltonian on a state of the form: $\ket{\psi}= \cdots \ket{f}_{\bm{x}} \ket{E}_{\bm{x}} \otimes \ket{f}_{\bm{x}+a\hat{\bm{x}}_1} \ket{E}_{\bm{x}+a\hat{\bm{x}}_1}\otimes \ket{f}_{\bm{x}+a\hat{\bm{x}}_2} \ket{E}_{\bm{x}+a\hat{\bm{x}}_2}\cdots$. For a system with $N_f$ fermion sites and $N_\ell$ links, the number of basis states spanning the full Hilbert space is, therefore, $2^{N_f} \times (2\Lambda+1)^{N_\ell}$. However, many of these states turned out to be unphysical, hence irrelevant. Let us see why.

Recall the QED Hamiltonian density in the continuum:
\begin{align}
\mathcal{H}_\text{QED}=-i\overline{\psi} \gamma^i D_i \psi +m \overline{\psi} \psi +\frac{1}{2}(\bm{E}^2+\bm{B}^2)-A_0(\partial_i E^i - g \psi^\dagger \psi),
\label{eq:QED-H}
\end{align}
where we have reverted to dimensionful fields momentarily. Clearly, $A_0$ is not dynamical and only serves as a Lagrange multiplier for the constant of motion, which is the Gauss-law equation of motion: $\partial_iE^i-g\psi^\dagger \psi=0$. Recall that we decided to work in a gauge in which $A_0=0$. This removes any knowledge of the Gauss-law constraint on the states. Thus, the constraint needs to be imposed on the Hilbert space \emph{a posteriori}. Nonetheless, the remainder Hamiltonian commutes with the Gauss-law operator, hence starting in a physical state, the evolution remains constrained to the physical sector (unless approximations made and the associated errors break the Gauss law).

\tcbset{colframe=black!10!black,colback=mygray,arc=1mm}
\begin{tcolorbox}[breakable]
\noindent
\begin{exercise}
\label{Exercise:H-QED} This exercise will refresh your memory on how to derive the QED Hamiltonian density from the QED Lagrangian density in the continuum. Consider the continuum QED Lagrangian density in ($d$+1)D dimensions for $d>1$:
\begin{align}
\mathcal{L}_\text{QED}=\overline{\psi} i\gamma^\mu D_\mu \psi -m \overline{\psi} \psi +\frac{1}{4}F_{\mu\nu}F^{\mu\nu},
\label{eq:xx}
\end{align}
where we have defined the covariant derivative $D_\mu=\partial_\mu+igA_\mu$ and the field-strength tensor $F_{\mu\nu}=\partial_\mu A_\nu-\partial_\nu A_\mu$.

\vspace{0.25 cm}
\noindent
\textbf{Part (a)} Determine the conjugate-momentum variable to the $A_i$ and $A_0$ fields.

\vspace{0.25 cm}
\noindent
\textbf{Part (b)} Use a Legendre transform to convert the Lagrangian density to a Hamiltonian density given the conjugate variables you determined in part (a).

\vspace{0.25 cm}
\noindent
\textbf{Part (c)} Using the definitions of the electric and magnetic fields, i.e., $\bm{E}=-\frac{\partial \bm{A}}{\partial t}-\bm{\nabla} A_0$ and $\bm{B}=\bm{\nabla} \times \bm{A}$, simplify the Hamiltonian such that it yields the form in Eq.~\eqref{eq:QED-H}.
\vspace{0.25 cm}
\end{exercise}
\end{tcolorbox}

In the Kogut-Susskind Hamiltonian formulation, the Gauss-law operator turns into:
\begin{align}
G(\bm{x})=\sum_{j}\left[E_j(\bm{x})-E_j(\bm{x}-a\hat{\bm{x_j}})\right]-Q(\bm{x}),
\label{eq:xx}
\end{align}
which restores the Gauss law in the continuum with
\begin{align}
Q(\bm{x})=-\psi^\dagger(\bm{x})\psi(\bm{x})+\frac{1-(-1)^{\sum_jx_j/a}}{2}.
\label{eq:electric-charge}
\end{align}
The rationale behind the form of $Q$ becomes clearer once we work out an example. Physical states are those that are annihilated by $G$ at all $\bm{x}$:
\begin{align}
G(\bm{x})\ket{\psi}_\text{phys}=0,~\forall \bm{x}.
\label{eq:xx}
\end{align}
Note that $[G(\bm{x}),H_\text{KS}^{d+1}]=0$ and that $G$ has integer eigenvalues.

\tcbset{colframe=black!10!black,colback=white,arc=1mm}
\begin{tcolorbox}[breakable]
\textbf{Example 1:} What are the physical states of the U(1) lattice gauge theory in (1+1)D coupled to staggered fermions with $N=2$ staggered-lattice sites, an electric-field cutoff $\Lambda=1$, and with periodic boundary conditions? What is the ground state of the theory in the limit of strong coupling, i.e., $ag \to \infty$?

\vspace{0.25 cm}
\textbf{Solution:} Naively, there are $2^2 \times 3^2 = 36$ states. Nonetheless, only 5 of these satisfy the Gauss laws (the site-index subscripts are in units of lattice spacing):
\begin{align}
\begin{cases}
\ket{\psi_1} = (\ket{0}_0 \ket{-1}_0) \otimes (\ket{1}_1 \ket{-1}_1), \\
\ket{\psi_2} = (\ket{0}_0 \ket{1}_0) \otimes (\ket{1}_1 \ket{0}_1), \\
\ket{\psi_3} = (\ket{0}_0 \ket{1}_0) \otimes (\ket{1}_1 \ket{1}_1), \\
\ket{\psi_4} = (\ket{1}_0 \ket{0}_0) \otimes (\ket{0}_1 \ket{1}_1), \\
\ket{\psi_5} = (\ket{1}_0 \ket{-1}_0) \otimes (\ket{0}_1 \ket{0}_1). \\
\end{cases}
\label{eq:xx}
\end{align}
Here, we have used the staggered-charge formula to assign a U(1) charge to each site. Clearly, even sites either hold an electron with charge -1 (if filled) or lack an electron (if unfilled). In contrary, the odd sites hold a positron with charge $1$ (if unfilled) ar lack one (if filled). The even and odd sites, therefore, represent fermion and antifermion sites. The total charge, which is a conserved quantity, is zero, which is the only charge sector allowed by periodic boundary conditions.

The strong-coupling ground state (or vacuum) is the lowest eigenstate in the limit in which the electric-field term dominates. No electric excitations are, therefore, preferred in the ground state. The mass term yields non-zero contribution to the energy in this limit. So the even sites better be unoccupied while the odd site occupied, such that the expectation value of $H_\text{KS,m}^{1+1}$ takes its least value, i.e., $m (-1)^0 \times 0 +m (-1)^1 \times 1 = -m$.  
This is consistent with no electron and no positron excitations on the even and odd sites, which is marked as $\ket{\psi_2}$ in the example above. So in the strong-coupling vacuum, no matter or gauge particles are present. The interacting vacuum, i.e., the ground state for any finite $g$, on the other hand, is a non-trivial superposition of several physical basis states, and has thus non-trivial particle content.
\end{tcolorbox}

The example above involves the case of periodic boundary conditions. How about open boundary conditions, in which the electric field entering the lattice is fixed while that leaving the lattice is left open? In a (1+1)D theory with such boundary conditions, gauge degrees of freedom are not dynamical and can be fully eliminated~\cite{hamer1997series,martinez2016real,davoudi2021search}. To see this, consider the Kogut-Susskind Hamiltonian terms in (1+1)D, Eq.~\eqref{eq:KS-1p1}. One can first apply a gauge transformation of the form:
\begin{align}
    &\varphi(x) \to \varphi'(x)=\left[\prod_{y<x}U(y)\right]\varphi(x),\\
    &\varphi^\dagger(x) \to {\varphi^\dagger}'(x)=\left[\prod_{y<x}U(y)\right]{\varphi^\dagger}'(x),\\
    &U(x) \to U'(x)= \left[\prod_{y<x}U(y)\right] U(x) \left[\prod_{z \leq x}U(z)\right]=\mathbb{I},
    \label{eq:xx}
\end{align}
which, therefore, sets all the gauge-link operators along the lattice to unity. The Hilbert space spanned by the transformed fermionic fields, $\varphi'$ and ${\varphi^\dagger}'$, occupation-number states is equivalent to that of untransformed fields, since these operators still have fermionic nature, and their associated occupation quantum numbers are either zero or one. We will, therefore, drop the prime notation in the following for simplicity. Next, the Gauss law can be used to eliminate the dependence on the electric-field operators $E(x)$ in exchange for fermionic operators:
\begin{align}
    &E(0)=\epsilon_0+Q(0), \nonumber\\
    &E(a)=E_0+Q(a)=\epsilon_0+Q(0)+Q(a),\nonumber\\
    & \vdots \nonumber\\
    &E(x)=\epsilon_0+\sum_{y=0}^xQ(y)=\epsilon_0+\sum_{y=0}^x\left[-\varphi^\dagger(y)\varphi(y)+\frac{1-(-1)^{y/a}}{2} \right],
    \label{eq:xx}
\end{align}
where $\epsilon_0$ is the incoming electric field and $Q(x)$ is the one-dimensional version of the staggered electric charge defined in Eq.~\eqref{eq:electric-charge}. The final Hamiltonian, in terms of the transformed fields, has the form:
\begin{align}
    H = \frac{i}{2a}\sum_{x=0}^{N-2}\big[\varphi^\dagger(x) \varphi(x+a)-&\varphi^\dagger(x+a)\varphi(x)\big]+m\sum_{x=0}^{N-1}(-1)^{x/a}\varphi(x)^\dagger \varphi(x)\,+\nonumber\\
    &\frac{g^2a}{2}\sum_{x=0}^{N-2}\left\{\epsilon_0+\sum_{y=0}^x\left[-\varphi^\dagger(y)\varphi(y)+\frac{1-(-1)^{y/a}}{2} \right]\right\}^2.
    \label{eq:fully-ferm-I}
\end{align}
Such a procedure can be applied to the (1+1)D theory with periodic boundary conditions, but a remnant dynamical gauge degrees of freedom survives~\cite{davoudi2025quantum}.

What other bases are there beside the electric-field basis?
Let us focus on the (2+1)D example of a U(1) lattice gauge theory, such that gauge-field degrees of freedom remain dynamical. We further consider a theory of pure gauge fields, i.e., in the absence of matter fields, to simplify the discussions below. The Hamiltonian is:
\begin{align}
H^{2+1}_{U(1)}=\frac{g^2}{2} \sum_l \bm{E}_l^2+\frac{1}{2a^2g^2}\sum_p(2-\mathcal{P}_p-\mathcal{P}_p^\dagger),
\label{eq:xx}
\end{align}
where the shorthand notation $l$ (link) and $p$ (plaquette) are used to replace the site and directionality dependence of the link and plaquettes. Define $\bm{E}_l=\bm{E}_l^T+\bm{E}_l^L$, where:
\begin{align}
\nabla\cdot \bm{E}_l^T=0,~\nabla \cdot \bm{E}_l^L=Q=0.
\label{eq:xx}
\end{align}
Here, the latter relation is due to the absence of charges in this example. Furthermore, define the rotor field $R_p$ via the relation $\bm{E}_l^L=\nabla \times R_p$. If we can rewrite the electric Hamiltonian only in terms of the rotor field, that is, if we solve for $E_l^L$ using the Gauss laws, we will be left with no redundancies\footnote{With the exception of a  magnetic Gauss’s law when periodic boundary conditions are imposed (which sets the total magnetic flux through the system to zero).} and can proceed more efficiently in subsequent computations. This is possible, and the Hamiltonian can be shown to turn to~\cite{bender2020gauge}:
\begin{align}
H^{2+1}_{U(1)}=\frac{g^2}{2} \sum_p (\nabla \times R_p)^2+\frac{1}{2a^2g^2}\sum_p(2-\mathcal{P}_p-\mathcal{P}_p^\dagger).
\label{eq:H-rotor}
\end{align}

To build a set of basis states, we first note that $R_p$ is the conjugate variable to $a^{1/2}gB_p$:
\begin{align}
[a^{1/2}gB_p,R_{p'}]=i\delta_{p,p'}.
\label{eq:xx}
\end{align}
Recall that $\mathcal{P}_p=e^{ia^{1/2}gB_p}$ and so $0 \leq a^{1/2}gB_p<2\pi$ is an angular variable. Therefore, $R$ is the generator of cyclic translations in variable $a^{1/2}gB$, hence the name ``rotor''. The rotor eigenvalues are discrete and unbounded. In the rotor basis:
\begin{align}
\begin{cases}
\hat R_p \ket{r}_p = r_p \ket{r}_p, \\
\hat{\mathcal{P}}_p \ket{r}_p =\ket{r+1}_p, \\
\hat{\mathcal{P}}_p^\dagger \ket{r}_p =\ket{r-1}_p,
\end{cases}
\label{eq:xx}
\end{align}
where the latter is a consequence of the commutation relation $[\mathcal{P}_p,R_p]=\mathcal{P}_p$ and $r_p \in \mathbb{Z}$.

We can also define a magnetic basis such that the action of the magnetic-field operator is diagonal in this basis:
\begin{align}
a^{1/2}g\hat B_p \ket{b}_p=b_p\ket{b}_p,~0 \leq b_p < 2\pi.
\label{eq:xx}
\end{align}

The magnetic (dual) and the rotor bases are related by a Fourier transform:
\begin{align}
\ket{b}_p=\sum_{r_p \in \mathbb{Z}}e^{ib_pr_p}\ket{r}_p.
\label{eq:xx}
\end{align}
The action of $R_p$ on $\ket{b}_p$ must, therefore, be defined by a derivative in the angular variable $b_p$. Working in the $\ket{r}_p$ basis needs a truncation on the discrete (integer) but unbounded eigenvalue $r_p$, and in the $\ket{b}_p$ basis requires a decimation of the continuous but bounded eigenvalue $b_p$. These truncations yield modified commutation algebra and operator definitions at the cutoff boundary, as was discussed in the case of the electric-field basis.

One can switch between the two $\ket{r}_p$ and $\ket{b}_p$ bases in the computation, via efficient quantum-Fourier-transform algorithms, so as to implement each term of the Hamiltonian in the basis it is diagonal in. Alternatively, depending on the value of the coupling, one or the other basis may me most economical. For example, when $g$ is large, the rotor Hamiltonian in Eq.~\eqref{eq:H-rotor} dominates, low-energy states are near the strong-coupling vacuum, which thus require a small cutoff. Therefore, in the strong-coupling limit, simulating physics in the rotor basis is more advantageous. When $g$ is small, the plaquette Hamiltonian in Eq.~\eqref{eq:H-rotor} dominates. In the rotor basis, this term is responsible for raising and lowering the rotor quantum number, implying that a large cutoff needs to be imposed on the rotor Hilbert space to keep the truncation errors small, yielding a large encoding overhead. One may, therefore, want to resort to the magnetic basis. In the magnetic basis, one needs to determine a suitable resolution for the angular variable $b_p$. In the weak-coupling limit, most of the wave function in the low-energy sector is concentrated around small magnetic fields. Thus, one could devise a dense sampling for small $b_p$ and can get away with a sparse one for larger $b_p$. Eventually, finding a suitable interpolation between the two regimes using the two bases may be the best strategy computationally, as demonstrated in Refs.~\cite{haase2021resource,bauer2023efficient}.

\subsubsection{SU(2) lattice gauge theory
\label{sec:SU(2)-basis}
}
The fermionic basis states in the SU(2) theory can be constructed as before: they are the eigenstates of the occupation-number operator associated with each SU(2) components of $\varphi$. Then:
\begin{align}
\begin{cases}
\varphi_1\left|f_{1}, f_{2}\right\rangle=\left(1-\delta_{f_1,0}\right)\left|f_{1}-1, f_{2}\right\rangle, \\
\varphi_1^\dagger\left|f_{1}, f_{2}\right\rangle=\left(1-\delta_{f_1,1}\right)\left|f_{1}+1, f_{2}\right\rangle, \\
\varphi_{2}\left|f_{1}, f_{2}\right\rangle=(-1)^{f_{1}}\left(1-\delta_{f_2,0}\right)\left|f_{1}, f_{2}-1\right\rangle, \\
\varphi_2^\dagger\left|f_{1}, f_{2}\right\rangle=(-1)^{f_{1}}\left(1-\delta_{f_2,1}\right)\left|f_{1}, f_{2}+1\right\rangle,
\end{cases}
\label{eq:fermion-opeators-SU(2)}
\end{align}
with $f_1,f_2 \in \{0,1\}$. Here and in the following, the site dependence of the operators, states, and quantum numbers is suppressed for brevity.

We observed in Sec.~\ref{sec:SU(2)} that the Hilbert space of the SU(2) link is that of a rigid-body rotor. In the irrep or $E$ basis, which is the angular-momentum basis for the body and space frames of the rotor, each link hosts a Hilbert space spanned by basis states $| J, m_{L}, m_{R}\rangle$, where $J^2=E_L^2=E_R^2$, $J=0,\tfrac{1}{2},1,\tfrac{3}{2}, \cdots$, and $ -J \leq m_{L/R} \leq J$. One can impose a cutoff of $J_{\text {max}} = \tfrac{\Lambda}{2}$ to render the on-link Hilbert space finite, with $\Lambda \in \mathbb{Z}^+$. The action of $H^{(d+1)}_\text{KS,E}$ an the link basis states can be understood by recalling the standard angular-momentum relation:
\begin{align}
(J_{L/R})^{2}\left|J,m_{L},m_{R}\right\rangle=J\left(J+1\right)\left|J, m_{L}, m_{R}\right\rangle.
\end{align}
The action of $H_\text{KS,h}$ depends on the action of the fermionic operators in Eqs.~\eqref{eq:fermion-opeators-SU(2)} and the action of $U$ in the fundamental representation ($\tfrac{1}{2}$): 
\begin{align}
U^{(\frac{1}{2})}_{\alpha,\beta}\left|J,m_L,m_R\right\rangle=\sum_{j=J\pm\frac{1}{2}}\sqrt{\frac{2J+1}{2j+1}} &\langle J,m_L;\frac{1}{2},\alpha\left|j,m_L+\alpha\right\rangle \times \nonumber\\
&\langle J,m_R;\frac{1}{2},\beta\left|j,m_R+\beta\right\rangle |j,m_L+\alpha,m_R+\beta\rangle.
\label{eq:U-SU(2)}
\end{align}
Here, $\alpha,\beta =\pm \frac{1}{2}$\footnote{$U_{1,1} \equiv U^{(\frac{1}{2})}_{\frac{1}{2},-\frac{1}{2}}$, $U_{1,2} \equiv U^{(\frac{1}{2})}_{-\frac{1}{2},-\frac{1}{2}}$, $U_{2,1} \equiv U^{(\frac{1}{2})}_{\frac{1}{2},\frac{1}{2}}$, and $U_{1,1} \equiv U^{(\frac{1}{2})}_{-\frac{1}{2},\frac{1}{2}}$.} and $\langle J,m_{L/R};\frac{1}{2},\alpha/\beta\left|j,m_{L/R}+\alpha / \beta\right\rangle$ are the SU(2) Clebsch-Gordan coefficients. From Eq.~\eqref{eq:U-SU(2)}, one concludes that the four $U_{\alpha,\beta}$ operators amounts to two terms, each associated with distinct raising/lowering of the link quantum numbers. Once again, at the edge of the Hilbert space, the action of $U$ must be modified to be consistent with the imposed cutoff. The action of the Casimir $E^2$ in the allowed space remains intact.

A generic basis state in this theory can be written as:
\begin{align}
\ket{\psi}= \cdots \ket{f_1,f_2}_{\bm{x}} \ket{J,m_L,m_R}_{\bm{x}} & \otimes \ket{f_1,f_2}_{\bm{x}+a\hat{\bm{x}}_1} \ket{J,m_L,m_R}_{\bm{x}+a\hat{\bm{x}}_1}\nonumber\\
&\otimes \ket{f_1,f_2}_{\bm{x}+a\hat{\bm{x}}_2} \ket{J,m_L,m_R}_{\bm{x}+a\hat{\bm{x}}_2} \cdots .
\label{eq:xx}
\end{align}
Out of $4^{N_d} \times \left[\sum_{J=0,\frac{1}{2},\cdots,\tfrac{\Lambda}{2}} 2 (2J+1)\right]^{N_\ell}$ states, only a finite set of linear combinations of these states are physical. Such physical states satisfy three non-Abelian Gauss laws of the form $G^a(\bm{x})\ket{\psi}_\text{phys}=0$ with $a \in \{1,2,3\}$ and for all $\bm{x}$, where:
\begin{align}
G^a(\bm{x}) = \sum_{i=1}^d \left(E_{L,i}^a(\bm{x})-E_{R,i}^a(\bm{x}-a\hat{\bm{x}}_i)\right)-\varphi^\dagger(\bm{x})T^a\varphi(\bm{x}).
\label{eq:xx}
\end{align}
These Gauss laws constitute non-commuting constraints, and their simultaneous solutions correspond to states that exhibit zero total angular momentum at any given site (count the SU(2) charge $\varphi^\dagger(\bm{x})T^a\varphi(\bm{x})$ as angular momentum $\tfrac{1}{2}$ if $f_1+f_2~\text{mod}~2 = 1$ and zero if $f_1+f_2~\text{mod}~2 = 0$).

\tcbset{colframe=black!10!black,colback=white,arc=1mm}
\begin{tcolorbox}[breakable]
\textbf{Example 2:} Consider a one-dimensional spatial lattice with $N=2$ staggered-lattice sites. Construct all the physical states in a sector with $\mathcal{Q}=\sum_{x=0}^a \varphi^\dagger(x)\varphi(x)=2$. (Note that $\mathcal{Q}$ is a conserved quantum number.) Consider open boundary conditions where the incoming electric field is set to zero (and the outgoing angular momentum is left open.)

\vspace{0.25 cm}
\textbf{Solution:} One can show that only the following four (linear combinations of) basis states satisfy the Gauss laws in this fermion-number sector (the site-index subscripts are in units of lattice spacing):
\begin{flalign}
&1)~\ket{0,0}_{-1}  \ket{0,0}_0 \ket{0,0}_0 
\ket{0,0}_0  \ket{1,1}_1 \ket{0,0}_1,
\nonumber
\end{flalign}
\begin{align}
&2)~
\frac{1}{2} 
\ket{0,0}_{-1}  \ket{1,0}_0 \ket{\frac{1}{2},-\frac{1}{2}}_0
\ket{\frac{1}{2},\frac{1}{2}}_0  \ket{0,1}_1 \ket{0,0}_1
\nonumber\\
&~
-\frac{1}{2}
\ket{0,0}_{-1}  \ket{1,0}_0 \ket{\frac{1}{2},-\frac{1}{2}}_0
\ket{\frac{1}{2},-\frac{1}{2}}_0  \ket{1,0}_1 \ket{0,0}_1
\nonumber\\
&~ -\frac{1}{2} 
\ket{0,0}_{-1}  \ket{0,1}_0 \ket{\frac{1}{2},\frac{1}{2}}_0
\ket{\frac{1}{2},\frac{1}{2}}_0  \ket{0,1}_1 \ket{0,0}_1
\nonumber\\
&~
+\frac{1}{2}
\ket{0,0}_{-1}  \ket{0,1}_0 \ket{\frac{1}{2},\frac{1}{2}}_0
\ket{\frac{1}{2},-\frac{1}{2}}_0  \ket{1,0}_1 \ket{0,0}_1,
\nonumber
\end{align}
\begin{flalign}
&3)~ \frac{1}{\sqrt{6}}
\ket{0, 0}_{-1} \ket{1, 0}_0 \ket{\frac{1}{2}, -\frac{1}{2}} _0
\ket{\frac{1}{2}, \frac{1}{2}}_0 \ket{1, 0}_1 \ket{1, -1}_1
\nonumber\\
&~
-\frac{1}{2 \sqrt{3}}
\ket{0, 0}_{-1} \ket{1, 0}_0 \ket{\frac{1}{2}, -\frac{1}{2}}_0
\ket{\frac{1}{2}, \frac{1}{2}}_0 \ket{0,1}_1 \ket{1, 0}_1 
\nonumber\\
&~ -\frac{1}{2 \sqrt{3}}
\ket{0, 0}_{-1} \ket{1, 0}_0 \ket{\frac{1}{2}, -\frac{1}{2}}_0
\ket{\frac{1}{2}, -\frac{1}{2}}_0 \ket{1, 0}_1 \ket{1, 0}_1 
\nonumber\\
&~
+\frac{1}{\sqrt{6}} 
\ket{0, 0}_{-1} \ket{1, 0}_0 \ket{\frac{1}{2}, -\frac{1}{2}}_0
\ket{\frac{1}{2}, -\frac{1}{2}}_0 \ket{0, 1}_1 \ket{1, 1}_1
\nonumber\\
&~ -\frac{1}{\sqrt{6}} 
\ket{0, 0}_{-1} \ket{0, 1}_0 \ket{\frac{1}{2},\frac{1}{2}}_0
\ket{\frac{1}{2}, \frac{1}{2}}_0 \ket{1, 0}_1 \ket{1, -1}_1 
\nonumber\\
&~
+\frac{1}{2 \sqrt{3}} 
\ket{0, 0}_{-1} \ket{0, 1}_0 \ket{\frac{1}{2}, \frac{1}{2}}_0 
\ket{\frac{1}{2}, \frac{1}{2}}_0 \ket{0, 1}_1 \ket{1,0}_1 
\nonumber\\
&~
+\frac{1}{2 \sqrt{3}} 
\ket{0, 0}_{-1} \ket{0, 1}_0 \ket{\frac{1}{2}, \frac{1}{2}}_0
\ket{\frac{1}{2}, -\frac{1}{2}}_0 \ket{1,0}_1 \ket{1, 0}_1 
\nonumber\\
&~
-\frac{1}{\sqrt{6}} 
\ket{0, 0}_{-1} \ket{0, 1}_0 \ket{\frac{1}{2}, \frac{1}{2}}_0 
\ket{\frac{1}{2}, -\frac{1}{2}}_0 \ket{0,1}_1 \ket{1, 1}_1 ,
\nonumber
\end{flalign}
\begin{flalign}
&4)~ 
\ket{0,0}_{-1}  \ket{1,1}_0 \ket{0,0}_0
\ket{0,0}_0  \ket{0,0}_1 \ket{0,0}_1.
\label{eq:KSN2nu2}
\end{flalign}
Here, the basis states should be read as $|J,m_R\rangle_{-1} |f_1,f_2\rangle_0 |J,m_L\rangle_0 |J,m_R\rangle_0 |f_1,f_2\rangle_1 |J,m_L\rangle_1$. 
\end{tcolorbox}

It is computationally costly to construct the physical Hilbert space, but once done, the Hamiltonian will be a matrix of much lower dimensionality. The difference in the dimensionality of Hilbert spaces is huge! For the SU(2) lattice gauge theory in (1+1)D in the Kogut-Susskind formulation, the number of basis states goes as $e^{c N}$, where $c \propto \log \Lambda$, while the number of physical basis states goes as $e^{\tilde{c} N}$, where $\tilde{c}$ is a constant~\cite{davoudi2021search}. So while the physical Hilbert still grows exponentially with the system size, it remains exponentially smaller than the full Hilbert space. 

Would it not be nice to have a formulation which involves only the physical states? It turned out that even if we construct such a formulation, the Hamiltonian, despite acting on a smaller Hilbert space, will become more complex and non-local. In other words, the presence of gauge redundancy is a blessing: it keeps the interactions local. We have already seen an example in Sec.~\ref{sec:U(1)-basis}, where solving gauss laws throughout the lattice in a (1+1)D theory with open boundary conditions led to a theory of only fermions but with nearly all-to-all interactions. In that case, the number of Hamiltonian terms scales as $O(N^2)$ for an $N$-site lattice while the Hilbert-space dimensionality is $2^N$. In the equivalent fermion-boson formulation, the number of Hamiltonian terms scales as $O(N)$ but the Hilbert-space dimensionality is $2^N\times(2\Lambda+1)^{N-1}$. These differences lead to different classical- and quantum-computational complexity of the Hamiltonian simulation. We come back to this point in Sec.~\ref{sec:U(1)-warm-up} where we compare the resource requirements of quantum simulating time dynamics in the two formulations. Such a fully fermionic construction can be obtained for the SU($N_c$) lattice gauge theories as well in a similar fashion~\cite{davoudi2021search,sala2018variational}.

\tcbset{colframe=black!10!black,colback=mygray,arc=1mm}
\begin{tcolorbox}[breakable]
\noindent
\begin{exercise}
\label{Exercise:fermionic-SU(2)}
The goal of this exercise is to extend the derivation of the fully fermionic form of the U(1) lattice gauge theory in (1+1)D with open boundary conditions (see Sec.~\ref{sec:U(1)-basis}) to the SU(2) case. This exercise should also make it clear that the same derivation goes for all SU($N_c$) gauge theories in (1+1)D with open boundary conditions.

\vspace{0.25 cm}
\noindent
\textbf{Part (a)} Propose a set of gauge transformations that transforms all the SU(2) gauge links to unity. This transformation should, therefore, eliminate the gauge link from the hopping Hamiltonian.

\vspace{0.25 cm}
\noindent
\textbf{Part (b)} Leverage the SU(2) [SU($N_c$)] relation $E_R(x)=U^\dagger(x) E_L(x) U(x)$ and the Gauss law to express the electric Hamiltonian only in terms of the fermionic fields and the incoming right electric field $\epsilon_0^a$ for $a \in \{1,2,3\}$.

\vspace{0.25 cm}
\noindent
\textbf{Part (c)} Explicitly work out all the Hamiltonian terms for the case of $N=4$, with $\epsilon_0^a=0$ for all $a$.
\end{exercise}
\end{tcolorbox}
Given this tradeoff between Hilbert-space dimensionality and Hamiltonian complexity, can we still find a sweet spot, where as much redundancy as possible is removed but without costing us locality? The answer is yes. This goal can be achieved, for example, by solving all non-Abelian Gauss laws \emph{a priori} as long as one retains the Abelian constraint $J_L^2=J_R^2$ on the link, and only imposes it on states explicitly. This is the essence of the loop-string hadron formulation of SU($N_c$) lattice gauge theories~\cite{raychowdhury2020loop,kadam2023loop,kadam2025loop}.

\begin{figure}[t!]
    \centering
    \includegraphics[scale=0.595]{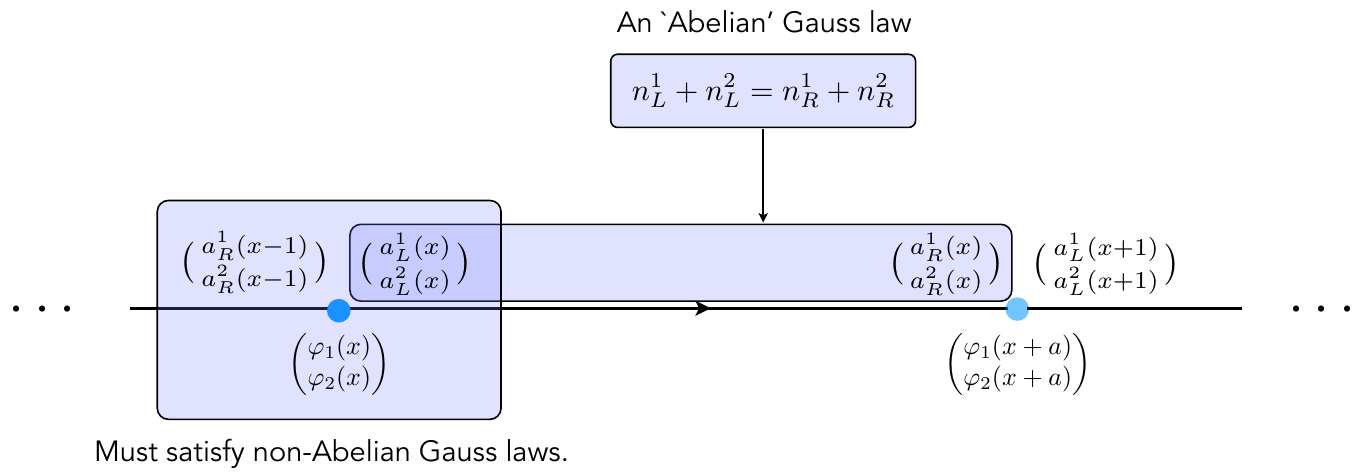}
    \caption{The building blocks of (1+1)D SU(2) lattice gauge theory in the Kogut-Susskind formulation using Schwinger bosons on two representative sites of a staggered lattice.}
    \label{fig:sb-lattice}
\end{figure}

In the nutshell, the stroy goes as follows. First recall that the angular momentum $J$ has a Schwinger-boson representation~\cite{schwinger2001angular}. One option is to work with the Schwinger-boson formulation itself. This amounts to adopting basis states $\ket{n^1_L}$, $\ket{n^2_L}$, $\ket{n^1_R}$, and $\ket{n^2_R}$, which are eigenbases of the occupation-number operator associated with each of the four types of quantum harmonic oscillators (two on the left and two on the right of the link), and cut off the corresponding bosonic Hilbert spaces to make subsequent computations finite. The total number of left and right oscillators should equal, i.e., $n^1_L+n^2_L=n^1_R+n^2_R$, on each link to restore the SU(2) gauge-theory algebra. The degrees of freedom on two representative sites of a 1D staggered lattice are depicted in Fig.~\ref{fig:sb-lattice}.

The other option is to proceed further, and form only gauge-invariant operators out of the three sets of SU(2) doublets, namely
\begin{align}
\begin{pmatrix}
\varphi_1 \\
\varphi_2
\end{pmatrix},~\begin{pmatrix}
a^1_L \\
a^2_L
\end{pmatrix},~\begin{pmatrix}
a^1_R \\
a^2_R
\end{pmatrix},
\end{align}
at each site. Here, $a^{1/2}_L$ and $a^{1/2}_R$ are lowering operators associated with the occupation-number bases $\ket{n^{1/2}_L}$ and $\ket{n^{1/2}_R}$, respectively. The options for gauge-invariant bilinears are: i) bilinears of only bosons (called loops), ii) bilinear of a boson and a fermion (called strings), and iii) belinears of only fermions (called hadrons, or mesons for this SU(2) example). These operators can only excite and de-excite locally gauge-invariant states. The only condition that should still be imposed on the Hilbert space is an Abelian one. The total number of outgoing loops and string out of a site needs to balance the total number of incoming loops or strings into the next site. The same strategy can be applied to higher dimensions, upon applying a useful procedure called point splitting, see Ref.~\cite{raychowdhury2020loop} for detail. The Kogut-Susskind Hamiltonian can be written in terms of only loop, sting, and hadron occupation-number and raising/lowering operators. The corresponding basis states span a much smaller Hilbert space without sacrificing locality. One only needs to apply an Abelian condition at each link, which is far more manageable than having to impose non-Abelian conditions, particularly in higher dimensions~\cite{davoudi2021search}.

All our discussions so far concern the irrep basis, which needs a high cutoff as one takes the continuum limit. Recall that both $H_\text {KS,h}$ and $H_\text{KS,B}$ in Eq.~\eqref{eq:KS-dp1} dominate as $a^{\frac{3-d}{2}}g \rightarrow 0$, hence wave functions with smaller $B$-field amplitudes, or larger $E$-field amplitudes, are most favored. It is, therefore, desired to avoid the irrep basis with a large-cutoff requirement in this limit, and instead use a basis using which hopping and magnetic terms are diagonal.

One option is the group-element basis. This is the group characterized by Euler angles of a rigid rotor. The challenge is to digitize the space of the continuous angles, for which various schemes are proposed, including several uniform partitioning of $S_{3}$ (3D sphere). The electric-field operator can be expressed as a discrete derivative operator in the digitized group manifold. For examples of this approach, see Refs.~\cite{jakobs2023canonical,romiti2024digitizing}. One can alternatively use a non-Abelian Fourier transform (by employing Wigner $D$-matrices~\cite{wigner1931gruppentheorie,wigner2012group}) to convert to the electric basis. However, one would need to develop a one-to-one mapping between the digitized group space and the truncated $J$-irrep space, which is a non-trivial problem.

The most ideal choice is a discrete subgroup of SU(2) that recovers the continuous SU(2) group as the subgroup dimensionality is increased (consider the Abelian example of $Z_{2 \Lambda+1} \to$ U(1) as $\Lambda \to \infty$), in which case, the group-element and the irrep bases could be connected in a systematic way. Unfortunately, the largest subgroup of SU(2) is $\mathbb{BI}$ (binary icosahedral group) with only 120 elements. As one takes the continuum limit, the gauge-group elements $U_{\mu}=e^{i a^{\frac{3-d}{2}} g A_{\mu}} \to 1$, requiring a much denser digitization around unity, which any finite-element subgroup cannot achieve. Significant work has been done in recant years to identify at what coupling (or lattice-spacing scale) the low-lying energy spectrum calculated within finite subgroups start to deviate from their counterparts calculated in the continuous SU(2) group, resorting to classical Monte-Carlo simulations~\cite{alexandru2019gluon,ji2020gluon,alexandru2022spectrum}.

What about the magnetic (or dual) basis for SU(2)? One approach is to use a gauge transformation to set to one as many links around a plaquette as possible. This procedure is called maximal-tree gauge fixing. Subsequently, gauge-invariant degrees of freedom are formed out of closed loops starting and ending at the origin. This makes the expression of magnetic Hamiltonian pretty simple but turns the electric Hamiltonian into a highly non-local form. The reason is, in setting the gauge-link degrees of freedom to one on most links, one would need to keep track of the corresponding gauge transformations along the lattice, which all connect to origin by construction. For progress in this direction in the case of the SU(2) Hamiltonian lattice gauge theory, see Refs.~\cite{mathur2016lattice,d2024new,grabowska2025fully,burbano2024gauge}.

We limit the discussion of various formulations and basis choices of the SU(2) lattice gauge theory in the Kogut-Susskind framework to what we briefly covered so far, and move an to a brief discussion of the SU(3) lattice-gauge-theory Hilbert space.

\subsubsection{SU(3) lattice gauge theory
\label{sec:SU(3)-basis}
}
The fermionic basis states are straightforward generalizations of the SU(2) case. One assigns an 8-dimensional on-site Hilbert space, spanned by basis states 
\begin{align}
\left|f_{1}, f_{2}, f_{3}\right\rangle,
\label{eq:SU(3)-site-basis}
\end{align}
with $ f_{1,2,3} \in \{0,1\}$. The link Hilbert space can be spanned in the irrep basis by basis states involving 8 quantum numbers~\cite{byrnes2006simulating}:
\begin{align}
\left|\Gamma, \lambda_L,\lambda_R\right\rangle \equiv \left|p, q, T_L, T_L^z, Y_L, T_{R}, T_R^z, Y_R\right\rangle,
\label{eq:SU(3)-link-basis}
\end{align}
where $p$ and $q$ are non-negative integers identifying the irrep, $T_{L/R}$ and $T_{L/R}^z$ are color isospin quantum numbers, and $Y_{L/R}$ are hypercharge quantum numbers. For a truncated irrep basis:
\begin{align}
& p, q \in \{0,1,2, \ldots, \Lambda \}, \\
& T_{L}, T_{R} \in \left\{0, \frac{1}{2}, 1, \ldots, \frac{1}{2}(p+q)\right\}, \\
& T_{L}^{z}, T_{R}^{z} \in\left\{-\frac{1}{2}(p+q),-\frac{1}{2}(p+q)+\frac{1}{2}, \ldots, \frac{1}{2}(p+q)\right\},\\
& Y_{L}, Y_{R} \in\left\{-\frac{1}{3}(q+2 p),-\frac{1}{3}(q+2 p)+\frac{1}{3}, \ldots, \frac{1}{3}(p+2 q)\right\}.
\end{align}
So the Hilbert-space dimensionality naively reads:
\begin{align}
8^{N^{d}} \times\left[(\Lambda+1)^{2}(2 \Lambda+1)^{2}(4 \Lambda+1)^{2}(6 \Lambda+1)^{2}\right]^{N_{l}},
\end{align}
where $N^d$ is the number of lattice sites and $N_l$ is the number of lattice links. This expression amounts to a huge number of states per link! Even with the most crude truncation $\Lambda=1$, there are 44,100 possible basis states per link (requiring $16$ qubits to store)! The Gauss laws constrain physics to a smaller subsector (in fact exponentially smaller), nonetheless solving these constraints is rather complicated. There are 8 non-commuting Gauss laws in the SU(3) gauge theory, which should be imposed simultaneously: 
$G^{a}(x)|\psi\rangle_{\text{phys}}=0$ with:
\begin{align}
\quad G^{a}(\bm{x})=\sum_{i=1}^{d}\left(E_{i}^{a}(\bm{x})-E_{i}^{a}(\bm{x}-a \hat{\bm{x}}_i)\right)-\varphi^\dagger(\bm{x}) \frac{\lambda^{a}}{2} \varphi(\bm{x}),
\label{eq:gauss-law-su3}
\end{align}
for $a \in \{1,2, \ldots, 8\}$ at all $\bm{x}$. Fortunately, formulations such as loop-string-hadron formulation are being developed for the SU(3) lattice gauge theory as well, so local gauge-invariant formulations with only Abelian conditions are entering the market~\cite{kadam2023loop,kadam2025loop}.

\tcbset{colframe=black!10!black,colback=mygray,arc=1mm}
\begin{tcolorbox}[breakable]
\noindent
\begin{exercise}
\label{Exercise:SU(3)-states}
The goal of this exercise is to put into practice your group-theory skills and build the physical basis states of a SU(3) lattice gauge theory in the irrep basis through a minimal example.

\vspace{0.25 cm}
\noindent
\textbf{Part (a)} Consider a staggered lattice of $N=2$ sites in (1+1)D with open boundary conditions such that on the incoming link to the lattice, $p_{-1}=q_{-1}=0$. Take the cutoff on $p$ and $q$ to be $\Lambda=1$. In the total fermion-number sector $\mathcal{Q}=\sum_{x}\varphi^\dagger(x)\varphi(x)=3$, how many naive basis states are there?

\vspace{0.25 cm}
\noindent
\textbf{Part (b) [Bonus]} For the example in part (a), find proper linear superpositions of basis states in the $\mathcal{Q}=3$ sector such that Gauss laws in Eq.~\eqref{eq:gauss-law-su3} are satisfied at both sites. How many such states do you find?
\end{exercise}
\end{tcolorbox}

As for the action of the Hamiltonian operators on the site and link basis states, we only need the following relations:
\begin{align}
\begin{cases}
\varphi_1\left|f_{1}, f_{2},f_3\right\rangle=\left(1-\delta_{f_1,0}\right)\left|f_{1}-1, f_{2},f_3\right\rangle, \\
\varphi_1^\dagger\left|f_{1}, f_{2},f_3\right\rangle=\left(1-\delta_{f_1,1}\right)\left|f_{1}+1, f_{2},f_3\right\rangle, \\
\varphi_{2}\left|f_{1}, f_{2},f_3\right\rangle=(-1)^{f_{1}}\left(1-\delta_{f_2,0}\right)\left|f_{1}, f_{2}-1,f_3\right\rangle, \\
\varphi_2^\dagger\left|f_{1}, f_{2},f_3\right\rangle=(-1)^{f_{1}}\left(1-\delta_{f_2,1}\right)\left|f_{1}, f_{2}+1,f_3\right\rangle,\\
\varphi_{3}\left|f_{1}, f_{2},f_3\right\rangle=(-1)^{f_{1}+f_2}\left(1-\delta_{f_3,0}\right)\left|f_{1}, f_{2},f_3-1\right\rangle, \\
\varphi_3^\dagger\left|f_{1}, f_{2},f_3\right\rangle=(-1)^{f_{1}+f_2}\left(1-\delta_{f_3,1}\right)\left|f_{1}, f_{2},f_3+1\right\rangle,
\end{cases}
\label{eq:fermion-opeators-SU(2)}
\end{align}
\begin{align}
\sum_{a=1}^{8}\left(E^{a}\right)^{2}|\Gamma, \lambda_{L}, \lambda_{R}\rangle=\frac{1}{3}\left(p^{2}+q^{2}+p q+3 p+3 q\right)|\Gamma, \lambda_{L}, \lambda_{R}\rangle,
\end{align}
\begin{align}
U_{\gamma_L,\gamma_R}^{(3)}|\Gamma, \lambda_{L}, \lambda_{R}\rangle=\sum_{\Gamma^{\prime}} \sum_{\lambda_{L}^{\prime}} \sum_{\lambda_{R}^{\prime}} \sqrt{\frac{\operatorname{dim}(\Gamma)}{\operatorname{dim}\left(\Gamma^{\prime}\right)}}\langle\Gamma^{\prime},\lambda_{L}^{\prime} | \Gamma, \lambda_{L} ; 3, \gamma_{L}\rangle \langle\Gamma^{\prime}, \lambda_{R}^{\prime}| \Gamma, \lambda_{R} ; 3, \gamma_{R}\rangle |\Gamma^{\prime}, \lambda_{L}^{\prime}, \lambda_{R}^{\prime}\rangle,
\label{eq:U-SU3}
\end{align}
where the site dependence of the operators, states, and quantum numbers is suppressed for brevity. Here, $\operatorname{dim}(\Gamma)=\frac{1}{2}(p+1)(q+1)(p+q+2)$, $\langle\Gamma^{\prime}, \lambda_{L, R}^{\prime} | \Gamma, \lambda_{L / R} ; 3, \gamma_{L/R}\rangle$ are the SU(3) Clebsch-Gordan coefficients, and $\gamma_{L,R} \in \{1,2,3\}$. The sum over $\Gamma^{\prime} \equiv \left(p^{\prime},q^{\prime}\right)$ runs for values $\{(p+1, q),(p-1, q+1),(p, q-1)\}$, and the sum over $\lambda_{L, R}^{\prime}$ runs from $\left|T_{L / R}-t_{L / R}\right|$ to $T_{L / R}+t_{L R}$. $\gamma_{L/R}$ is the collection of $(t_{L/R}, t^z_{L/R} ,y_{L/R})$ quantum numbers associated with the $3 \equiv(1,1)$ irrep of SU(3), specified by the $U$ labels `1', `2', and `3', corresponding to $\left(\frac{1}{2}, \frac{1}{2}, \frac{1}{3}\right),\left(\frac{1}{2},-\frac{1}{2}, \frac{1}{3}\right)$, and $\left.(0,0,-\frac{2}{3}\right)$, respectively. Despite the SU(2) case where the action of $U$ and $U^\dagger$ is the same (as SU(2) is self-adjoint), in the SU(3) case, $U^\dagger$ should be implemented by $U^{(3^{*})}_{\gamma_L,\gamma_R}$. In this case, $\left(p^{\prime},q^{\prime}\right)$ sum runs over $\{(p,q+1),(p+1, q-1), (p-1, q)\}$. $\gamma_{L/R}$ labels are the same as with $U^{(3)}$ upon $y_{L/R} \rightarrow -y_{L/R}$. The action of $U^{(3)}_{\gamma_L,\gamma_R}$ and $U^{(3^{*})}_{\gamma_L,\gamma_R}$ each constitutes up to 12 terms. Each term raises/lowers various quantum numbers depending of the $\gamma_L$ and $\gamma_R$ values. For details, check out Ref.~\cite{byrnes2006simulating}.

The Hamiltonian framework for the SU(3) lattice gauge theory is developed in the group-element basis too. For examples, there are studies of the largest discrete (crystal-like) subgroup of SU(3), which has 1,080 elements~\cite{alexandru2019gluon,ji2020gluon,alexandru2022spectrum}. Yet, such groups can fail to describe the physics of the continuous group toward the continuum limit, as discussed before.

\subsection{On Hamiltonian-simulation methods
\label{sec:KS}
}
With a Hamiltonian at hand, and the chosen basis states with which to express the Hamiltonian matrix, one can go ahead and compute physical observables of interest. The first set of quantities are the Hamiltonian-matrix eigenvalues and eigenstates, namely eigenenergies and energy eigenstates. While the eigenstates' expression depends on the chosen basis states, eigenenergies are basis independent. Other quantities include correlation functions of static and dynamical observables. Various entanglement quantities can also be accessed once one has access to the state. For large quantum many-body Hamiltonian matrices, however, exact methods are out of the picture. Let us briefly enumerate some of the Hamiltonian simulation methods and their pros and cons. By the end of this discussion, it should become clear why quantum simulation emerges as the only viable contender when it comes to generic Hamiltonian-simulation problems.

\vspace{0.25 cm}
\noindent
\emph{Perturbative methods}: Obviously, this method applies to a particular regime of parameters. For example, in the strong-coupling regime of the (1+1)D U(1) and SU($N_c$) lattice gauge theories, that is when $ga \gg 1$, one can apply the standard time-independent perturbation theory to obtain perturbative corrections to the strong-coupling vacuum and its higher excitations. This method, which goes under the name of strong-coupling expansion, was developed and advanced in early days of Hamiltonian lattice gauge theory~\cite{banks1976strong,kogut1983lattice}. Perturbative calculations were consistent with exact analytical predictions, where e.g., the fermion mass vanishes. Such methods are applicable to gauge theories in higher dimensions as well. For example, in pure gauge theory, one may start from the strong-coupling vacuum and evaluate the expectation value of the first-order Hamiltonian perturbation in the strong-coupling (large-$a^{(3-d)/2}g$) limit, which amounts to plaquette excitations over the lattice volume. Nonetheless, such analyses quickly became inaccurate for larger $a^{(3-d)/2}g$ values, where a non-perturbative all-order analysis is required. Note that small-$a^{(3-d)/2}g$ limit corresponds to physics near the continuum limit. Alternatively, one can perform a weak-coupling expansion, which demands perturbing around the weak-coupling vacuum, a state that in general is complicated in the irrep basis. Hence, one may need to start from alternative bases.

\vspace{0.25 cm}
\noindent
\emph{Exact diagonalization (ED)}: Ideally, one can use numerical linear-algebra-based methods to exactly diagonalize the Hamiltonian matrix for arbitrary coupling and mass values. While ED is always the method of choice for small, tractable systems, it unfortunately becomes quickly intractable for larger systems. The Hamiltonian-matrix dimensionality grows exponentially with system size and even the largest supercomputers struggle encoding a moderate-size problem. With a relatively fast desktop computer, you can manage to find the lowest eigenenergies and eigenstates of a system equivalent to 25 qubits or so using efficient classical algorithms for matrix manipulations. As we already saw in the lattice-gauge-theory examples in this section, the Hilbert space of a small few-site theory becomes comparable to tens of qubits, particularly with gauge-boson degrees of freedom at play. Exercise~\cref{Exercise:ED-Schwinger} will guide you through the example of exact diagonalization for the (1+2)D U(1) lattice gauge theory, namely the lattice Schwinger model~\cite{schwinger1962gauge}, so that you would get to practice some ED tasks. Efficient packages have been developed to handle ED for quantum many-body systems, such as the \texttt{Python} package \texttt{QuSpin}~\cite{weinberg2017quspin,weinberg2019quspin}. You can check them out and compare their output against your from-scratch ED code.

To get a sense of computational complexity of exact diagonalization (or more accurately nearly-exact classical algorithms), let us consider two cases (see Ref.~\cite{davoudi2021search} for more details). First, applying the Lanczos algorithm~\cite{arnoldi1951principle}, a generalization of
the Arnoldi algorithm~\cite{lanczos1950iteration}, to obtain the first $k$ eigenvalues of a Hermitian matrix of dimensionality $M \times M$ and density $\rho_H$ (which is the ratio of the number of non-zero elements to $M^2$), requires $O(\xi \rho_H M^2)$ operations, where empirically $\xi \approx \frac{3}{2}k$~\cite{ojalvo1970vibration}. For time evolution, that is applying the exponentiated Hamiltonian matrix on a state vector of density $\rho_v$, a common algorithm is the truncated Taylor expansion~\cite{moler2003nineteen}. The order $k$ at which the series should be truncated to reach certain accuracy depends on time, Hamiltonian’s dimensionality, and the Hamiltonian norm. The time complexity to obtain such a $k$-th order approximation of the time-evolved state is then $O(k\rho_H\rho_vM^2)$~\cite{bell2009implementing}. Both these examples show that our best classical algorithms for Hamiltonian-matrix manipulation depends polynomially on the matrix size (hence exponentially on the system size). Note that most lattice-gauge-theory Hamiltonians are local and hence sparse, but their density is at best $O(M^{-1})$.

\tcbset{colframe=black!10!black,colback=mygray,arc=1mm}
\begin{tcolorbox}[breakable]
\noindent
\begin{exercise}
\label{Exercise:ED-Schwinger} The goal of this exercise is to obtain, numerically via exact diagonalization, the energy spectrum of the Schwinger model with open boundary conditions for a given set of parameters, and to evaluate the continuous time evolution under the Schwinger-model Hamiltonian.

\vspace{0.25 cm}
\noindent
\textbf{Part (a)} Consider the Schwinger-model Hamiltonian in Eq.~\eqref{eq:fully-ferm-I}. First rescale the Hamiltonian by $H \to \frac{2}{g^2a}H$, such that the new Hamiltonian is dimensionless, and the coefficients of the hopping, mass, and electric Hamiltonians become, retrospectively, $i\rm{x} \coloneq \frac{i}{a^2g^2}$, $\mu \coloneq \frac{2m}{g^2a} $ and $1$. Consider a system of $N=4$ staggered-lattice sites. Construct the Hamiltonian matrix, i.e., find all the matrix elements in terms of $\rm{x}$, $\mu$, and $\epsilon_0$. Now numerically diagonalize the Hamiltonian to find the energy eigenvalues for the following model parameters: $\epsilon_0 = 0$, $\rm{x} = 0.6$, and $\mu = 0.1$.

\vspace{0.25 cm}
\noindent
\textbf{Part (b)} Consider the strong-coupling vacuum (i.e., the eigenstate of the Hamiltonian in the limit $\rm{x} = 0$) and call it $\ket{\psi(0)}$. Apply the time-evolution operator $e^{-itH}$ onto this state for the total evolution time $t=5$. For all other parameters, use the values given in Part (a). This procedure gives you $\ket{\psi(t)}=e^{-itH}\ket{\psi(0)}$. Evaluate and plot, as a function of time, the Loschmidt echo, i.e., the survival probability of the initial state, defined as
\begin{align}
    \mathcal{P}(t) \coloneq \left | \langle \psi(0) | \psi(t) \rangle \right|^2.
\end{align}
What do you learn from this quantity?

\vspace{0.25 cm}
\noindent
\textbf{Part (c)}
For the same procedure and parameters as in the previous part, evaluate and plot, as a function of time, the particle-number density defined as
\begin{align}
    \nu(t) \coloneq ~\frac{1}{N}\sum_{x=0}^{N-a}{\nu(x;t)}~~\text{with}~~
    \nu(x;t) \coloneq (-1)^{x/a+1} \langle \psi(t) | \left(E(x)-E(x-a)\right) | \psi(t) \rangle.
\label{eq:nu-def}
\end{align}
How many electron-positron pairs are there in the initial state? (Recall the mapping discussed in Sec.~\ref{sec:U(1)-basis}.) Do the dynamics generate electron-positron pairs? The phenomenon of pair production out of ``vacuum'' is a hallmark of a relativistic quantum field theory, which can be seen in non-equilibrium dynamics as simple as those you just studied!
\end{exercise}
\end{tcolorbox}

\vspace{0.25 cm}
\noindent
\emph{Tensor networks (TNs)}: TNs are a class of variational wave functions characterizing many-body quantum systems~\cite{orus2014practical,biamonte2017tensor,montangero2018introduction}. Consider a system of $N$ lattice sites where each site hosts a $d$-dimensional Hilbert space. A generic state of this system can be expressed as: $\ket{\Psi} = \sum_{\{s\}} \psi_{s_1s_2\cdots s_N} \ket{s_1s_2\cdots s_N}$, where the wave function $\psi_{s_1s_2\cdots s_N}$ is a $d^N$-dimensional tensor. To reduce this form to one that is computationally tractable, one can replace this tensor with a product of $N$ tensors each with a dimensionality much smaller than $d^N$. In (1+1)D, this form is called a matrix-product-state (MPS) ansatz~\cite{verstraete2008matrix,schollwock2011density}. Explicitly, the MPS has the form $\ket{\Psi}^{\text{(MPS)}} = \sum_{\{s\}} A^{s_1}_{\alpha_1} A^{s_2}_{\alpha_1,\alpha_2} \cdots A^{s_N}_{\alpha_{N-1},\alpha_N} \ket{s_1s_2\cdots s_N}$ (assuming open boundary conditions), where $\alpha_i \in \{1,\cdots,\chi\}$, with $\chi$ called the bond dimension. These tensors can be further constrained based on system's symmetries. So while the original generic ansatz contained $d^N$ parameters, the MPS ansatz involves at most $Nd\chi^2$ parameters, which is a dramatic improvement. Such an ansatz is reasonable, e.g., for ground states of local Hamiltonians, whose entanglement entropy is known to scale with the system's area not the volume. Such a property holds for the MPS ansatz, and in (2+1)D for what is called the projected entangled-pair state (PEPS) ansatz~\cite{verstraete2004renormalization,verstraete2008matrix}.

Once the ansatz is constructed, one can use a variational method to optimize the parameters' values. A powerful algorithm is density-matrix renormalization group (DMRG), which leverages an entanglement-based truncation to iteratively solve for converged solutions as the Hilbert space is systematically enlarged from one iteration to the next~\cite{white1992density,ostlund1995thermodynamic}. Expectation values can be computed by contracting the MPS bra and ket with the matrix product operators (MPOs). Both real-time and imaginary-time evolutions can also be performed using algorithms such as time-evolving block decimation (TEBD) and the time-dependent variational principle (TDVP) methods~\cite{paeckel2019time}. Once again, useful software packages are developed to allow access to many of these functionalities out of the box. A popular one is the \texttt{ITensor} Software Library~\cite{itensor}.

The challenge with TN methods is two-fold. First, the bond dimension required to capture a time-evolved state with fixed accuracy grows exponentially in evolution time~\cite{osborne2006efficient,schuch2008entropy}. Second, computations using high-dimensional ansatzes, such as PEPS, become highly complex. The reason is that more tensor contractions are needed in higher dimensions plus the bond dimension would need to roughly scale exponentially in $L^{d-1}$ to accurately capture an area-law state, such as ground states of local Hamiltonians in ($d$+1)D.\footnote{The bond dimension, or the highest Schmidt value retained in the Schmidt decomposition of the state in a TN ansatz, is proportional to $e^S$, where $S$ is the bipartite entanglement entropy. $S$ grows as $L^{d-1}$ for ground states of local Hamiltonians.} Despite these challenges, TN methods have been successfully applied to study low-lying and equilibrium states of lower-dimensional\footnote{Cheaper versions of TNs, such as tree TNs~\cite{shi2006classical}, have made possible simulations of (2+1)D and (3+1)D gauge theories in recent years~\cite{felser2020two,cataldi2024simulating,magnifico2021lattice}. For conciseness, we leave out discussion of these ansatzes.} gauge theories, as well as their time dynamics, see e.g., Refs.~\cite{banuls2020review,banuls2023tensor,meurice2022tensor,magnifico2024tensor,zohar2022quantum} for recent reviews.

\vspace{0.25 cm}
\noindent
\emph{Quantum simulation}: Given the barriers in large-scale reliable Hamiltonian simulation of gauge theories of nature using the methods outlined so far, one may want to consider quantum-computing methods instead. As will become clear in the next section, quantum simulation can require resources that scale polynomially in system size. This is especially true when it comes to time-evolution algorithms. Preparation of arbitrary quantum states can remain inefficient even quantumly~\cite{kempe2006complexity}, but efficient heuristic algorithms can enable progress. Several algorithms and strategies will be reviewed in Sec.~\ref{sec:simulation} to make these points clear. The general idea is that while the Hilbert-space dimension is still exponential in system size, its encoding into qubits' Hilbert space (or other quantum degrees of freedom in a quantum simulator) is exponentially more compact than the classical counterpart. This means that if we can find polynomial-size algorithms that leverage the encoding advantage, the classically intractable Hamiltonian problems can turn into tractable ones on a quantum hardware.

The statement above by no means indicates that quantum computing lattice gauge theories of relevance to nature will be cheap. In fact, based on the discussions of Sec.~\ref{sec:QCD}, one concludes that QCD simulations of relevance to phenomenology may require supercomputers of enormous capacity and capability. This fact should not discourage us. The lattice-gauge-theory community was in the same position at early stages of the development of the field, with classical computers that were at least ten orders of magnitude smaller and slower than the current classical supercomputers. Nonetheless, patience, algorithmic advances, and hardware progress eventually paid off! With this hope in the quantum-computing era, let us begin the discussion of quantum computation of Hamiltonian lattice gauge theories.

\section{Quantum computation of gauge theories
\label{sec:simulation}
}
\noindent
The goal of this section is to learn about quantum-simulation algorithms and apply them to the Hamiltonian lattice gauge theories we studied in the previous section. The section starts with a gentle introduction to quantum-computing basics, then introduces the concept of quantum simulation and its various steps, along with a few qualitative descriptions of the relevant algorithms. Ultimately, we focus our attention to time evolution, and present concrete examples of time-evolution algorithms in gauge theories. The section concludes with a discussion of our current understanding of the quantum-computing cost of time evolution in QCD.

\subsection{Quantum-computing basics
\label{sec:QC-basics}
}

Quantum computing, primarily concerning a qubit-based, gate-based approach to computing, is an expansive subject. I try to present only a few basics such that we can carry on with our goal---that is, to find out how to simulate gauge theories using quantum computers. For a more comprehensive coverage of the topic, check out popular textbooks in the subject, such as the book by Nielsen and Chuang~\cite{nielsen2010quantum}.

\subsubsection{Qubits, gates, and circuits
\label{sec:qubits}
}
\begin{figure}
\begin{center}
\includegraphics[scale=0.675]{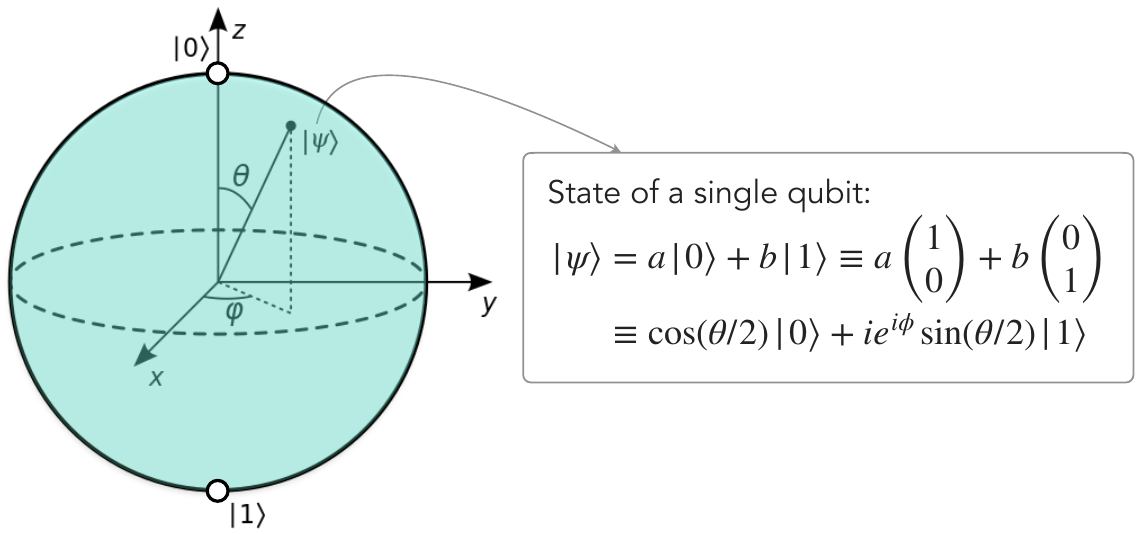}
\end{center}
\caption{The Bloch sphere representation of the state of a qubit corresponding to Eq.~\eqref{eq:1q-Bloch}.}
\label{eq:Bloch}
\end{figure}

Information can be stored in the quantum wave function of a physical system with an isolated but accessible two-state Hilbert space. Such a physical system is our \emph{qubit}. There are several different qubit technologies, including trapped ions, superconducting circuits, Rydberg atoms, and quantum dots, see e.g., a useful review in Ref.~\cite{altman2021quantum}. For algorithmic purposes, we only need to assume a qubit can be prepared and retained in a given quantum state:
\begin{align}
|\psi\rangle_{1q} & =\alpha|0\rangle+\beta|1\rangle \equiv \alpha\binom{1}{0}+\beta\binom{0}{1},
\end{align}
with complex $\alpha$ an $\beta$ satisfying $|\alpha|^{2}+|\beta|^{2}=1$. Alternatively, using two real parameters $\theta$ and $\phi$, the same single-qubit state can be written as a vector spanning the qubit's Bloch sphere, see Fig.~\ref{eq:Bloch}:
\begin{align}
|\psi\rangle_{1q} = \cos\left(\frac{\theta}{2}\right)|0\rangle+e^{i \phi} \sin\left(\frac{\theta}{2}\right)|1\rangle,
\label{eq:1q-Bloch}
\end{align}
since an overall phase of the state can be dropped.

Two- and multi-qubit states can be similarly expressed as:
\begin{align}
|\psi\rangle_{2 q} &=\gamma|00\rangle+\delta|01\rangle+\rho| 10\rangle+\sigma|11\rangle
\equiv \gamma \left(\begin{array}{l}
1 \\
0 \\
0 \\
0
\end{array}\right)+\delta\left(\begin{array}{l}
0 \\
1 \\
0 \\
0
\end{array}\right)+\rho\left(\begin{array}{l}
0 \\
0 \\
1 \\
0
\end{array}\right)+\sigma\left(\begin{array}{l}
0 \\
0 \\
0 \\
1
\end{array}\right),
\end{align}
with $|\gamma|^2+|\delta|^2+|\rho|^2+|\sigma|^2=1$. Similarly, a general $n$-qubit state can be decomposed into $2^n$ complex amplitudes, along with a normalization condition. These state are expressed in the so-called computational basis, i.e., the Bloch $\mathbf{z}$-basis of all qubits, which is most commonly used, but other bases can be used too when appropriate. 

Logical gates are unitary operations on the state of one or multiple qubits. Examples of single-qubit gates are the Pauli gates $\sigma^{\mathbf{x}}=\left(\begin{array}{cc}0 & 1 \\ 1 & 0\end{array}\right)$,  $\sigma^{\mathbf{y}}=\left(\begin{array}{cc}0 & -i \\ i & 0\end{array}\right)$, and $\sigma^{\mathbf{z}}=\left(\begin{array}{cc}1 & 0 \\ 0 & -1\end{array}\right)$, Hadamard gate $\text{H}=\frac{1}{2}\left(\begin{array}{cc}1 & 1 \\ 1 & -1\end{array}\right)$, phase gate $S=\left(\begin{array}{ll}1 & 0 \\ 0 & i\end{array}\right)$, and the T gate $T=\left(\begin{array}{cc}1 & 0 \\ 0 & e^{i \pi / 4}\end{array}\right)$. An example of an entangling two qubit gate is the controlled-not gate $\text{CNOT}=\left(\begin{array}{llll}1 & 0 & 0 & 0 \\ 0 & 1 & 0 & 0 \\ 0 & 0 & 0 & 1 \\ 0 & 0 & 1 & 0\end{array}\right)$: it flips the state of the second qubit only if the first qubit is in state $\ket{1}$. An important theorem by Solovay and Kitaev~\cite{kitaev1997quantum,dawson2005solovay} states that any unitary on a finite set of qubits can be approximated to precision $\varepsilon$ by only $O\left(\log ^{c} (1 / \varepsilon)\right)$ gates, which is efficient. In fact, only a finite universal set of single and two-qubit gates is sufficient to achieve this. Two common choices of universal gate sets are:
\begin{itemize}
  \item $R^{\textbf{x}}(\theta) \coloneq e^{-i \theta \sigma^{\textbf{x}} / 2}$, $R^{\textbf{y}}(\theta) \coloneq e^{-i \theta \sigma^{\textbf{y}} / 2}$, $R^{\textbf{z}}(\theta) \coloneq e^{-i \theta \sigma^{\textbf{z}} / 2}$,  $P(\phi) \coloneq \left(\begin{array}{cc}1 & 0 \\ 0 & e^{i \phi}\end{array}\right)$, and CNOT.\footnote{Since a global phase is irrelevant in quantum computation, $R^{\textbf{x}}(\theta)$, $R^{\textbf{z}}(\theta)$, and CNOT gates suffice to construct all unitaries of relevance to quantum computing.}
  \item H, $S$, CNOT, and $T$.\footnote{$S=T^2$ is not strictly needed but it is more economical to include it in this set.}
\end{itemize}
In the near-term era of quantum computing, qubits are scarce resources. Furthermore, CNOT gates are costly resources, as their operation is slower and results in more decoherence. In the far-term era, qubit resources are abundant. Nonetheless, the gates need to be synthesized fault-tolerantly, and doing so is costlier for $R^{\textbf{x}/\textbf{y}/\textbf{z}}$, $P$, and $T$ gates than the Clifford gates (H, CNOT, $S$). As a result, algorithm design for near-term era focuses on minimizing qubit and CNOT-gate counts while in the far-term era, the focus is shifted to minimizing non-Clifford gates such as the $T$ gate.\footnote{Not surprisingly, while Clifford gate are simulatable efficiently using classical computing, the non-Clifford gate are not. Therefore, non-Clifford gates are true quantum resources.}

A quantum circuit is a collection of qubits and gates. The gates implement a unitary on the state of the qubits. In general, finding the gate decomposition of a $2^{n} \times 2^{n}$ unitary on the space of $n$ qubits is hard. A classical algorithm that finds such a decomposition can scale exponentially with the number of qubits, and can generate a circuit that involve exponentially many operations in the number of qubits, which is inefficient. The art is to find a quantum circuit whose cost scales only polynomially with the number of qubits. We will later see examples of quantum-circuit design for the time-evolution unitary in gauge theories that satisfy this efficiency requirement.

\subsubsection{Mapping fermions and bosons
\label{sec:mapping}
}

To simulate fermions and bosons, one has to first map their Hilbert space to that of qubits' Hilbert space, which is of hard-core bosons. Bosonic and fermionic operators should also turn into qubit operators (gates) with correct boson and fermion commutation or anticommutation relations.

\vspace{0.25 cm}
\noindent
\emph{Fermions:} The on-site Hilbert space of (single-component) fermions, such as a non-relativistic spinless fermions or staggered fermions, is two dimensional (no fermion or one fermion present). This matches the Hilbert-space dimensionality of a qubit. Nonetheless, a naive mapping of fermionic operators according to:
\begin{align}
\begin{cases}
\varphi(\bm{x}) \rightarrow \sigma_{n}^{+} , \\
\varphi^\dagger(\bm{x}) \rightarrow \sigma_{n}^{-},
\end{cases}
\label{eq:naive-map}
\end{align}
with $\sigma^\pm_n \coloneq \frac{1}{2}(\sigma^{\mathbf{x}}_n\pm i \sigma^{\mathbf{y}}_n)$, does not yield all the required anticommmutation relations (i.e., $\left\{\varphi(\bm{x}),\varphi^\dagger(\bm{x}')\right\}=\delta_{\bm{x}, \bm{x}'}$ and $\left\{\varphi(\bm{x}),\varphi(\bm{x}')\right\}=\left\{\varphi^\dagger(\bm{x}),\varphi^\dagger(\bm{x}')\right\}=0$). Here, each lattice-site coordinate $\bm{x}$ is assigned a qubit index, hence $n \in \{0,1,\cdots L^d\}$, where $L$ is the number of lattice sites per dimension and $d$ is the space dimensionality. One way to fix the issue is through a (non-local) Jordan-Wigner map~\cite{jordan1993paulische}:
\begin{align}
\begin{cases}
\varphi(\bm{x}) \rightarrow \left(\prod_{m<n} \sigma_{m}^{\mathbf{z}}\right)  \sigma_{n}^{+} , \\
\varphi^\dagger(\bm{x}) \rightarrow \left(\prod_{m<n} \sigma_{m}^{\mathbf{z}}\right) \sigma_{n}^{-}.
\end{cases}
\label{eq:jw-map}
\end{align}
The fermions' Fermi statistic is an intrinsically non-local property. No matter how distant two fermions are, exchanging them introduces a non-trivial negative sign (hence the non-local Jordan-Wigner Pauli-$Z$ strings to keep track of this exchange-symmetry sign).

\tcbset{colframe=black!10!black,colback=mygray,arc=1mm}
\begin{tcolorbox}[breakable]
\noindent
\begin{exercise}
\label{Exercise:JW}
This exercise encourages you to compute the anticommutation relations explicitly and appreciate how the Jordan-Wigner map restores the correct relations.  

\vspace{0.25 cm}
\noindent
\textbf{Part (a)} Show that the map in Eq.~\eqref{eq:naive-map} does not satisfy all the required fermionic anticommutation algebra. 

\vspace{0.25 cm}
\noindent
\textbf{Part (b)} 
Show that in contrary, the map in Eq.~\eqref{eq:jw-map} satisfies all the anticommutation relations.

\vspace{0.25 cm}
\noindent
\textbf{Part (c)} What are the expectations for commutation relations? Does the Jordan-Wigner map meet those expectations?
\end{exercise}
\end{tcolorbox}

While the Jordan-Wigner map is straightforward to implement in (1+1)D, in higher dimension, it requires first assigning a linear map to points in an $N$-site lattice of fermions. One consequence of this map is that the nearest-neighbor hopping of fermions can turn into maximally non-local interactions, as shown in the example below. In general, the \emph{Pauli weight}, i.e., the number of Pauli operators, of the Jordan-Wigner-transformed operator scales as $O(N^{d-1})$, where $d$ is the spatial dimension of the lattice.
\tcbset{colframe=black!10!black,colback=white,arc=1mm}
\begin{tcolorbox}[breakable]
\textbf{Example 3:} Consider one-component fermions living on the sites of a 2D lattice. Through a minimal example, demonstrate that the nearest-neighbor hopping of the fermions turns into Pauli strings of length proportional to the lattice extent.

\vspace{0.25 cm}
\textbf{Solution:} The figure below shows one such minimal example.
\begin{center}
\includegraphics[scale=0.65]{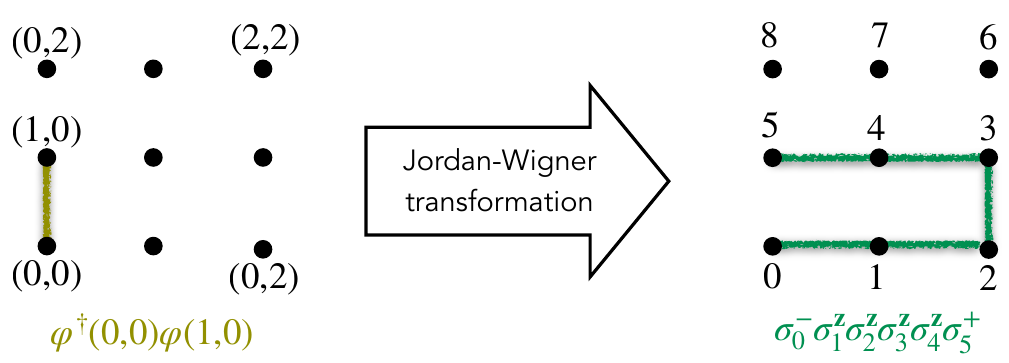}
\end{center}
Here, $\varphi(x,y)$ denotes the fermion operator at site $(x,y)$ of the 2D lattice. These Cartesian points are mapped to a linear index in the right. As shown, a two-body operator is turned into a six-body operator, and the Pauli string spans the lattice extent. Note that an additional $\sigma^{\mathbf{z}}_0$ is dropped from the string in the right since the action of $\sigma^-_0\sigma^{\mathbf{z}}_0$ is equivalent to the action of $\sigma^-_0$ alone.
\end{tcolorbox}

One can in turn choose alternative mappings of fermions to qubits that retain locality. To achieve this, we need a set of ancillary qubits, and a set of local constraints that relate the state of physical and ancillary qubits at each site---a scenario that resembles the Gauss laws in gauge theories. As with gauge theories, redundancies help with a local formulation of an intrinsically non-local problem (in this case, Fermi statistic)! For these simulations to restore Fermi statistics, however, they should necessarily start in the correct Gauss-law sector and remain there throughout the simulation. There are efficient polynomial algorithms to prepare the initial state of the extended fermion-ancillary resisters, and various symmetry protection ideas can be used to constrain the dynamics to the proper sector. For examples of such local mappings, see Refs.~\cite{bravyi2002fermionic,verstraete2005mapping,setia2019superfast,derby2021compact,jiang2019majorana,chen2020exact,chen2023equivalence}; for examples of balancing the mapping overhead and simulation cost of fermionic theories considering the hardware architecture, see Refs.~\cite{chien2020custom,chien2022optimizing,bringewatt2023parallelization}; and for applications of local fermionic encodings in gauge theories and nuclear effective field theories, see Refs.~\cite{rhodes2024exponential,watson2023quantum}.

\vspace{0.25 cm}
\noindent
\emph{Bosons:}
Bosons share the same statistics with qubits, but they have an infinite-dimensional Hilbert space. In order to encode them onto a finite set of qubits, one can cut off their occupation number at each site (link) by an integer $\Lambda >0$. A variety of encodings allow mapping such a finite-dimensional Hilbert space to that of qubits. For example:
\begin{itemize}
  \item Binary encoding amounts to using $\eta \equiv\lceil\log (\Lambda+1)\rceil$ number of qubits. A boson Fock state with occupation $n_b$ can be represented as: $\left|n_{b}\right\rangle=\bigotimes_{j=0}^{\eta-1}\left|n_{b, j}\right\rangle$ with $n_{b}=\sum_{j=0}^{\eta-1} 2^{j}n_{b,j}$, where $n_{b,j} \in \{0,1\}$.
 \item Unary encoding amounts to using $\eta \equiv \Lambda+1$ number of qubits. A boson Fock state with occupation $n_b$ can be represented as: $\left|n_{b}\right\rangle=|0\rangle \otimes|0\rangle \ldots |0\rangle \otimes|1\rangle \otimes|0\rangle \ldots |0\rangle$ with the qubit in the $\left(n_{b}+1\right)$-th position being in state $\ket{1}$, and the remaining qubits being in state $\ket{0}$. \\
\end{itemize}
The binary encoding is more qubit efficient, but operations on the boson register can be more complex than in unary encoding. For example, single-boson addition only amounts to operation on two nearby qubits with unary encoding, while in the binary encoding, it often involves operations on more than two qubits (and at most on all qubits in the register).

\begin{figure}[t!]
    \centering
    \includegraphics[scale=0.6]{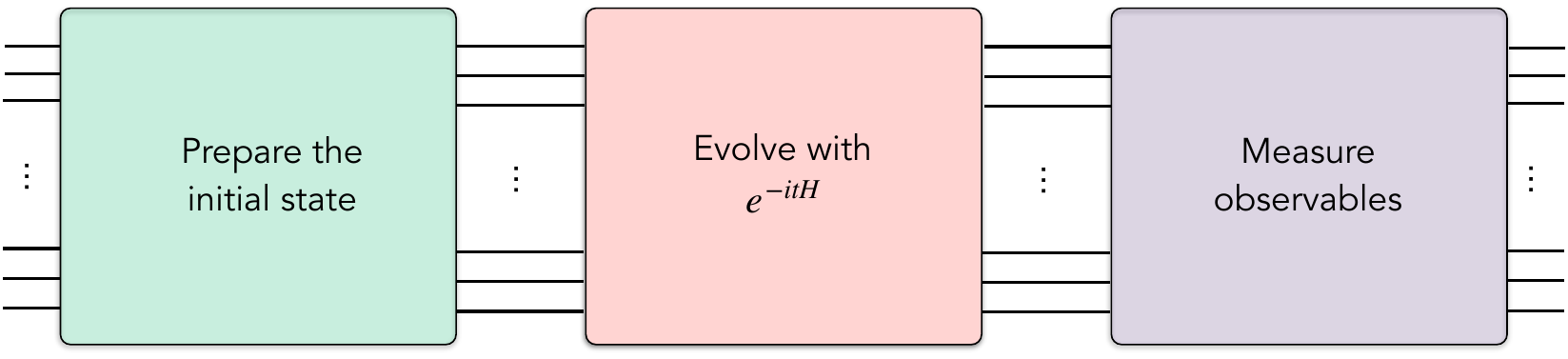}
    \caption{Schematic of quantum-simulation steps. Each of these simulation blocks involves different algorithms depending on the problem at hand.}
    \label{fig:steps}
\end{figure}

When encoding bosonic degrees of freedom which are fractional integers, one can first multiply them out by proper integers to make them whole integers, then encode those integers as prescribed above (this would be useful for encoding the SU(2) and SU(3) link quantum numbers in the irrep basis). Similarly, when dealing with negative-integer quantum numbers, one can first shift them by a proper positive integer, then encode a non-negative integer onto the qubit register, so as to avoid dealing with signed arithmetic on quantum computers (this would be useful for encoding the electric or rotor fields in the U(1) lattice gauge theory, as we will discuss in Sec.~\ref{sec:U(1)-warm-up}). Encoding irrational numbers in quantum computer is a possibility but they can only be encoded with finite precision as with classical computers.

\subsection{Quantum-simulation steps in nutshell
\label{sec:QS-steps}
}
Any quantum simulation proceeds in three steps: state preparation, time evolution, and observable measurement, as schematically depicted in Fig.~\ref{fig:steps}. One or more of these steps may be trivial but often that is not the case. Let us briefly describe some of the strategies for each of these simulation tasks.

\subsubsection{State preparation
\label{sec:state-prep}
}
Preparing eigenstates of physical quantum Hamiltonians is generally hard. For $k$-local spin Hamiltonians, this task is QMA-Complete~\cite{kempe2006complexity}---the complexity class that is the quantum analog of the classical (probabilistic) NP-Complete~\cite{bookatz2012qma}. Such problems are believed to be difficult to solve, even with a quantum computer, but a quantum computer would easily be able to verify whether a proposed solution is correct. So there is no proven quantum advantage in quantum-state preparation in general. This has not stopped the development of efficient, often heuristic, quantum algorithms that prepare the ground state with a given fidelity goal and/or success rate. Classical knowledge of certain properties of the state, and of its symmetries, can also aid in optimizing the various algorithms. Below, we qualitatively describe some of the most popular approaches to ground-state preparation:
\begin{itemize}
\item \emph{Adiabatic state preparation:} One can first prepare the ground state of a simpler Hamiltonian, then slowly change the Hamiltonian along some evolution path to approach the target-system's Hamiltonian, as schematically depicted in Fig.~\ref{fig:adiabatic-imaginary}(a). If the rate of change in the Hamiltonian is smaller than the smallest energy gap encountered as one traverses the path, the system will not leave the ground state~\cite{born1928beweis,kato1950adiabatic,farhi2000quantum,albash2018adiabatic}. The process, therefore, can be slow and requires time-dependent Hamiltonian evolution. It also breaks down if a phase transition occurs along the path, at which point the gap closes exponentially in system size. For QCD, this procedure can be problematic if one starts from the ground state of free quarks and gluons, since it is not guaranteed that the bound hadronic states of the confined phase are obtained in any finite time. For examples in the context of quantum field theories, see Refs.~\cite{jordan2012quantum,chakraborty2022classically,pederiva2021quantum,kaikov2025phase,cohen2024efficient,farrell2024quantum}.
\begin{figure}[t!]
\begin{center}
\includegraphics[scale=0.695]{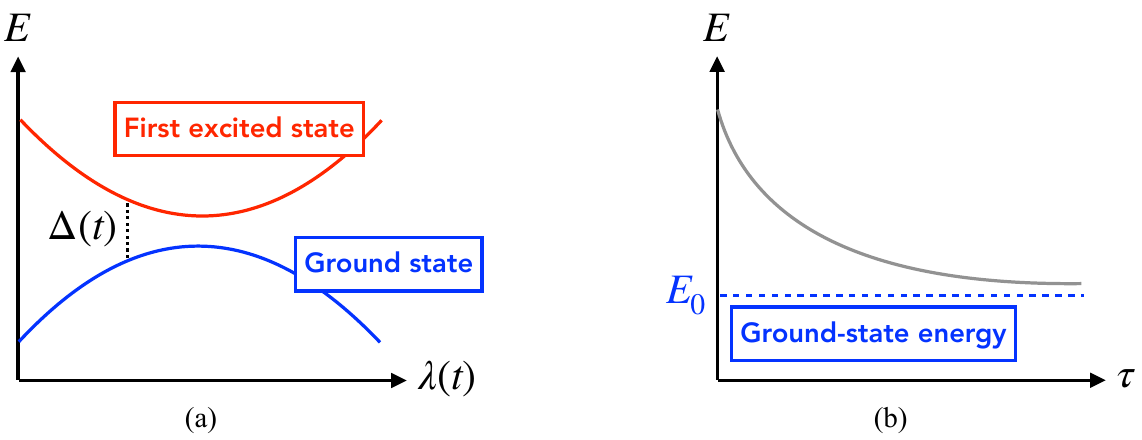}
\end{center}
\caption{(a) A schematic of the lowest-energy spectrum of a system undergoing adiabatic evolution in a Hamiltonian parameter $\lambda(\tau)$. $E_0$ and $E_1$ denote the instantaneous ground- and the first excited-state energies. If driven slowly, the system does not leave the ground-state manifold (blue line), and will not populate the excited state (red line). (b) Imaginary-time evolution $e^{-\tau H}$ results in the decay of the energy of an arbitrary state, driving the energy toward the lowest-energy state of the Hamiltonian, assuming an nonzero overlap between the initial state and the target ground state.}
\label{fig:adiabatic-imaginary}
\end{figure}
\item \emph{Imaginary-time evolution:} This algorithm starts in an arbitrary state and evolves the state in imaginary time, as schematically depicted in Fig.~\ref{fig:adiabatic-imaginary}(b). The excitations die off at a rate exponentially fast in the excitation gaps, and only the ground state survives at late times (assuming non-zero overlap between the initial state and the target state). The challenge is implementing imaginary-time evolution, which is a non-unitary operation, and hence less natural for quantum computers. Such non-unitary operations can still be performed but often using ancillary registers to encode unitary dynamics on an extended Hilbert space. Upon tracing out such ancilla qubits, the system can be shown to have evolved under non-unitary dynamics. For some quantum-field-theory examples, see Refs.~\cite{yeter2022quantum,turro2022imaginary,davoudi2023towards}.
\item \emph{Variational quantum eigensalvers (VQE):} The idea, as depicted in Fig.~\ref{fig:VQE}, is to prepare a non-trivial state using a parametrized quantum circuit, measure its energy, minimize the energy with respect to those parameters using classical computing, then repeat until a global minimum is found. This is, therefore, a hybrid classical-quantum algorithm and fully heuristic. The challenge is when the algorithm is stuck in a local minimum, or if there is a manifold with exponentially small gradient in system size~\cite{larocca2025barren}. VQE, however, has been very popular in the current era of quantum computing~\cite{cerezo2021variational,tilly2022variational}, since one can get away with shallow quantum circuits, plus the algorithm can be pretty resilient to hardware noise. For examples of VQE applications in the context of gauge theories, see Refs.~\cite{klco2018quantum,atas20212,farrell2024scalable,crippa2024analysis,fromm2024simulating,farrell2024quantum,davoudi2024scattering,davoudi2025quantum,than2024phase,xie2022variational}.
\begin{figure}[t!]
\begin{center}
\includegraphics[scale=0.75]{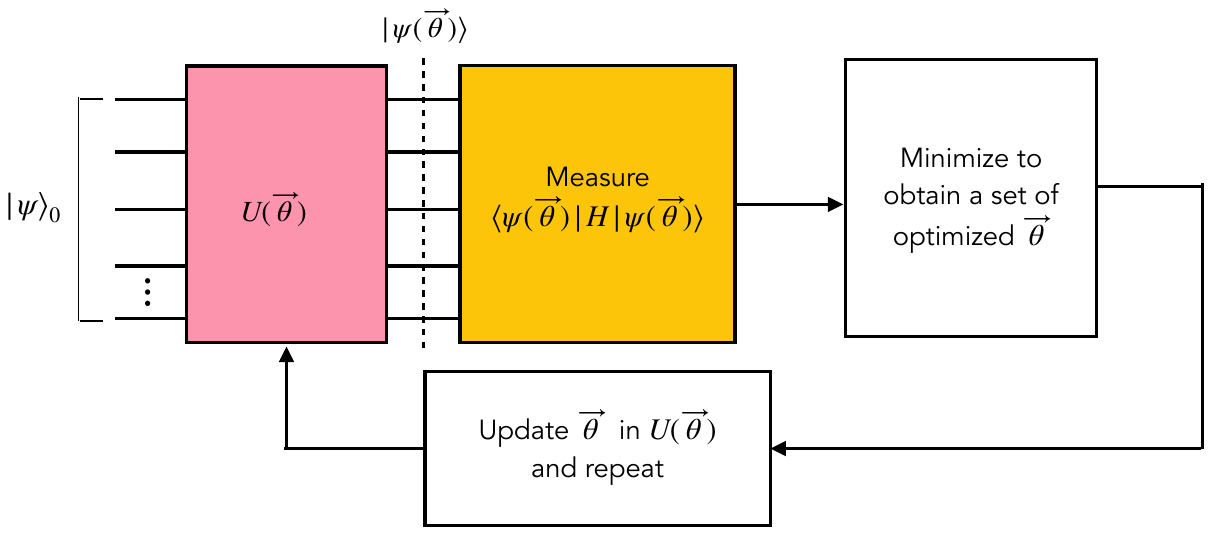}
\end{center}
\caption{The schematic of a variational quantum circuit. A trial state $\ket{\psi}_0$ is turned into a non-trivial state $\ket{\psi(\vec{\theta})}$ using a parametrized unitary $U(\vec{\theta})$ with the set of parameters $\vec{\theta}=\{\theta_1, \theta_2, \cdots\}$. Measurement of the Hamiltonian $H$ in this state, and minimizing the value using a classical optimizer result in an optimized set of parameters $\vec{\theta}$ characterizing an approximate ground state. The classical steps are shown in the uncolored boxes while the (unitary and non-unitary) quantum steps are shown in the colored boxes.}
\label{fig:VQE}
\end{figure}
\item \emph{Filtering-based algorithms:} The idea here is to encode given functions of Hamiltonian using, e.g., near-optimal block-encoding schemes such that repeated applications of those functions can filter out excited states and push the state close to the ground state (like Krylov methods in classical computing~\cite{gowers2010princeton}). Here, block encoding refers to encoding a non-unitary operator such as Hamiltonian or functions of, as a block of a larger unitary, which can be implemented efficiently.\footnote{Imaginary-time evolution can be implemented in this fashion too.} Filtering-based methods, therefore, often need additional ancillary qubits which will be plentiful in the far-term era of quantum computing. These algorithms are probabilistic: the ancillary register must be repeatedly measured, with a success rate of implementing the desired function that depends on a number of factors, including the overlap of the initial trial state with the true state that is being prepared. For recent examples of such algorithms, see Refs.~\cite{ge2019faster,lin2020near,dong2022ground,choi2021rodeo,stetcu2023projection}, including an application to quantum-field-theory state preparation in Ref.~\cite{kane2024nearly}.

\item \emph{Classically-prepared states:} When the state can be efficiently prepared classically, one can use classical routines to find the quantum circuit that encodes that state. Unfortunately, for an exact description of a state, one needs an exponentially large number of amplitudes in system size to be determined and encoded, which is not classically feasible. Hence, one may have to resort to approximate polynomial-size ansatzes. These ansatzes can be inspired by TNs and be optimized using, e.g., DMRG, in lower dimensions, or be guided by other physical considerations such as symmetries, correlation, and Gaussianity, and be optimized using classical Monte Carlo or other classical methods. For some examples in the context of lattice gauge theories, see Refs.~\cite{klco2020minimally,harmalkar2020quantum,gustafson2021toward,gupta2025euclidean}.
\end{itemize}
Several other methods exist too for preparing ground and lowest-lying excited states of generic quantum Hamiltonians. There are also methods for thermal-state preparation. We refrain from expanding this discussion and move to the next stage of quantum simulation: time evolution.

\subsubsection{Time evolution
\label{sec:evolution}}
Time-evolution unitary $e^{-itH}$ with $t \in \mathbb{R}$ is an important operator in quantum simulation. We are often interested in how a non-trivial state evolves is time (consider scattering and collision processes, or thermalization and non-equilibrium dynamics). Time evolution or its imaginary-time partner $e^{-\tau H}$ with $\tau \in \mathbb{R}$ , are often important routines in state-preparation algorithms (consider adiabatic and imaginary-time state preparations), or in observation-estimation schemes (consider the quantum phase estimation for energy spectroscopy or time-dependent correlation functions). Time evolution enjoys proven quantum advantage~\cite{lloyd1996universal}: for $k$-local Hamiltonians, there exists quantum algorithms for estimating the time-evolution operator which scale polynomially in system size, and are efficient in other parameters, such as the simulation time and the approximation error.

\begin{figure}[t!]
    \centering
    \includegraphics[scale=0.675]{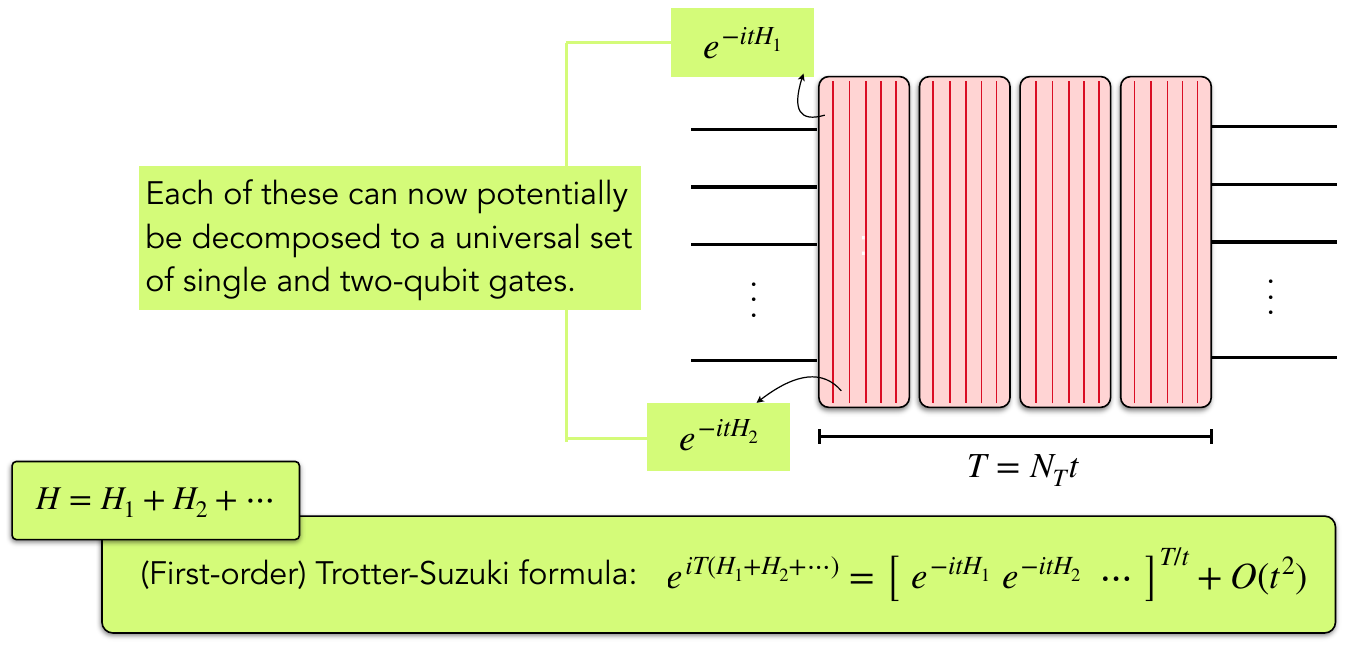}
    \caption{Schematic of a time-evolution circuit via a Trotter-Suzuki product formula.}
    \label{fig:trotter-circuit}
\end{figure}

The most common approach to digitize the time-evolution operator, as depicted in Fig.~\ref{fig:trotter-circuit}, is via product formulas, i.e., the familiar (Lie-)Trotter-Suzuki expansion of the exponential of the sum of operators~\cite{lie1880theorie,trotter1959product,suzuki1976generalized}. The first-order product formula is given by
\begin{align}
V_{1}(t) \coloneq e^{-it H_{1}} e^{-it H_{2} } \ldots e^{-it H_{r}},
\label{eq:V1-def}
\end{align}
and its error is bounded by~\cite{childs2021theory}
\begin{align}
\left\|V_{1}(t)-e^{-i t H}\right\| \leq \frac{t^{2}}{2} \sum_{i=1}^{\Gamma}\Big\|\Big[\sum_{j=i+1}^{\Gamma} H_{j}, H_{i}\Big]\Big\|.
\label{eq:p1-formula}
\end{align}
Here, $H=\sum_{i=1}^{\Gamma}H_{i}$, and is assumed to be Hermitian, as is the case with Hamiltonians. The second-order formula is given by 
\begin{align}
V_{2}(t) \coloneq \left(e^{-i\frac{t}{2} H_{\Gamma}} \cdots e^{-i\frac{t}{2} H_{2}} e^{-i \frac{t}{2}H_{1}}\right) \left( e^{-i\frac{t}{2} H_{1}} e^{-i \frac{t}{2} H_{2}} \cdots e^{-i \frac{t}{2} H_{\Gamma}}\right),
\end{align}
and its error is bounded by~\cite{childs2021theory}
\begin{align}
\left\|V_{2}(t)-e^{-it H}\right\| \leq \frac{t^{3}}{12} \sum_{i=1}^{\Gamma} \Big\|\Big[\sum_{k=i+1}^{\Gamma} H_{k},\Big[\sum_{j=i+1}^{\Gamma} H_{j}, H_{i}\Big]\Big]\Big\|+
\frac{t^{3}}{24} \sum_{i=1}^{\Gamma}\Big\|\Big[H_{i},\Big[H_{i}, \sum_{j=i+1}^{\Gamma} H_{j}\Big]\Big]\Big\|.
\label{eq:p2-formula}
\end{align}
Therefore, at the cost of doubling the number of exponentials to be implemented, the error can be reduced by a factor of $t$.

To estimate the time-evolution operator for the full simulation time $T$ using a $p$-th order product formula, one can implement $V_{p}(t)$ for $N_{T} \coloneq T/t$ times, where $N_T$ is called the number of Trotter steps. The bound on the full error for the $p$-th order product-formula thus becomes:
\begin{align}
\left\|\big({V_p(T/N_T)}\big)^{N_T}-e^{-iTH}\right\| \leq N_{T}\left\|V_{p}\left(T / N_{T}\right)-e^{-i \frac{T}{N_{T}}H }\right\| \leq N_{T}\left(\frac{T}{N_{T}}\right)^{p+1} \alpha_{p}.
\label{eq:bound-for-T}
\end{align}
Here, $\alpha_{p}$ is the (nested) commutator of Hamiltonian terms, which can be read off from Eqs.~\eqref{eq:p1-formula} and \eqref{eq:p2-formula} for the $p=1$ and $p=2$ formulas, respectively, and be found for general $p$ in Ref.~\cite{childs2021theory}.

In order for this error to be bounded by $\epsilon_{T}$, one can set:
\begin{align}
N_{T}=\left(\frac{\alpha_{p}}{\epsilon_{T}} T^{p+1}\right)^{1 / p}.
\end{align}
So for the first-(second-)order formula $N_{T} \propto \frac{T^{2}}{\epsilon_{T}}$ ($N_{T} \propto\frac{T^{3/2}}{\sqrt{\epsilon_{T}}}$). Asymptotically as $p \rightarrow \infty$, one approaches the (optimal) linear $T$ scaling,~\cite{childs2019nearly} but at the cost of $2 \times 5^{\frac{p-2}{2}}$ repetitions of the exponentiated operators in each Trotter step (for even $p$)~\cite{childs2021theory}. So a balance must be found between the error incurred and the cost of each Trotter step. Nonetheless, estimating the algorithmic error will be challenging, since it requires bounding the norm of increasingly higher-order nested commutators of Hamiltonian terms hidden in $\alpha_{p}$. Another challenge is to find the best way to decompose the Hamiltonian to $H_i$ terms. One needs to split the terms to the point that a given $e^{-i t H_{i}}$ is implementable with an elementary set of gates. Splitting further than needed may cause more Trotter error, and potentially break symmetries that better be retained as much as possible.

Once the product-formula scheme is set in place, one needs to implement $e^{-it H_i}$ for each~$i$. Assume the number of qubits a given $H_{i}$ acts on is $n_i$. Decomposing $H_{i}$ into $4^{n_{i}}$ number of Pauli operators in the complete set $\bigotimes_{k=0}^{n_i-1} \{\mathbb{I}_k,\sigma_k^{\mathbf{x}},\sigma_k^{\mathbf{y}},\sigma_k^{\mathbf{z}}\}$ may sound the most straightforward path, but it is not the most computationally feasible: one needs to find the decomposition classically first, which incurs an exponential cost in $n_i$, then implement $4^{n_{i}}$ exponentiated Pauli terms, which will need to be further split and implemented using, e.g., another round of Trotter-Suzuki decomposition, which introduces even more Trotter error. So we may need smarter strategies to get around these problems. Often a better path is to first diagonalize $e^{-i t H_i}$ in the computational basis using unitary $\mathcal{U}_i$, i.e.,
\begin{align}
e^{-it H_{i}}=\mathcal{U}_{i} e^{-it \mathcal{D}_{i}} \mathcal{U}_{i}^{\dagger},
\end{align}
then implement each $\mathcal{U}_{i}$ and diagonal $e^{-it \mathcal{D}_{i}}$ separately. Finding the diagonalizing transformation $\mathcal{U}_i$ may be non-trivial and need classical computing, but sometime tricks based on, e.g., singular-value decomposition of the operator can aid us to do this efficiently~\cite{davoudi2023general}. $e^{-it \mathcal{D}_{i}}$ may be computed by computing the function $\mathcal{D}_{i}$ quantumly in a register, then be exponentiated (via an algorithm broadly known as phase kickback, see the examples in Sec.~\ref{sec:U(1)-warm-up}). This route is asymptotically polynomial in $n_i$ but will need (a polynomial-size) ancillary register, which could be costly in near term. The last resort is to Pauli decompose $\mathcal{D}_{i}$ to $2^{n_{i}}$ terms (only $2^{n_{i}}$ since only $\mathbb{I}$ and $\sigma^{\mathbf{z}}$ operators contribute to the decomposition of diagonal operators in the computational basis). In this case, there would be no Trotter error when implementing the terms separately. The exponential scaling with $n_{i}$ remains though, and it may still be best to take the former route for a sizable $n_i$.\\

Other simulation algorithms exist too. These sometimes go under the name \emph{near-optimal} algorithms. The idea is to find algorithms whose complexity goes as $T$ and $\text{polylog} (1/\epsilon)$. Note that better than linear scaling with time is not generally possible~\cite{atia2017fast,gu2021fast}, as that would mean fast forwarding time evolution in quantum mechanics, for which there is a no-go theorem! Several near-optimal algorithms have been developed is recent years~\cite{childs2012hamiltonian,berry2014exponential,berry2015simulating,berry2015hamiltonian,low2017optimal,low2019hamiltonian,low2018hamiltonian,haah2021quantum}, but they often are more complex and require abundant ancilla qubits (although the number of ancilla qubits still scales polynomially in system size). For example, an algorithm called qubitization~\cite{low2017optimal,low2019hamiltonian} first implements a block encoding of $H$, then the function $e^{-it H}$ of $H$, without the need to break down the $e^{-it H}$ operator to smaller pieces as in product formulas. The algorithm cost, however, goes as $\|H\|$. so if there are large terms in $H$ (recall the $E^{2}$ term in lattice-gauge-theory Hamiltonians, which go as $\Lambda^{2}$ in the irrep basis), then the algorithm becomes costly! In such situations, one can transform to a rotating frame with respect to the large term in the Hamiltonian, hence simulate dynamic in a frame that does not contain the large contribution~\cite{rajput2022hybridized}. This method, nonetheless, requires simulating dynamics of a time-dependent Hamiltonian (given the transformation to the rotating frame), which is slightly more complicated~\cite{low2018hamiltonian,berry2020time,an2022time,fang2025time}. We will not discuss these developments here, but mention an application of a near-optimal algorithm to QCD simulations in the next section.

\subsubsection{Observable measurement
\label{sec:measurements}
}
While quantum computers encode the full wave function, it is not computationally feasible to measure the entire state's amplitudes (i.e., to do full state tomography). This will amount to exponentially many measurements in system size: one needs to perform $2^{n}$ basis rotations to access amplitude of non-diagonal basis states, where $n$ is the number of qubits encoding the state to be measured. These rotations involve applying Clifford operators containing $\sigma^{\mathbf{x}/\mathbf{y}}$ before measuring the state of the $n$ qubits in the computational basis. Such full-state tomography would also require exponentially large classical storage to record the accessed information. Furthermore, often we spend significant quantum resources to create the state and, if we measure it, it will collapse to a set of classical information, losing the ability to perform further quantum tasks on the state. In such situations, it is best to entangle the state with that of some ancillary register, and learn something about the state by measuring the ancilla, without completely destroying the state we have prepared, or at least project it to a particular subspace.

For most physical purposes though, we only need to learn certain information about the state, and so we can get away with having to learn the full state. Quantities one can measure efficiently on quantum computers include:
\begin{figure}
\begin{center}
\includegraphics[scale=0.75]{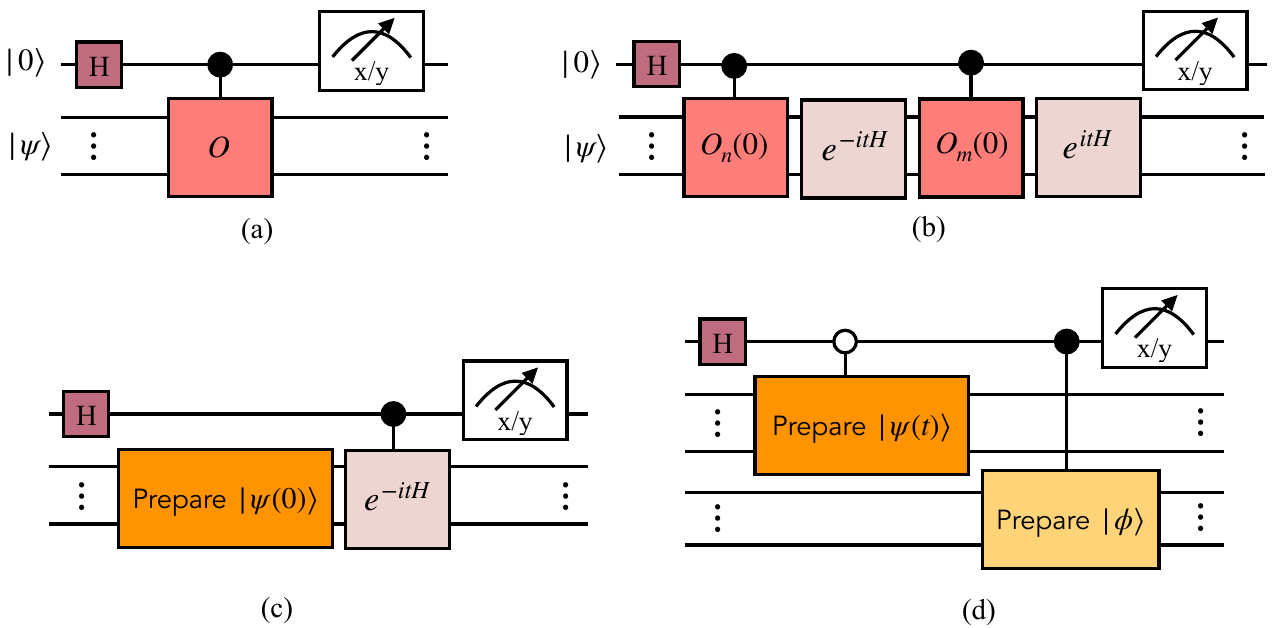}
\end{center}
\caption{Quantum-circuit schematics for measuring (a) the expectation value of operator $O$ in state $\ket{\psi}$, (b) the expectation value of $O_n(t)O_m(0)$ in state $\ket{\psi}$ where $t$ is time and $n$ and $m$ are arbitrary lattice-site indices, (c) the overlap $\langle \psi(0) | \psi(t)\rangle$, and (d) the overlap $\langle \phi | \psi(t)\rangle$. Controlled operation with filled (unfilled) circle means the gate is applied only if the control qubit is in state $\ket{1}$ ($\ket{0}$). The preparation boxes can be non-trivial in general and depend on the state to be prepared. If the operator $O$ is not unitary, additional operations/ancillas are needed to encode it unitarily, see e.g., a trick in Ref.~\cite{jordan2012quantum}. The final uncolored box in all circuits indicates measurement in the $\sigma^{\textbf{x}}$ or $\sigma^{\textbf{y}}$ basis of the ancilla.}
\label{fig:meas-circuits}
\end{figure}
\begin{itemize}
\item Energy and momentum, particle number and charges (both locally and globally): These often amount to measuring a set of local or semi-local operator expectation values. Energy determination can also proceed via more far-term algorithms with the aid of ancilla qubit(s), e.g., via an algorithm called quantum phase estimation~\cite{kitaev1995quantum,nielsen2010quantum}.
\item Various correlation functions (both equal and non-equal time): A useful algorithm, called the Hadamard test~\cite{aharonov2006polynomial}, allows for computing such correlation functions. Let us first see how the expectation value of an operator $O$ in state $\ket{\psi}$ is obtained.  The requisite quantum circuit is depicted in Fig.~\ref{fig:meas-circuits}(a) and involves a single ancilla qubit. Starting from the state $\ket{0}_a\otimes \ket{\psi}$, the circuit yields:
\begin{align}
|0\rangle_{a} \otimes|\psi\rangle & \rightarrow \frac{1}{\sqrt{2}}(|0\rangle_{a}+|1\rangle)_{a} \otimes|\psi\rangle \rightarrow \frac{1}{\sqrt{2}}(|0\rangle_{a} \otimes |\psi\rangle+|1\rangle_{a} \otimes O|\psi\rangle),
\end{align}
which upon measuring the ancilla in the $\sigma^{\mathbf{x}}$ basis gives:
\begin{align}
& \to \frac{1}{2}\Big[|0\rangle_{a} \otimes(\mathbb{I}+O)|\psi\rangle+|1\rangle_{a} \otimes(\mathbb{I}-O)|\psi\rangle\Big].
\end{align}
Therefore, measuring the ancilla obtains $|0\rangle_{a}$ with the probability $\frac{1}{4}\langle\psi|(I+O)^\dagger(I+O)|\psi\rangle$, and $|1\rangle_{a}$ with probability $\frac{1}{4}\langle\psi|(I-O)^\dagger(I-O)|\psi\rangle$. Subtracting the two probabilities gives $\text{Re}(\langle\psi| O|\psi\rangle)$. Similarly, if one instead measures the outcome in the $\sigma^{\mathbf{y}}$ basis, one can reconstruct $\text{Im}(\langle\psi| O|\psi\rangle)$. For two- and higher-point functions, as well as for non-equal time corelation functions, a similar strategy can be applied. For example, the circuit in Fig.~\ref{fig:meas-circuits}(b) implements $\langle \psi | O_{n}(t) O_{n'}(0) | \psi\rangle \equiv \langle \psi | e^{itH} O_{n}(0) e^{-itH} O_{n'}(0) | \psi\rangle $. The number of measurements to reach precision $\epsilon$ in the expected value can be shown to be $O\left(\frac{1}{\epsilon^{2}}\right)$ (using Chernoff bound)~\cite{chernoff1952measure}. This cost can be improved down to $O\left(\frac{1}{\epsilon}\right)$ (using quantum-amplitude estimation~\cite{brassard2000quantum}).

\item Asymptotic $S$-matrices: If the asymptotic final state is reached in a quantum simulation of a scattering process, and the aim is to find the overlap with a particular final state, as in exclusive processes in experiment, then one simply quantum-computes the overlap ${}_\text{in}\langle \psi|\psi\rangle_\text{out}$. The quantum circuit that enables this computation is not too different from that of $\langle\psi|O| \psi\rangle$ explained before. One needs to prepare $|\psi \rangle_\text{out}$ in a separate qubit register and find its overlap with $|\psi \rangle_\text{in}$, which is prepared in another qubit register. Two examples are depicted in Fig.~\ref{fig:meas-circuits}: the circuit that obtains the survival probability (a diagonal element of the $S$-matrix) defined as $\langle \psi(t) | \psi(0) \rangle$ in Fig.~\ref{fig:meas-circuits}(c), and the overlap $\langle \psi(t) | \phi \rangle$ in Fig.~\ref{fig:meas-circuits}(d).

\item Inclusive processes and hadron tensor: Consider the optical theorem applied to, e.g., a current-current correlation function (i.e., hadron tensor), from which inclusive cross sections can be deduced~\cite{peskin2018introduction}. Then recall that non-equal-time amplitudes can be computed with quantum computers, using methods discussed above. So such inclusive amplitudes are accessible in quantum computers, as is the hadron tensor itself. Semi-inclusive amplitudes are more challenging to compute, since they require finding an overlap with a state which is only partially projected to an identified state.

\item Entanglement measures, including entanglement Hamiltonian: Given that quantum computers encode the entire quantum state, one can measure more novel quantities, such as entanglement measures. For example, bipartite von Neumann entanglement entropy, defined as $S_A \coloneq -\text{Tr}[\rho_A \log \rho_A]$ can be computed. Here, $\rho_A$ is the reduced density matrix of the subsystem $A$. Nonetheless, such a computation is costly. One can efficiently measure approximate entanglement spectra instead. Entanglement spectrum is defined as the eigenvalues of the subsystem's entanglement Hamiltonian:
\begin{align}
H_{E} \coloneq -\log \rho_A.
\end{align}
It turned out that $H_{E}$ can sometimes be estimated with polynomial-size resources in system size, despite the fact that $\rho_A$ is an exponentially large matrix. In general, techniques based on randomized measurements~\cite{elben2023randomized} augmented by polynomial-size ansatzes for the entanglement Hamiltonian~\cite{bisognano1976duality,giudici2018entanglement}, can be used to enable learning classical snapshots of the entanglement Hamiltonian, and reconstruct its low-lying spectrum approximately~\cite{kokail2021entanglement}. For non-equilibrium states, the statistical properties of the entanglement spectrum gives access to information such as early onset of chaos and thermalization in the system, see e.g., Refs.~\cite{mueller2022thermalization,mueller2025quantum} for a lattice-gauge-theory example. Entanglement Hamiltonian can also be related to thermodynamical quantities such as the dissipated work in non-equilibrium processes is strongly-interacting systems~\cite{davoudi2024quantum}. 
\end{itemize}
As already mentioned, fidelities and full state tomography are not computationally viable, requiring an exponential number of measurements in system size.

\subsection{Time evolution in gauge theories
\label{sec:LGT-evolution}
}
Time evolution constitutes an important step in quantum simulation, and often is a subroutine to other simulation steps. It would, therefore, be important to learn how to implement time evolution in lattice gauge theories. For concreteness, and to keep the discussions simple and accessible, we first focus on the example of time evolution via product formulas in the U(1) lattice gauge theory in (1+1)D. Here, we provide as much detail as possible to demonstrate the simulation algorithms and circuitry strategies, as well as strategies for performing a rigorous cost analysis. This exercise will set us up to continue with a review of quantum-resource requirements of QCD time dynamics based on state-of-the-art algorithms and analyses that have emerged in recent years.

\subsubsection{In-depth study: (1+1)D U(1) lattice gauge theory
\label{sec:U(1)-warm-up}
}
In Sec.~\ref{sec:KS}, we introduced two equivalent forms of the (1+1)D U(1) lattice gauge theory in the Kogut-Susskind formulation (the lattice Schwinger model), when open boundary conditions are imposed. Furthermore, in Sec.~\ref{sec:mapping}, we introduced the Jordan-Wigner transformation that maps fermionic operators to qubit operators. The fermion-boson form of the Schwinger Hamiltonian after Jordan-Wigner transformation is:
\begin{align}
    H = &~ \text{x}\sum_{n=0}^{N-2}{\left(\sigma^-_nU_n\sigma^+_{n+1}+\text{h.c.}\right)} + \frac{\mu}{2} \sum_{n=0}^{N-1}{(-1)^{n+1}\sigma^{\text{z}}_n} + \sum_{n=0}^{N-2}E_n^2 \\
    \coloneq &~ \sum_{n=0}^{N-2} H_{n,n+1}^{(\text{x})}+H^{(\mu)}+H^{(\text{E})},
    \label{eq:fermion-boson-fomr-repeat}
\end{align}
while the fully fermionic form is
\begin{align}
    H = &~ \text{x}\sum_{n=0}^{N-2}{\left(\sigma^+_n\sigma^-_{n+1}+\text{h.c.}\right)} + \frac{\mu}{2} \sum_{n=0}^{N-1}{(-1)^{n+1}\sigma^{\text{z}}_n} + \sum_{n=0}^{N-2}{\left\{ \sum_{m=0}^n\frac{\sigma^{\text{z}}_m-(-1)^m}{2} \right\}^2}
    \label{eq:fully-ferm-I-repeat}
    \\
    \coloneq &~ \sum_{n=0}^{N-2}H_{n,n+1}^{(\rm XX)}+\sum_{n=0}^{N-2}H^{(\rm YY)}_{n,n+1} + H^{(\rm ZZ)} + H^{(\rm Z)},
    \label{eq:fully-ferm-II-repeat}
\end{align}
where the electric field entering the lattice is set to zero. $H_{n,n+1}^{(\text{x})}$, $H^{(\mu)}$, and $H^{(\text{E})}$ in Eq.~\eqref{eq:fermion-boson-fomr-repeat} constitute all the terms proportional to the coupling $\text{x}$ (involving fermion qubits $n$ and $n+1$), proportional to $\mu$, and the electric Hamiltonian, respectively. In Eq.~\eqref{eq:fully-ferm-II-repeat}, $H_{n,n+1}^{(\text{XX}/YY)}$  constitute all terms proportional to the product of two Pauli-X/Y matrices on qubits $n$ and $n+1$, $H^{(\rm ZZ)}$ are all terms proportional to the product of two Pauli-Z matrices on two distinct qubits, while $H^{(\rm Z)}$ are all terms proportional to a Pauli-Z matrix on a single qubit. The form in Eq.~\eqref{eq:fully-ferm-II-repeat} does not  require boson-encoding overhead but involves long-range interactions among the fermions (qubits).

In this section, we will study the quantum-circuit implementation of the time-evolution operator using product formulas in both formulations, and will compare and contrast their resource requirements. This example make it clear that different Hamiltonian formulations of lattice gauge theories may have different quantum-resource requirements. It also teaches us how to approach the question of resource analysis in general.

Let us start from the fully fermionic form in Eq.~\eqref{eq:fully-ferm-II-repeat}. Various choices exist for the decomposition of Hamiltonian terms into $H=\sum_i H_i$ for the use in the product formulas. For example, one may consider the following first-order product-formula decomposition:
\begin{equation}
    V_1(T) = \prod_{i=1}^{N_T=T/t} \left( e^{-i t H^{(Z)}}e^{-i t H^{(ZZ)}}  \prod_{n=0}^{N-2}e^{-i t H^{(XX)}_{n,n+1}} \prod_{n=0}^{N-2}e^{-i t H^{(YY)}_{n,n+1}}\right).
    \label{eq:V-1}
\end{equation}
Before learning how to circuitize this form, the following exercise guides you to obtain an expectation for the Trotterized time dynamics through numerical methods. 

\tcbset{colframe=black!10!black,colback=mygray,arc=1mm}
\begin{tcolorbox}[breakable]
\noindent
\begin{exercise}
\label{Exercise:ED-Trotter}
The goal of this exercise is to obtain, numerically via exact diagonalization, the Trotterized time evolution of the Schwinger model, to compare with the continuous time evolution you studied in Exercise~\cref{Exercise:ED-Schwinger}. As in that exercise, consider a system of $N=4$ staggered-lattice sites and the following model parameters: $\epsilon_0 = 0$, $\rm{x} = 0.6$, and $\mu = 0.1$.

\vspace{0.25 cm}
\noindent
\textbf{Part (a)} Divide the time evolution for duration $T=5$ into $N_T = 10$ Trotter steps. Hence, the evolution time for each step is $t=0.5$. Apply the first-order Trotter-Suzuki approximation in Eq.~\eqref{eq:V-1} to evolve the strong-coupling vacuum state $\ket{\psi(0)}$ as in Exercise~\cref{Exercise:ED-Schwinger} for time $T$. Plot both the Loschmidt echo, $\mathcal{P}(t)$, and the particle-number density, $\nu(t)$, defined in Exercise~\cref{Exercise:ED-Schwinger} as a function of Trotterized time. Overlay your plots with the continuous time evolution you obtained previously to clearly see any deviation from exact result. Change the Trotter step size to explore how Trotter error responds to this change.

\vspace{0.25 cm}
\noindent
\textbf{Part (b)} Numerically obtain the spectral norm of the difference between the exact and Trotterized evolution for times $t$ and $T=N_Tt$, i.e., $\left\|V_1(t)-e^{-i t H}\right\|$ and $\left\|(V_1(t))^{N_T}-e^{-iT H}\right\|$. Furthermore, compute the error bound, i.e., the right-hand side of Eq.~\eqref{eq:p1-formula} and of Eq.~\eqref{eq:bound-for-T}. Is the exact spectral norm close to the bound value?

\vspace{0.25 cm}
\noindent
\textbf{Part (c)} First show that the total charge operator
\begin{align}
Q \coloneq \sum_{x=0}^{N-a} \left(E(x)-E(x-a) \right)
\end{align}
commutes with the Schwinger-model Hamiltonian. What are the eigenvalues of this operator? Since $[H,Q]=0$, the eigenvalues of $Q$ are conserved quantities: if you start the evolution in a given $Q$-eigenvalue sector, Hamiltonian time dynamics should not take you out of this sector. Does the Trotter-Suzuki expansion we picked in this exercise conserve the total charge $Q$? If yes, argue why. If no, can you come up with a decomposition that conserves the total charge? This example shows that approximate algorithms can break the symmetries of the model!
\end{exercise}
\end{tcolorbox}
%
\begin{figure}[b!]
\begin{center}
\includegraphics[scale=0.505]{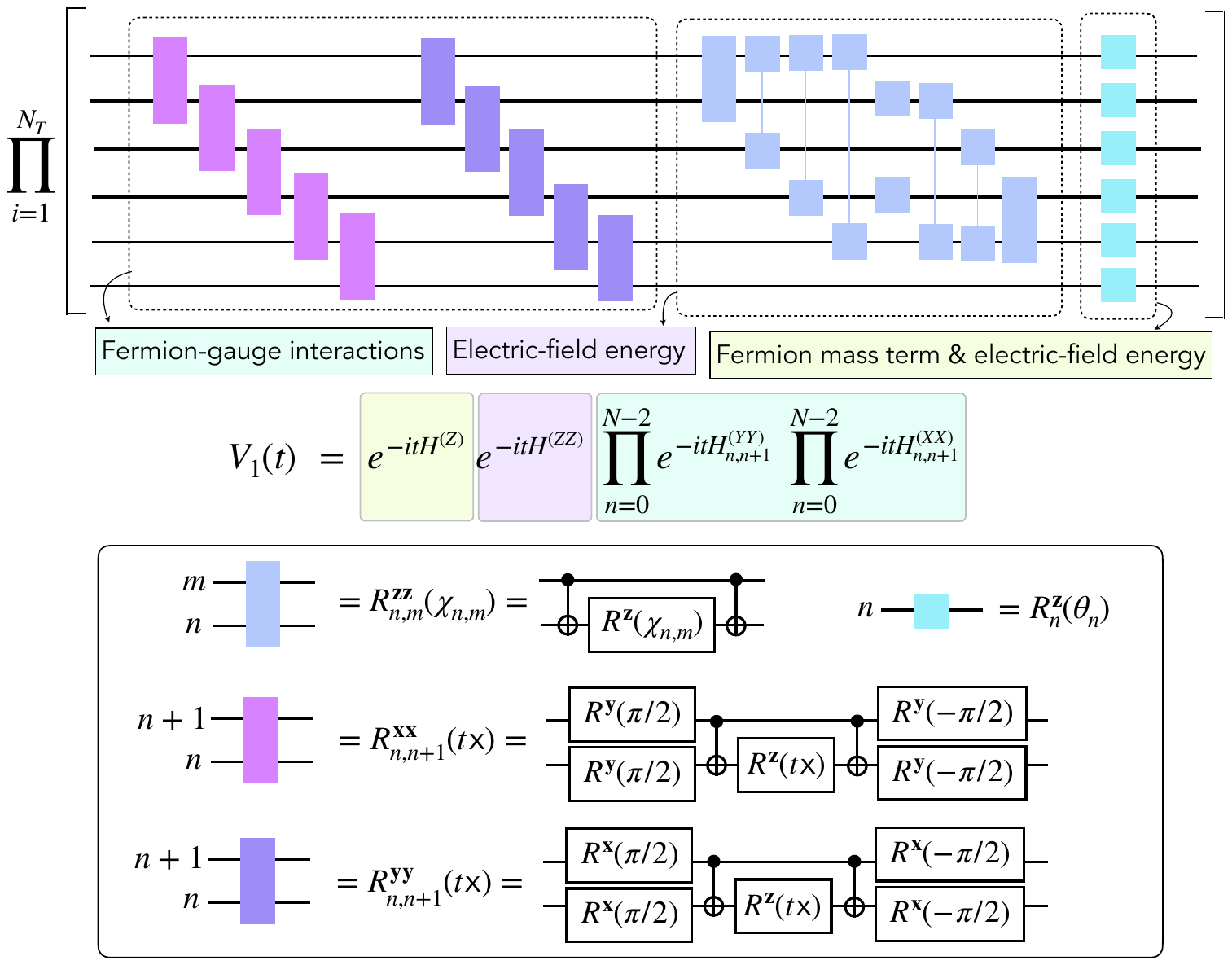}
\end{center}
\caption{The circuit schematic for the first-order product-formula evolution of the fully fermionic form of the lattice Schwinger model in Eq.~\eqref{eq:fully-ferm-II-repeat}. The rotation gates are defined in Sec.~\ref{sec:qubits}, $\theta_n=(-1)^{n+1}t\mu+t\lfloor\frac{N-n}{2}\rfloor$, and $\chi_{m,n}=t(N-1-|m-n|)$.}
\label{fig:schwinger-trotter}
\end{figure}
A schematic of the quantum circuit that implements $V_1(t)$ is shown in Fig.~\ref{fig:schwinger-trotter}. There are only a few types of gates required to implement this decomposition:
\begin{itemize}
\item Single-qubit rotations $R_n^\textbf{z}(\theta_n)$ with $\theta_n \coloneq (-1)^{n+1}t\mu+t\lfloor\frac{N-n}{2}\rfloor$. 
\item Two-qubit rotations $R_{n,m}^\textbf{zz}(\chi_{m,n})$ with $\chi_{m,n}\coloneq t(N-1-|m-n|)$, which can be implemented using two CNOT gates and one single-qubit rotation $R_{n}^\textbf{z}(\chi_{m,n})$, as demonstrated in Fig.~\ref{fig:schwinger-trotter}.
\item Two-qubit rotations $R_{n,n+1}^\textbf{xx/yy}(t\text{x})$, which can be turned to $R_{n,n+1}^\textbf{zz}(t\text{x})$ rotations upon appropriate single-qubit rotations around the $\mathbf{y}/\mathbf{x}$ axes of the Bloch sphere.
\item Single-qubit rotations $R_n^\textbf{y}(\pm\frac{\pi}{2})$ and $R_n^\textbf{x}(\pm\frac{\pi}{2})$ required to do the basis transformation in the previous item, as demonstrated in Fig.~\ref{fig:schwinger-trotter}.
\end{itemize}

Clearly this form requires $N$ qubits to encode the quantum circuit, $O(N^2)$ two-qubit rotation due to the (nearly) all-to-all nature of the interactions in $H^{(\text{ZZ})}$, and similarly $O(N^2)$ one-qubit rotations per Trotter step (if the two-qubit rotations are converted to the CNOT and single-qubit gates). Exercise~\cref{Exercise:circuit-qiskit} allows you to put into practice what you have learned so far to simulate, on a quantum-hardware emulator, time dynamics of the lattice Schwinger model in its fermionic form.

\tcbset{colframe=black!10!black,colback=mygray,arc=1mm}
\begin{tcolorbox}[breakable]
\noindent
\begin{exercise}
\label{Exercise:circuit-qiskit}
For this exercise, we continue to work with the qubit Hamiltonian in Eq.~\eqref{eq:fully-ferm-II-repeat}, associated with the lattice Schwinger model with open boundary conditions. We further consider Trotterized time evolution via the first-order Trotter-Suzuki formula using the decomposition in Eq.~\eqref{eq:V-1}. The initial state, $\ket{\psi(0)}$, of this evolution is the strong-coupling vacuum, and the time-evolved state is $\ket{\psi(t)} \coloneq V(t)\ket{\psi(0)}$. The system size and parameters are as in the previous exercise: $N=4$, $\epsilon_0 = 0$, $\rm{x} = 0.6$, $\mu = 0.1$, $t=5$, $\delta t=0.5$ (or $N_T=10$).

\vspace{0.25 cm}
\noindent
\textbf{Part (a)} Write a \texttt{Qiskit}~\cite{qiskit2024} code that implements your circuit in part (a). If you are not familiar with \texttt{Qiskit}, use the IBMQ tools and documentations~\cite{qiskit}.

\vspace{0.25 cm}
\noindent
\textbf{Part (b)}
Evaluate the Loschmidt echo $\mathcal{P}(t)$ and the particle-number density $\nu(t)$ as defined in Exercise~\cref{Exercise:ED-Schwinger}. To do this, run the circuit 1,000 times on an emulator (not the actual hardware) and measure the outcome in the computational basis (the Pauli-Z basis). Then reconstruct the observables using these measurements. Note that the labeling in \texttt{Qiskit} is such that the first ($0$-th) qubit represents the least significant digit. Explore the dependence of observables on the number of shots by changing the shot count. In particular, you should be able to see the convergence to the numerically evaluated Trotterized time evolution in the previous exercise by increasing the shot count. Think about how you would you assign statistical uncertainty to the observables.

\vspace{0.25 cm}
\noindent
\textbf{Part (c) [Bonus]} Make appropriate measurements to obtain the energy of the time-evolved state: 
\begin{align}
    E(t) \coloneq \langle \psi(t) | H | \psi(t) \rangle,
\end{align}
at $t=5$. To do this, you need to measure all the Hamiltonian operators and sum them up. For non-diagonal operators in the computational basis, i.e., $H_{n,n+1}^{(\rm XX/YY)}$, you need to first perform a change of basis, then measure the associated qubits. Compare your results with the numerically evaluated result. Importantly, note that energy is a conserved quantity, so any deviation from the initial state's energy should be attributed to the errors in the implementation. Explore the dependence of the error on the shot count and Trotter error.
\end{exercise}
\end{tcolorbox}

Simulation in the fermion-boson form requires dealing with bosonic degrees of freedom. We work in the electric basis and encode the integer electric-field eigenvalues $-\Lambda \leq E_n \leq \Lambda$ in binary. To do so, one can first shift the eigenvalues by $-\Lambda$ such that $E'_n \coloneq E_n+\Lambda \geq 0$. Using $\eta \coloneq \lceil \log(2\Lambda+1) \rceil$ qubits per lattice link $n$, the shifted electric-field operator
\begin{align}
\hat{E}'_n = \frac{1}{2}(2^\eta-1)\hat{\mathbb{I}}-\frac{1}{2}\sum_{j=0}^{\eta-1}2^j\hat{\sigma}^z_{n,j} 
\end{align}
acting on $\left|E'_{n}\right\rangle=\bigotimes_{j=0}^{\eta-1}\left|E'_{n, j}\right\rangle$ returns $E_n+\Lambda$. Here, $E'_{n}=\sum_{j=0}^{\eta-1}2^j E'_{n, j}$ with $E'_{n, j} \in \{0,1\}$. The index $j$ refers to the $j$-th qubit within the $\eta$-sized register, and the index $n$ identifies the link associated with that register.

To simulate the Hamiltonian in Eq.~\eqref{eq:fermion-boson-fomr-repeat}, one can adopt the following first-order Trotter decomposition:
\begin{equation}
    V_1(T) = \prod_{i=1}^{N_T=T/t} \left( e^{-i t H^{(\mu)}} e^{-i t H^{(E)}}  \prod_{n=0}^{N-2}e^{-itH_{n,n+1}^{(\text{x})}}\right).
    \label{eq:V-1-fb}
\end{equation}
\begin{figure}[t!]
\begin{center}
\includegraphics[scale=0.625]{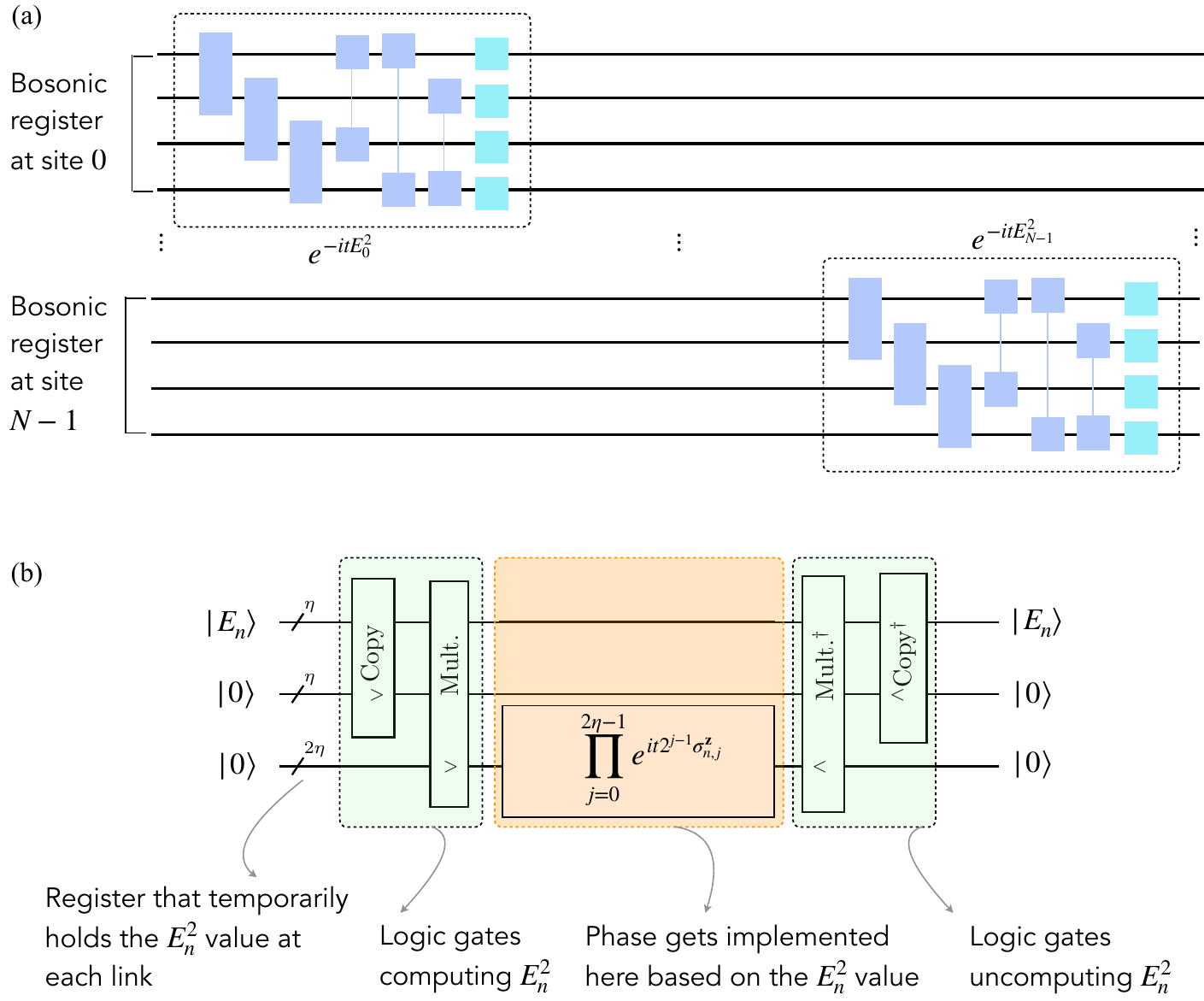}
\end{center}
\caption{Quantum-circuit implementation of $e^{-itE_n^2}$ in the lattice Schwinger model in the fermion-boson form using (a) a near-term and (b) a far-term strategy. The exact gate angles in (a) can be obtained considering Eq.~\eqref{eq:E-squared}. The notation for the gate is as that introduced in Fig.~\ref{fig:schwinger-trotter}. The Copy and Multi. subcircuits can be constructed from the discussions in Lemma A.3 of Ref.~\cite{davoudi2023general}. The $>$ ($<$) notation indicates that the result of the corresponding circuit block is stored in (erased from) the qubit register indicated by $>$ ($<$).}
\label{fig:electric-circuit}
\end{figure}
Simulating $e^{-i t H^{(\mu)}}$ is trivial and amounts to $O(N)$ single-qubit $R^{\mathbf{z}}$ rotations. Simulating $e^{-i t H^{(E)}}=\prod_{n=0}^{N-2}e^{-i t E_n^2}$ can proceed by noting that $E_n^2$ at each link is
\begin{align}
\hat{E}^2_n &= 
\left[-\Lambda \mathbb{I}+
\frac{1}{2}(2^\eta-1)\hat{\mathbb{I}}-\frac{1}{2}\sum_{j=0}^{\eta-1}2^j\hat{\sigma}^z_{n,j}\right]^2\nonumber\\
&= A\hat{\mathbb{I}}+(\Lambda-2^{\eta-1}+2^{-1})\sum_{j=0}^{\eta-1}2^j\sigma_j+\sum_{j=0}^{\eta-1}\sum_{j' \neq j}2^{j+j'-2}\hat{\sigma}^z_{n,j}\hat{\sigma}^z_{n,j'},
\label{eq:E-squared}
\end{align}
where $A$ is an immaterial constant (contributing a constant phase to the evolution). Implementing an exponentiation of this operator, therefore, requires $O(\eta)$ single-qubit $R^\textbf{z}$ gates and $O(\eta^2)$ $R^\textbf{zz}$ gates, as schematically shown in Fig.~\ref{fig:electric-circuit}(a). The total cost amounts to $O(N\eta)$ single-qubit $R^\textbf{z}$ gates and $O(N\eta^2)$ $R^\textbf{zz}$ gates, since there are $N-1$ such $e^{-itE_n^2}$ to be implemented. Converting the $R^\textbf{zz}$ rotations to a combination of CNOT gates and single-qubit $R^\textbf{z}$ rotations, as shown in Fig.~\ref{fig:schwinger-trotter}, results in an $O(N\eta^2)$ $R^\textbf{z}$-gate cost for the circuit. These circuits obviously do not require any ancilla qubits, hence the qubit-encoding cost is $N\eta$

It is possible to reduce the $O(N\eta^2)$ cost for the $R^{\mathbf{z}}$ gates to $O(N\eta)$, but at the expense of increasing the qubit cost by a factor of four. The procedure goes as follows. We can use ancillary qubits to evaluate and store the value of $E_n^2$ on a $2\eta$-sized register, then compute $e^{-itE_n^2}$ directly on this register, on which $E_n^2$ have the operator form
\begin{align}
\hat{E}_n^2 = 
-\Lambda \mathbb{I}+
\frac{1}{2}(2^{2\eta}-1)\hat{\mathbb{I}}-\frac{1}{2}\sum_{j=0}^{2\eta-1}2^j\hat{\sigma}^z_{n,j}. 
\end{align}
The exponentiated operator on each link, therefore, only requires $O(\eta)$ $R^\textbf{z}$ gates. A schematic circuit is shown in Fig.~\ref{fig:electric-circuit}(b). One first copies the value of the electric field in an ancillary $\eta$-sized register, then uses a multiplication routine to obtain $E^2_n$. Once the phase $e^{-itE_n^2}$ is evaluated and imprinted in the circuit's wave function, one proceeds to rest the ancilla registers to their original all-$\ket{0}$ state. Such an irreversible quantum circuit is a useful strategy since the ancilla qubits can be reused in subsequent computation, for example, to implement the next $e^{-itE_m^2}$ operator for $m \neq n$. The exact details of the copy and multiplication circuits do not matter for this discussion. Their explicit form can be found in literature, see, e.g., Lemma A.3 of Ref.~\cite{davoudi2023general}. They require at most $O(\eta^2)$ T gates per $e^{-itE_n^2}$ implementation. Nonetheless, this is a tolerable cost compared with the $R^{\textbf{z}}$ gate-synthesis cost, which requires $O(\log(1/\epsilon))$ T gates to be implemented with precision $\epsilon$~\cite{paetznick2013repeat}.

This examples above demonstrate how near- and far-term considerations can change our circuit-design strategies. In the near term, we prioritized simpler circuits that do not demand ancilla qubits, while in the far term, we reduced the costly continuous-angle single-qubit rotations but allowed for the use of ancilla qubits. The ancillary register's size is still polynomial in $\eta$, and the ancilla qubits can be reused in subsequent simulations.

Next, and most non-trivial, is the implementation of each $e^{-itH^{(\text{x})}_{n,n+1}}$. This is a non-diagonal operator in the electric-field basis, and hence cannot be decomposed to only strings involving Pauli-Z operators. We can adopt the algorithm of Ref.~\cite{davoudi2023general} to first transform this operator to a diagonal form using a singular-value decomposition, then implement the diagonal operator using the same strategy as with the electric Hamiltonian. Explicitly, noting that $U_n=\sum_{E_n=-\Lambda}^{\Lambda-1} |E_n+1\rangle\langle E_n |$, it is straightforward to show that $H$ can be decomposed into:
\begin{align}
H^{(\text{x})}_{n,n+1}= \sigma_n^+U_n\sigma_{n+1}^-+\text{h.c} =\mathcal{U}_n^\dagger \,\left(\sigma^{\textbf{x}}_n\mathcal{P}^{(0)}_{n+1}\mathcal{D}_n\right)\,\mathcal{U}_n,
\label{eq:Hx-SVD}
\end{align}
with the diagonalizing transformation
\begin{align}
\mathcal{U}_n \coloneq \mathsf{H}_n\left(\mathcal{P}^{(0)}_n \sigma^{\textbf{x}}_{n+1}\lambda^+_n+\mathcal{P}^{(1)}_n\right),
\label{eq:diag-transform}
\end{align}
and diagonal bosonic operator at link $n$,
\begin{align}
\mathcal{D}_n \coloneq \sum_{E_n=-\Lambda}^\Lambda |E_n\rangle\langle E_n |.
\label{eq:Dn-def}
\end{align}
Here, $\mathcal{P}^{(0)/(1)}_n$ is the projector to the $\ket{0/1}$ state of the fermion qubit at site $n$. Furthermore, we have defined the cyclic incrementer $\lambda_n^+\ket{E_n}=(1-\delta_{E_n,\Lambda}) \ket{E_n+1}+\delta_{E_n,\Lambda}\ket{-\Lambda}$ (and decrementer $\lambda_n^-\ket{E_n}=(1-\delta_{E_n,-\Lambda}) \ket{E_n-1}+\delta_{E_n,-\Lambda}\ket{\Lambda}$). These definitions lead to an  unphysical transition between the UV modes $\ket{\pm\Lambda}$, and can be canceled out in the quantum circuit using additional CNOT operations, see Ref.~\cite{davoudi2023general} for details. Such definitions are, however, useful in quantum-circuit decomposition of incrementers.

\begin{figure}[t!]
\begin{center}
\includegraphics[scale=0.625]{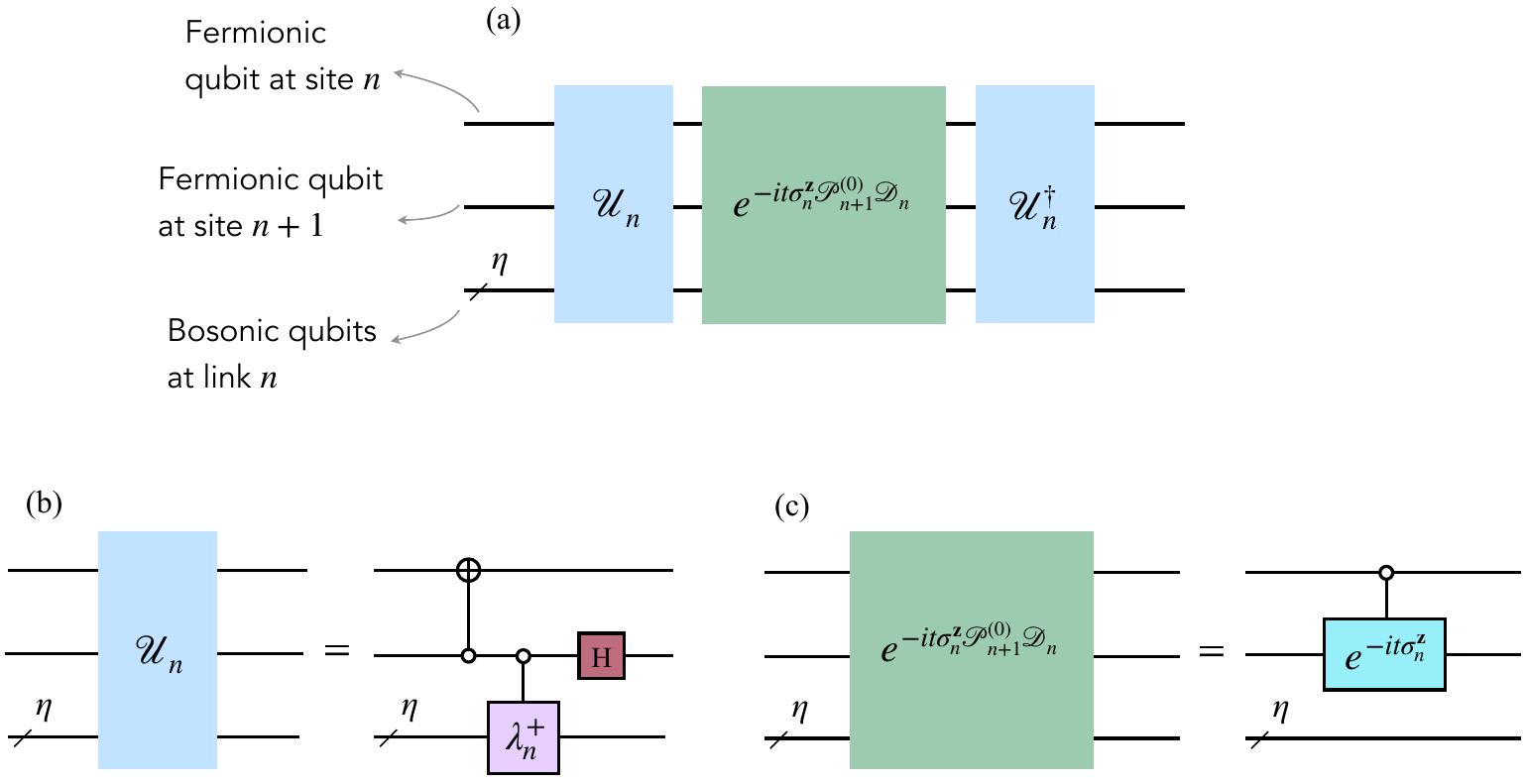}
\end{center}
\caption{(a) Quantum-circuit implementation of $e^{-itH^{(\text{x})}_{n,n+1}}$ in the lattice Schwinger model in the fermion-boson form using a singular-value-decomposition of the operator, i.e., Eq.~\eqref{eq:Hx-SVD}. (b) The unitary transformation defined in Eq.~\eqref{eq:diag-transform}. (c) The diagonal piece of the operation, with the operator $D_n$ defined in Eq.~\eqref{eq:Dn-def}. H denotes the Hadamard gate and $\lambda_n^+$ is an integer incerementer in binary, which can be implemented using algorithms described in Lemmas A.1 and A.2 of Ref.~\cite{davoudi2023general}.}
\label{fig:hopping-circuit}
\end{figure}

Given the decomposition in Eq.~\eqref{eq:Hx-SVD}, one can implement 
\begin{align}
e^{-itH^{(\text{x})}_{n,n+1}}=\mathcal{U}_n^\dagger \,e^{-it\sigma^{\textbf{z}}_n\mathcal{P}^{(0)}_{n+1}\mathcal{D}_n}\,\mathcal{U}_n
\end{align}
according to the schematic circuit shown in Fig.~\ref{fig:hopping-circuit}. There are at least two ways one can implement the binary incrementer in a quantum circuit. One is with a near-term approach, which involves no ancilla qubit and uses a quantum-Fourier-transform routine~\cite{coppersmith2002approximate,nielsen2010quantum}. It requires $O(\eta^2)$ CNOT gates. The more far-term approach uses only $O(\eta)$ T gates but requires $\eta-3$ ancilla qubits. For details of these algorithms, see Lemmas A.1 and A.2 of Ref.~\cite{davoudi2023general}. Finally, there are $N-1$ such terms to be implemented per Trotter step.

With the circuit analysis of the Schwinger model in each of the formulations complete, we now turn to the question of the relative cost of the simulation in each formulation. Two factors determine the cost: the cost of each Trotter step, and the total number of Trotter steps. We have already determined the Trotter-step cost from the circuit analyses above. Let us now determine the number of Trotter steps in each formulation. For concreteness, consider a second-order product formula, and recall that the number of Trotter steps required to reach accuracy $\epsilon$ for total evolution time $T$ is $N_T=(\alpha_2T^3/\epsilon)^{1/2}$. Here, $\alpha_2$ is the second-order commutator error bound from Eq.~\eqref{eq:p2-formula}. To arrive at an estimate of the size of $\alpha_2$, one can consider only the commutators that contribute the largest norm in the large-$N$ limit, where $N$ is the lattice size.

For the fully fermionic form, there are $O(N^2)$ Hamiltonian terms in the product-formula decomposition, and these arise from the non-local fermion-fermion interactions. It is then easy to see that the largest norm of the non-commuting nested commutator of Hamiltonian terms arises roughly from those of the form $[H^{(\text{ZZ})},[H^{(\text{ZZ})},H_{n,n+1}^{(\text{XX/YY})}]]$. This yields a norm of $O(N^5)$. So the number of Trotter steps $N_T$ goes as $O(N^{5/2}T^{3/2}\epsilon^{-1/2})$. Considering that each Trotter step in the fully fermionic form has a gate complexity $O(N^2)$, we arrive at a full gate complexity of $O(N^{9/2}T^{3/2}\epsilon^{-1/2})$.

\begin{figure}[t!]
\begin{center}
\includegraphics[scale=0.675]{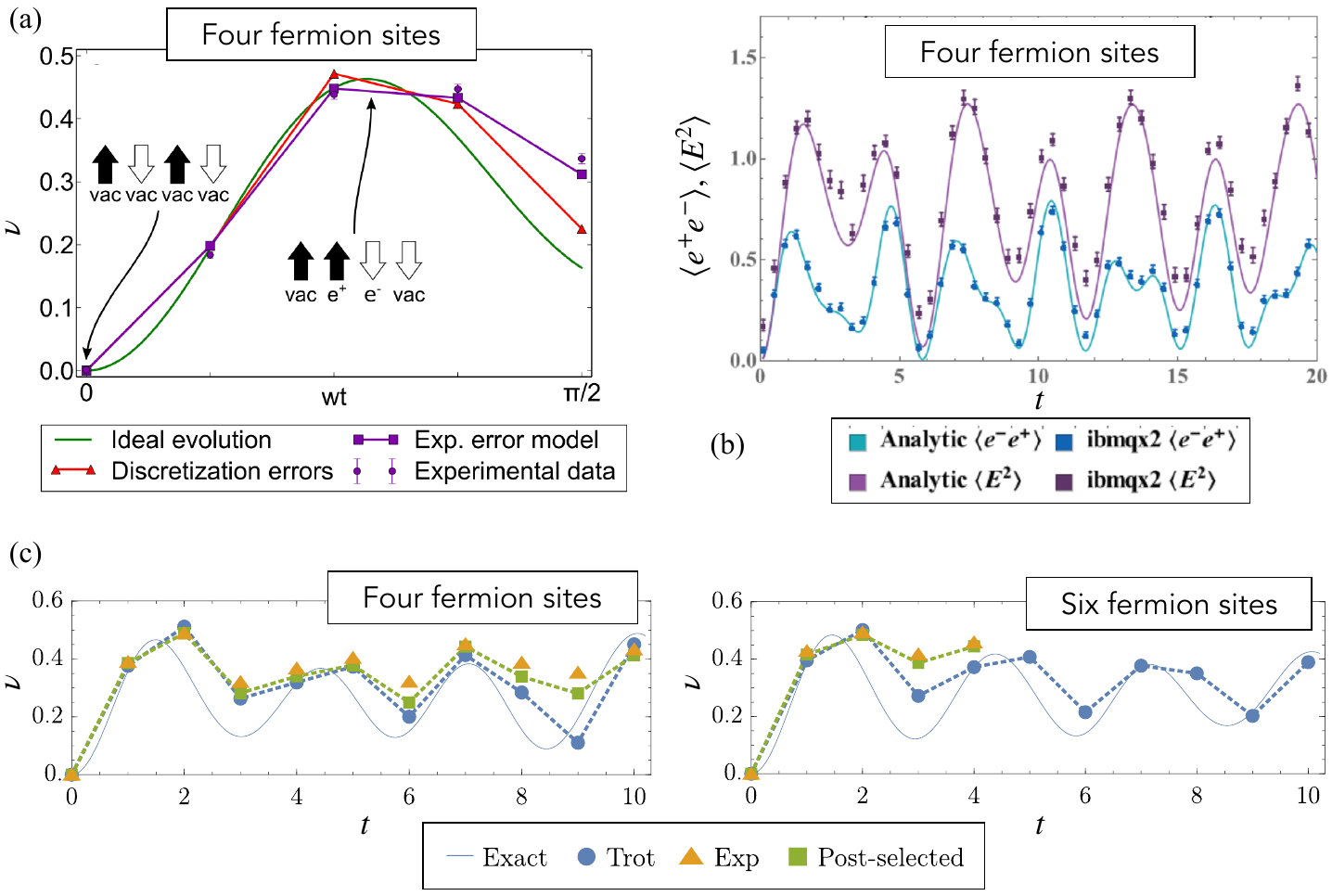}
\end{center}
\caption{Examples of the quantum simulations of time dynamics in the lattice Schwinger model on a quantum computer. The initial state in all plots is the strong-coupling vacuum, the evolution is performed via a first-order product formula, and the quantity plotted is the fermion density defined in Eq.~\eqref{eq:nu-def}. The expectation value of the electric operator is further plotted in (b). The plot in (a) is obtained using a trapped-ion quantum computer and is based on the fully fermionic form of the Schwinger model~\cite{martinez2016real}. The plot in (b) is obtained using a superconducting quantum computer and involves a simulation of truncated lattice Schwinger model with periodic boundary conditions in the physical Hilbert space~\cite{klco2018quantum}. The plots in (c) are obtained using a trapped-ion quantum computer with the fully fermionic form of the model~\cite{nguyen2022digital}. For further details, including the description of the legends, see Refs.~\cite{martinez2016real,klco2018quantum,nguyen2022digital} from which these plots are reproduced. The four-fermion-site plot in part (c) corresponds to the exact same simulation parameters as in Exercises~\cref{Exercise:ED-Schwinger}, \cref{Exercise:ED-Trotter}, and \cref{Exercise:circuit-qiskit}, hence it should serve as a reference for your solutions!}
\label{fig:schwinger-sim}
\end{figure}

For the fermion-boson form, on the other hand, there are only $O(N)$ Hamiltonian terms in the product-formula decomposition, since the Hamiltonian is local. It is then easy to see that the commutator error bound has a norm of $O(N^3)$. So the number of Trotter steps $N_T$ goes as $O(N^{3/2}T^{3/2}\epsilon^{-1/2})$. Considering that each Trotter step in the fermion-boson form has a gate complexity $O(N)$, we arrive at a full gate complexity of $O(N^{5/2}T^{3/2}\epsilon^{-1/2})$, which is clearly an improvement compared to the fully fermionic form. Note that in this asymptotic scaling analysis, we have assumed that $\eta \ll N$, and have hence dropped dependence on $\eta$. With open boundary conditions, the largest required electric-field cutoff $\Lambda$ to recover the full theory is $N/2$ assuming a zero background field. Since in the binary encoding $\eta=O(\log\Lambda)$, the above assumption is valid.

A few comments are in order. First, the above asymptotic scaling ignores dependence on other parameters, such as lattice spacing, couplings, and electric-field cutoff (for the fermion-boson formulation). For example, in the continuum limit which (in our dimensionless units) corresponds to $\text{x} \to \infty$, gate complexity's dependence on $\text{x}$, which asymptotically goes as $x^{3/2}$, matters too. A complete error analysis bounds the full commutator norm $\alpha_2$ to obtain dependence on all model's parameters, as carried out in Refs.~\cite{shaw2020quantum,kan2021lattice} for the fermion-boson form. These analyses, nonetheless, do not account for theory errors, such as finite-size, discretization, and truncation effects. Second, the commutator error-bound analysis can be quite loose, as it relies on repeated application of triangle (Cauchy–Schwarz) inequalities. It is known that empirically, the product formulas perform much better than the analytical bounds on them indicate~\cite{childs2019nearly,stetina2022simulating,nguyen2022digital}. For example, Ref.~\cite{nguyen2022digital} finds an empirical dependence on the system size for the second-order product formula that is roughly $O(N^2)$ smaller than the bounds we described above, using exact numerical evaluation of the bound for feasible system sizes in the fully fermionic form of the Schwinger model. Analytical error bounds, nonetheless, are still useful in most situations, since empirical bounds are not often known or are hard to obtain.

The example of the lattice Schwinger model discussed in this section allowed us to learn the art of quantum-circuit design, practice quantum-resource analysis, and understand the importance of considering equivalent formulations of the same theory and assessing their relative simulation costs. The lattice Schwinger model has indeed served as a rich testing ground for many quantum-simulation algorithms and experiments in recent years. To honor some of the pioneering quantum-simulation experiments of time evolution in the Schwinger model, we showcase some of their results in Fig.~\ref{fig:schwinger-sim}.

\subsubsection{An overview: (3+1)D SU(3) lattice gauge theory
\label{sec:QCD}
}
What does it take to simulate QCD on a digital quantum computer? We do not have a full answer to this question yet, but we have some partial answers. A full answer requires understanding resource requirements for preparing non-trivial states in QCD. As already mentioned, many state-preparation algorithms are heuristic, and performance guarantees are hard to achieve in general. The success probability of most far-term, near-optimal algorithms, in turn, depends on the overlap between the initial state and the target state under preparation. While we know how to prepare vacuum, hadronic, wave packets, and thermal states in simpler low-dimensional field theories using primarily near-term algorithms in smaller systems, knowing the costs of preparing, e.g., a proton at TeV energy scales for Large-Hadron-Collider simulations, or of an atomic nuclei for nuclear-reaction simulations, are so far unknown to us. The simulation cost also depends on the type of observables we are after, and the precision goals.

Furthermore, we have not yet fully quantified all systematic theory and algorithmic uncertainties. Theory uncertainties include finite-volume effects, discretization effects, truncation and digitization effects, etc. These have been studied so far in limited contexts (see examples in Refs.~\cite{carena2021lattice,carena2022improved,briceno2021role,burbano2025real,jordan2012quantum,tong2022provably,kane2025obtaining}), but need to be properly and holistically applied to the QCD Hamiltonian-simulation problem. Algorithmic errors include time-digitization errors in product formulas, or errors relevant to other beyond-Trotter algorithms, function-evaluation errors, observable-estimation and measurement (finite statistic) errors, etc. In the end, all theory and algorithmic errors must be combined rigorously. Importantly, setting algorithmic accuracy without regard for theory systematics can lead to unrealistic resource estimations.\\

For all these reasons, let us only focus of one aspect of the simulation algorithm, time evolution, for which concrete algorithms with bounded errors exist. Time evolution is also a routine in several state-preparation and observable-estimation algorithms. Therefore, costing this step can give us a hint at how costly these other simulation tasks are. Obviously, the simulation cost also depends the Hamiltonian formulation one chooses for QCD. Here, we focus on the Kogut-Susskind formulation of Hamiltonian lattice gauge theory in the electric-field basis, for which more comprehensive cost analysis is achieved in literature. Simulations are mostly based an product-formula algorithms, but an analysis based on near-optimal algorithms has also appeared recently, which is included in the following comparative review. For some of related algorithmic development, see Refs.~
\cite{byrnes2006simulating,shaw2020quantum, ciavarella2021trailhead,kan2021lattice,lamm2019general,haase2021resource,davoudi2023general,murairi2022many,rhodes2024exponential,lamm2024block,balaji2025quantum,halimeh2025universal}. Here, we focus only on a few works that allow for a more apple-to-apple comparison (extracted partly from an upcoming paper~\cite{davoudi2025qcd}).

\vspace{0.25 cm}
\noindent
\emph{Product-formula-based algorithms.} Recall that in the Kogut-Susskind formulation in the electric-field basis, the plaquette term is a complicated operator with off-diagonal matrix elements. Its cost asymptotically dominates the full simulation cost of $e^{-it H_{\text{KS}}^{\text{QCD}}}$, so it suffices to only focus on this term to get an estimate of the overall cost.
 
In a pioneering work~\cite{byrnes2006simulating}, Byrnes and Yamamoto discussed a possible mapping of the U(1) and SU($N_c$) lattice gauge theories to qubit-based quantum computers, and mentioned binary and unary mapping of the gauge-boson degrees of freedom, and the Jordan-Wigner mapping of the fermions. While they considered the product-formula simulations of the time-evolution operator, their analysis fell short of a full resource estimation. They, nonetheless, correctly stated that time evolution of QCD using quantum algorithms demands resources that are only polynomial in system's volume.

At a first attempt, one may adapt a naive Pauli decomposition of the operators in $e^{-itH_i}$ in the Trotter-Suzuki expansion of the full Hamiltonian. There are $N_p$ plaquettes to be implemented, and further, each plaquette $\mathcal{P}$ involves $\text{Tr}(UUU^\dagger U^\dagger)+\text{h.c.}$, which constitutes $N_{c}^{4}$ terms, with $N_{c}=3$ for a SU(3) lattice gauge theory. Each $U$ acts on a register of size $O\left(\left(N_{c}^{2}-1\right) \log \Lambda\right)$, hence $\mathcal{P}$ acts on an $O\left(4\left(N_{c}^{2}-1\right) \log \Lambda\right)$-qubit register. The number of Pauli strings that span this space is, therefore, $O\left(4^{4\left(N_c^{2}-1\right) \log\Lambda}\right)$, giving $O\left(\Lambda^{8\left(N_{c}^{2}-1\right)}\right)$ exponentiated Pauli strings to be implemented. Consider as an example a small cutoff $\Lambda=4$. The number of exponentiated Pauli strings to be implemented for each plaquette term is $4^{64} \simeq 10^{38}$, which is gigantic! Even more challenging is the need for the classical computing involved in finding the decomposition at the first place, which would be intractable. Additionally, one should not forget about the number of Trotter steps, which is set by the accuracy goal, and depends on the amount of non-commutations among the exponentiated Hamiltonian terms. Implementing Pauli strings requires additional decomposition, hence Trotter error. In the worst case,
a second-order product formula requires $N_{T} \propto \nu^{5/2}$ Trotter steps, where $\nu$ is the number of terms to be implemented [recall the error-norm bound in Eq.~\eqref{eq:p2-formula}]. This inflates the cost of the simulation to $O(10^{95})$ operations for the example above!

To improve the situation, Kan and Nam adopt in Ref.~\cite{kan2021lattice} the block-diagonalization technique of Shaw et al~\cite{shaw2020quantum} to implement each $e^{-it H_i}$ as a whole, instead of first Pauli decompose the $H_{i}$ operators. Here, each $H_{i}$ is a term that uniquely changes the $N_c^2-1=8$ quantum numbers $\{\Gamma,\lambda_L,\Lambda_R\}\equiv\left\{p, q, T_{L}, T_{L}^{z}, T_{R}, T_{R}^{z}, Y_{L}, Y_{R}\right\}$ on each link of a plaquette (recall the discussions in Sec.~\ref{sec:SU(3)-basis}). It turned out that there are at most $\left(12 \times 3 \times 2^{N_{c}^{2}-1}\right)^{4} \simeq 2^{52} \simeq 10^{15}$ terms, independent of $\Lambda$. The factor of $(12 \times 3)^{4}$ is the number of terms in the trace in the plaquette term, and the factor of $\left(2^{N_{c}^{2}-1}\right)^{4}$ comes from the fact that the block diagonalization algorithm of Shaw et al~\cite{shaw2020quantum} requires breaking the operators acting on each of the $N_{c}^{2}-1$ bosonic Hilbert space to two pieces such that each piece only acts on half of the Hilbert space of each quantum number. The total time complexity of the second-order product formula goes as $N_{T} \propto \nu^{3/2}$ at best~\cite{kan2021lattice}, giving a factor of $10^{23}$. Finally, each term's implementation cost still depends on $\Lambda$ but any as $\text{polylog}(\Lambda)$, which is less severe than the naive Pauli-decomposition approach.

Unfortunately, this is still not the full cost. One needs to implement all the terms in the lattice volume. Importantly, each of the terms requires its own dedicated quantum circuit. The most expensive ones are the ones requiring $e^{-it f(U)}$, where $f$ is some function of the link operator $U$ (this could arise from a hopping term or a plaquette term). These involve SU(3) Clebsch-Gordan coefficients, see Eq.~\eqref{eq:U-SU3}. They can be precomputed classically and stored in a quantum memory (CROM), and accessed as needed. This method yields a large overhead since there are $O\left(\Lambda^{5}\right)$ terms in each $U$ term, hence $O\left(N^{20}\right)$ calls to the memory to implement the plaquette term, which is significant~\cite{rhodes2024exponential}. Alternatively, these Clebsch-Gordan coefficients can be evaluated on the fly, using quantum arithmetic, up to function-evaluation errors (think about Newton's method for evaluating functions)~\cite{kan2021lattice,davoudi2023general,rhodes2024exponential}. This turns out to be only requiring $\text{polylog}(\Lambda)$, with a small polynomial exponent, but still a non-negligible cost, given that this evaluation has to be done for every non-diagonal operator, and every Trotter step.

All carefully analyzed and put together, Kan and Nam obtain~\cite{kan2021lattice}:
\begin{align}
O\left(\nu^{3/2} V^{3/2} T^{3/2} \epsilon^{-1/2} \Lambda \, \text{polylog} \left(V^{3/2} T^{3/2} \Lambda \epsilon^{-3/2}\right)\right).
\end{align}
This is the cost of time evolution with a second-order product formula for both SU(2) and SU(3) lattice gauge theories in the Kogut-Susskind formulation in the electric-field basis with bosonic cutoff $\Lambda$, total evolution time $T$, spatial volume $V$, and accuracy $\epsilon$. How does this cost depend on the Hamiltonian parameters $m$, $g$, and $a$? This dependence is hidden in various places, including in the $\alpha_{p}$ function, which is the commutator hound on the Trotter formula, see Eq.~\eqref{eq:p2-formula}. As one takes the continuum limit, some of the Hamiltonian-term coefficients grow large, increasing the cost, but only polynomially in $\frac{1}{a}$. For example, in the limit $a \to 0$ and $g(a) \to 0$ for QCD, the dependence of the number of Trotter steps on these parameters goes as $a^{-3/2}g^{-3}$~\cite{kan2021lattice}.

How many costly T gate does this all amount to? Let us consider a $V=10 \times 10 \times 10$ simulation. For $\epsilon$ ranging from $10^{-1}$ to $10^{-3}$, lattice spacing ranging from 1 to $10^{-2}, \Lambda=10$, $m=g=10$, and an evolution time comparable to the lattice extent, the T-gate complexity ranges from $10^{49}-10^{53}$~\cite{kan2021lattice}! The total number of qubits, which is also an important factor, is $\sim 10^{10}$~\cite{kan2021lattice}! Quantum-chemistry simulations (precise molecular-energy computations) have been estimated to have a $10^{10}-10^{15}$ T-gate complexity~\cite{dalzell2023quantum} while nuclear effective field theory simulations ( dynamics of medium-mass atomic nuclei) demand $10^{12}-10^{24}$ T-gate complexity~\cite{watson2023quantum,roggero2020quantum}, so we need to improve on the QCD estimates significantly to be competitive with other disciplines. One should, however, keep in mind though that despite quantum-chemistry and nuclear-physics problems, quantum-field-theory problems are relativistic in nature, and significantly more demanding computational resources for QCD simulations should not be surprising.\\

It turned out that this is not the end of the story, and a significant cost reduction is possible with improved algorithms. For example, one can already strip off $\sim 14$ orders of magnitude from the simulation cost of Kan and Nam. The idea is to use the singular-value-decomposition algorithm of ZD, Shaw, Stryker~\cite{davoudi2023general} (in place of Shaw et al~\cite{shaw2020quantum}) to block diagonalized the off-diagonal Hamiltonian terms~\cite{davoudi2025qcd}. This would require only splitting one of the $N_c^2-1$ bosonic quantum-number's Hilbert space to two pieces (where each piece acts on disjoint subspaces). This yields only up to a factor of $(12 \times 3)^{4} \times 2 \simeq 10^{6}$ terms, or a total cost $\nu^{3/2} \simeq 10^{9}$, which is to be compared with a factor of $\simeq 10^{23}$ in Kan and Nam, hence a $\sim 10^{14}$ improvement in gate complexity~\cite{davoudi2025qcd}.

Another approach is to only encode physical degrees of freedom. This can be particularly useful in the case of a SU(3) gauge theory, where there are $N_{c}^{2}-1$ registers of size $\log \Lambda$ (with binary encoding), but such a large Hilbert space is mostly redundant. Nonetheless, such an approach amounts to classical preprocessing to obtain physical degrees of freedom, which even if developed, would lead to a highly non-local Hamiltonian, which may give rise to higher T-gate complexity (recall the cost comparison in the Schwinger-model example in the two formulations in Sec.~\ref{sec:U(1)-warm-up}). Alternatively, one can find and encode only physical transitions locally, as proposed by Ciavarella, Klco, and Savage~\cite{ciavarella2021trailhead}. This means that physics can be encoded only in terms of the irrep quantum numbers $(p,q)$, which is far more efficient. This approach, nonetheless, amounts to classically finding and storing the values of physical matrix elements, then hardcoding them in the quantum circuit. For a pure SU(3) lattice gauge theory and with only trivalent vertices (a ribbon of 1D plaquettes), it turned out that the are $O\left(\Lambda^{12}\right)$ number of such transitions to be encoded~\cite{ciavarella2021trailhead,ciavarella2022some}. A full cost analysis of the time-evolution operator given an accuracy goal still needs to be done to enable comparison with Refs.~\cite{kan2021lattice,davoudi2025qcd}.

\vspace{0.25 cm}
\noindent
\emph{Near-optimal algorithms.}
Can the QCD simulation cost be improved by resorting to post-Trotter methods, such as near-optimal algorithms? The answer is yes. In Ref.~\cite{rhodes2024exponential}, Rhodes, Kreshchuk, and Pathak employ a qubitization algorithm~\cite{low2017optimal,low2019hamiltonian} for QCD time dynamics. Qubitization amounts to first block encoding the Hamiltonian, then evaluating functions of Hamiltonian, such as the time-evolution operator. Since one needs to only circuitize the Hamiltonian as apposed to the exponentiated Hamiltonian, the maximum number of terms is only $O(N_{c}^{4})$ for the plaquette term (which is roughly the number of terms in the trace), avoiding the large exponential factor in $N_{c}^{2}$ in the Kan and Nam approach. One first uses a sparse-Hamiltonian evaluation approach, which amounts to finding where the non-zero elements of the Hamiltonian matrix are, then evaluating those elements. This latter requires computing the Clebsch-Gordan coefficients on the fly. Both steps scale poly-logarithmically in $\Lambda$ (in binary encoding of the bosonic degrees of freedom). \\

Another technicality in this algorithm is that it involves working in a rotating frame with respect to $H_{\text {KS,E}}$ [defined in Eq.~\eqref{eq:H-KS-E}]. One then has to simulate a time-dependent-Hamiltonian dynamics: $e^{-i\int^t H_{\text{rot}}(t') dt'}$ with $H_{\text{rot}}(t)=e^{itH_{\text{KS},E}}\left(H_\text{KS}-H_{\text{KS,E}}\right)e^{-itH_{\text{KS,E}}}$. This can be done via a truncated Dyson-series expansion~\cite{low2018hamiltonian,berry2020time}. Such an approach prevents the algorithm error to scale as $\Lambda^2$, which is an important factor in reducing the cost. Rhodes, Kreshchuk, and Pathak further adopt a local encoding of fermions~\cite{verstraete2005mapping,setia2019superfast} to avoid the need for long non-local Jordan-Wigner Pauli strings.\\

All put together, Rhodes, Kreshchuk, and Pathak report the T-gate complexity:
\begin{align}
O\left(N_c^4 V T \text{polylog}\left(V T \epsilon^{-1}\right)\right)
\end{align}
for simulating time evolution in QCD for evolution time $T$, system volume $V$, and error tolerance $\epsilon$. When applied to the example outlined above, i.e., a $V=10 \times 10 \times 10$ simulation, with $\epsilon$ ranging from $10^{-1}$ to $10^{-3}$, lattice spacing ranging from 1 to $10^{-2}, \Lambda=10$,, $m=g=10$, and an evolution time comparable to the lattice extent, one obtains $10^{27}-10^{28}$ T gates, which is a great improvement compared to Kan and Nam $\left(10^{49}-10^{53}\right)$ and compared to ZD and Stryker $\left(10^{35}-10^{40}\right)$. The qubit costs remains nearly the same between these approaches.

These estimates, unfortunately, point to unfeasibly large qubit and gate resources for simulating QCD time dynamics. The above examples, nonetheless, show that the field of quantum-algorithm development for QCD (and Standard-Model) simulations is vibrant and timely, and has led to many orders of magnitude improvement in cost estimates compared to early analyses. We need to continue to develop increasingly more efficient algorithms, not only for time evolution, but also for state preparation and observable estimation, and to do so for a range of Hamiltonian-formulation choices, to find the most efficient routes forward. 

As a last word to end this section, one needs to recall that QCD simulations using classical computers (for problems not suffering from a sign problem) consume fractions of Exascale machines ($10^{18}$ floating-point operations per second) per year and petabytes ($10^{15}$ bytes) of memory in current times. It should, therefore, not be a surprise that quantum simulations of the same theory would eventually require quantum Exascale machines and beyond! Large-scale fault-tolerant quantum computers are not here yet, but once they arrive, QCD physicists should be ready to execute their algorithms. Therefore, shedding off a few more orders of magnitudes from our algorithms' time complexity remains essential in a likely competitive quantum-computing era.

\section{Summary and conclusions
\label{sec:summary}}
\noindent
These notes constituted a gentle introduction to, and in places a light review of, the topic of quantum computing gauge theories, with a focus on $U(1)$, SU(2), and SU(3) lattice gauge theories. These notes should have helped building sufficient knowledge and skill set to appreciate the differences between path-integral-based and Hamiltonian-based methods; to work with at least one popular choice of Hamiltonian formulation for U(1) and SU($N_c$) lattice gauge theories, namely the Kogut-Susskind staggered form, and to identify its Hilbert space and basis states; to practice numerical Hamiltonian methods for small instances of the problems; to recognize the various steps of a quantum simulation and different simulation modes; to become familiar with a few leading state-preparation, time-evolution, and observable-estimation techniques; to realize the types of states that can be prepared on quantum computers and the types of observables that can be accessed for nuclear- and particle-physics applications; to circuitize time evolution in simple gauge theories using product formulas, analyze their algorithmic error, and obtain an estimate of their resource requirements in both near- and far-term eras of quantum computing; and finally to become aware of the state-of-the-art algorithmic studies of QCD.

These notes, nonetheless, only scratch the surface of the extensive, vibrant, and multi-faceted field of quantum simulation/computing for nuclear and particle physics. On the theoretical side, we did not touch on the various other formulations of Hamiltonian gauge theories distinct from the Kogut-Susskind formulation, and we did not give a comprehensive account of basis choices and Hamiltonian processing. We also did not discuss systematic uncertainties, including finite-volume, discretization, and boson-truncation uncertainties, renormalization and continuum limit, and matching to physical observables. On the algorithmic side, we only focused on product-formula algorithms for time evolution and did not present details of post-Trotter methods. Furthermore, we left out a complete discussion of, and examples for, interacting vacuum, hadronic, wave-packet, and thermal states' preparation in gauge theories. We did not offer explicit examples of accessing spectrum, expectation values, and $S$-matrices in gauge theories. Importantly, we did not review a plethora of explorations on various quantum hardware in recent years, from scattering and structure quantities, to thermalization and other aspect of non-equilibrium dynamics. Last but not least, we left out discussions of analog and hybrid analog-digital simulations of gauge theories, and of an exciting hardware co-design program for gauge-theory applications. For starters, the interested reader can consult a range of reviews in these topics in recent years, such as Refs.~\cite{banuls2020simulating,klco2022standard,bauer2023quantumnature,bauer2023quantum,di2024quantum,beck2023quantum}.

These notes, nonetheless, should have placed an interested reader in a good position to start taking the next steps in learning and mastering this topic, to read and appreciate the relevant literature, and to engage in research in various aspects of this fast-moving program. Many open questions still remain, on the theoretical, algorithmic, and implementation and co-design fronts. Fresh minds and new ideas are, therefore, much welcome at this exciting frontier of nuclear- and particle-physics research!

\section*{Acknowledgment}
\noindent
I am grateful to all the students who participated in the lectures covered in these notes over the past year, asked numerous thoughtful and curious questions, and shared their comments and feedback on the lecture material, which helped me converge to a more focused and coherent story over time. I especially thank Navya Gupta, Chung-Chun Hsieh, and Vinay Vikramaditya for their help in co-presenting these lectures at the LGT4HEP Traineeship course in Fall 2025, and for their contribution to the development of some of the exercises in these notes. I further thank Christopher Kane and Jinghong Yang, who reviewed these notes carefully and provided valuable feedback. Last but not least, I am grateful to the organizers of the schools and lecture series in which these lectures were presented, and for their support and encouragement to publish the notes. These colleagues include Nathaniel Craig, Tongyan Lin, Jesse Thaler, Oliver DeWolfe, Ethan Neil, and Tom DeGrand (TASI 2024); Justus Tobias Tsang, Michele Della Morte, Matteo Di Carlo, Felix Erben, Andreas Jüttner, Simon Kuberski, and Alexander Rothkopf (CERN LFT School 2024); Huey-Wen Lin (LGT4HEP 2024), Nora Brambilla and Alice Smith-Gicklhor (Origins Cluster Visitor Lecture Series 2025); Morten Hjorth-Jensen and Alessandro Roggero (ECT* DTP-TALENT 2025).

My research in quantum simulation for gauge theories, which enabled the development of these lecture series, has been supported over the years by the following grants: the U.S. Department of Energy (DOE), Office of Science, Early Career Award (award no. DESC0020271); the National Science Foundation (NSF) Quantum Leap Challenge Institutes (QLCI) (award no. OMA2120757); the U.S. DOE, Office of Science, Office of Nuclear Physics, via the program on Quantum Horizons: QIS Research and Innovation for Nuclear Science (award no.  DESC0021143 and DESC0023710); the U.S. DOE, Office of Science, Office of Advanced Scientific Computing Research Quantum Computing Application Teams program, under fieldwork proposal number ERKJ347; the U.S. DOE, Office of Science, Office
of Advanced Scientific Computing Research, Accelerated Research in Quantum Computing program award DESC0020312 and Fundamental Algorithmic Research toward Quantum Utility (FARQu); the Simons Foundation through the Simons Foundation Emmy Noether Fellows Program at the Perimeter Institute for Theoretical Physics; and the Department of Physics, Maryland Center for Fundamental Physics, and College of Computer, Mathematical, and Natural Sciences at the University of Maryland, College Park.

I finally thank the hospitality of the Kavli Institute for Theoretical Physics at the University of Santa Barbara, and the Excellence Cluster ORIGINS at the Technical University of Munich, where parts of these lectures notes were drafted in LaTeX form and finalized. Research at KITP was supported in part by the NSF grant PHY-2309135. Research at ORIGINS was supported by the Deutsche Forschungsgemeinschaft
(DFG, German Research Foundation) under Germany’s
Excellence Strategy (EXC-2094—390783311).

\bibliography{bibi.bib}


\end{document}